\documentclass[aps,pra,twocolumn,showpacs,superscriptaddress,groupedaddress]{revtex4-2}

\usepackage{amsmath,braket,amssymb,natbib,graphicx,float,amsthm,xcolor,outlines,mathrsfs,multirow,hhline}
\usepackage{multirow}
\usepackage{mathtools}
\usepackage{bm}

\usepackage[english]{babel}
\usepackage[utf8x]{inputenc}
\usepackage[T1]{fontenc}
\usepackage{listings}
\usepackage[colorinlistoftodos]{todonotes}
\usepackage[colorlinks=true, allcolors=blue]{hyperref}
\usepackage{pifont}

\begin{document}
\title{Tripartite multiphoton Jaynes-Cummings model: Analytical solution and Wigner nonclassicalities}

\author{Pradip Laha}
\email{plaha@uni-mainz.de}
\affiliation{Institute of Physics, Johannes-Gutenberg University of Mainz,  Staudingerweg 7, 55128 Mainz, Germany}
\author{P. A. Ameen Yasir}
\email{apooliab@uni-mainz.de}
\affiliation{Institute of Physics, Johannes-Gutenberg University of Mainz,  Staudingerweg 7, 55128 Mainz, Germany}
\author{Peter van Loock}
\email{loock@uni-mainz.de}
\affiliation{Institute of Physics, Johannes-Gutenberg University of Mainz,  Staudingerweg 7, 55128 Mainz, Germany}


\begin{abstract}
We investigate a generic tripartite quantum system featuring a single qubit interacting concurrently with two quantized harmonic oscillators via nonlinear multiphoton Jaynes-Cummings (MPJC) interactions. Assuming the qubit is initially prepared in a superposition state and the two oscillators are in arbitrary Fock states, we analytically trace the temporal evolution of this tripartite pure initial state. We identify four broad cases, each further divided into two subcases, and derive exact analytical solutions for most cases. Notably, we obtain perfect transfer of excitations between the oscillators by carefully selecting system parameters. In addition, we extensively examine the manner in which the nonclassicalities of various initial oscillator Fock states, quantified by the volume of negative regions in the associated Wigner functions, evolve under the MPJC Hamiltonian, considering diverse system parameters including environmental effects. Besides producing substantial enhancements in the initial value for higher photon number states, our analysis reveals that driven solely by the initial qubit energy, with both oscillators initialized in the vacuum state, the nonlinear MPJC interaction yields a significant amount of nontrivial Wigner negativity in the oscillators. The additional nonlinearity introduced by the multiphoton process plays a pivotal role in surpassing the initial nonclassicalities of the photon number states. 
\end{abstract}
\maketitle



\section{\label{intro}Introduction}
Since its introduction to depict the nonlinear dynamics of a two-level atom interacting with a single quantized mode of a cavity field~\cite{jaynes_cummings_1963}, the paradigmatic Jaynes-Cummings (JC) model has transcended its original scope to become a fundamental framework for understanding various phenomena in quantum optics and quantum information processing (see Ref.~\cite{Larson_JCM_2021} for a recent comprehensive review). The significance of the JC model lies in its capacity to accurately describe and predict a plethora of phenomena in the aforementioned areas~\cite{Shore_JCM_1993,mm2013,bina2012,sasaki96}. Its elegant formulation not only provides deep theoretical insights into the dynamics of light-matter interactions at the most fundamental level but also serves as a cornerstone for the development of quantum technologies, including quantum computing, and quantum simulations~\cite{GAF2014,azuma2011,Li2022}. Experimentally, the JC model has been successfully demonstrated across various quantum platforms that use, for instance, atoms and optical cavities~\cite{Rempe_PRL_1987,Boca_PRL_2004,Birnbaum_nature_2005}, Rydberg atoms and microwave cavities~\cite{Brune_PRL_1996,Walther_2006,Lee_PRA_2017}, superconducting qubits and microwave resonators~\cite{Deppe_nat_phys_2008,Fink_nature_2008,Hofheinz_nature_2009} or acoustic resonators~\cite{Martin_science_2014,Manenti_nat_commun_2017,Bienfait_science_2019}, ion traps~\cite{Meekhof_PRL_1996,Leibfried_RMP_2003,Lara_PRA_2005}, quantum dots~\cite{delValle_PRB_2009,Kasprzak_nat_mat_2010,Basset_PRB_2013}, and graphene~\cite{Dora_PRL_2009}.

Researchers have extended their inquiries beyond the standard bipartite JC model to explore its multiphoton version, characterized by an interaction Hamiltonian featuring terms proportional to powers of bosonic creation and annihilation operators. This extension enables the investigation of diverse quantum phenomena, including field statistics and squeezing dynamics, thereby offering valuable insights into quantum systems' responses to multiphoton processes (see Refs.~\cite{SUKUMAR_pla_1981,surendra_pra_1982,SHUMOVSKY_pla_1987,Kien_PRA_1988,LuHuai_Chin_phys_2000,El_Orany_job_2003,El_Orany_jpa_2004,Villas_PRL_2019} for details). Recently, an existing analogy between supersymmetric quantum mechanics and the standard bipartite JC model has been generalized to include multiphoton interactions~\cite{Rodriguez_sci_rep_2021}.

Another straightforward approach to generalizing the bipartite JC model involves incorporating additional degrees of freedom. A well-studied example is the standard tripartite JC model, where a single two-level system (qubit) interacts simultaneously with two bosonic modes (oscillators)~\cite{Dutra_PRA_1993,Dutra_PRA_1994,Benivegna_JPhysA_1994,Abdalla_opt_commun_2002,Messina_j_mod_opt_2003,Wildfeuer_PRA_2003,larson_jmo_2006,Strauch_PRL_2010,Li_PRA_2012,Ma_PRA_2014,Rodriguez_jphysa_2016,Rodriguez_sci_rep_2018,Alderete_2021,laha_thermally_2022}. The model has been used to generate various two-mode entangled photon number states, including NOON states $\frac{1}{\sqrt{2}}\left(\ket{n, 0}+\ket{0, n}\right)$ and maximally entangled $n$-photon states $\frac{1}{\sqrt{n+1}}\sum_{k=0}^n\ket{k, n-k}$~\cite{Wildfeuer_PRA_2003,Strauch_PRL_2010}, as well as maximally entangled coherent states~\cite{larson_jmo_2006,Ma_PRA_2014}. The spectra and eigenstates of the system have also been analyzed~\cite{Rodriguez_jphysa_2016}. The presence of an additional constant of motion due to symmetry enables a canonical transformation of the degenerate Hamiltonian, reducing it to a form where only one JC interaction remains involving the symmetric normal mode, while the antisymmetric mode decouples, a feature leveraged in several prior studies~\cite{Dutra_PRA_1993,Dutra_PRA_1994,Benivegna_JPhysA_1994,Abdalla_opt_commun_2002,Messina_j_mod_opt_2003,Wildfeuer_PRA_2003,larson_jmo_2006,Rodriguez_jphysa_2016,Rodriguez_sci_rep_2018,Alderete_2021}.

In this article we present a generalization of the standard tripartite JC model, where the qubit simultaneously interacts with both oscillators via $m$-photon JC interactions. Notably, this generic Hamiltonian configuration has not been examined in the existing literature within the scope of our investigations. Our primary focus lies in understanding the temporal evolution of a tripartite pure initial state under such a highly nonlinear Hamiltonian. Specifically, we analyze scenarios where the qubit exists in an arbitrary superposition of its two basis states, while the oscillators occupy arbitrary photon number states. Interestingly, we outline four broad cases, each further divided into two subcases, and provide exact analytical solutions for the majority of these scenarios. This comprehensive examination sheds light on the intricate dynamics governed by this generalized multiphoton Jaynes-Cummings (MPJC) model.

With the rapid advancement of quantum technologies, there is a critical need to devise methods and architectures that enable the transfer of excitations, crucial for quantum information processing and communication~\cite{Kuzmich_PRL_2000,Wang_PRL_2012,Palomaki_nature_2013,Takeda_PRL_2015,Maleki_optics_commun_2021}. Specifically, developing quantum SWAP gates is essential for faithfully transferring arbitrary quantum states between different nodes in a quantum network. In recent years, various powerful and elegant methods have emerged to enable robust quantum-state transfer in a variety of two-level qubits~\cite{Matsukevich_science_2004,Northup_nature_2014,Kurz_nat_commun_2014,Kurpiers_nature_2018,Li_PRApplied_2018,Bienfait_science_2019,Liu_jetp_lett_2023}. However, compared to optomechanical systems~\cite{Wang_PRL_2012,Weaver_nature_commun_2017,Ventura_sci_rep_2019,Qi_opt_lett_2020,Lei_appl_phys_b_2023}, relatively few studies have explored continuous variable approaches demonstrating the transfer of arbitrary bosonic states in optical systems~\cite{Maleki_optics_commun_2021} and superconducting circuits~\cite{Sun_PRA_2006,Mei_PRA_2018}. 
In this work, by analytically tracking the temporal evolutions of the two oscillator states in a deterministic manner, i.e., by tracing over the relevant subsystems, we show the possibility of perfect excitation transfer between two oscillators. The quantum model and the ensuing excitation transfer scheme discussed herein can be realized in optical~\cite{laha_spinboson_2024} and superconducting circuits.

It is well known that the higher photon number states are highly nonclassical in the sense that the associated Wigner functions traverse negative regions in the phase space. Previous works have shown that the degree of nonclassicality of a quantum state can be quantified by the volume of the negative region of the associated Wigner function~\cite{Anatole_joptb_2004,Arkhipov_sci_rep_2018,rosiek_arxiv_2023}. Therefore, the manner in which the nonclassicality of the initial Fock states dynamically evolves under such a nonlinear MPJC Hamiltonian remains an interesting task. In the latter part of this article, we address this issue in detail, considering diverse system parameters, including environmentally induced effects.
Our analysis reveals substantial enhancements in the initial volume of the Wigner negativities for higher photon number states for some specific cases. Notably, driven solely by the initial qubit energy, with both oscillators initialized in the vacuum state, the nonlinear MPJC interaction induces nontrivial Wigner negativities in the oscillators. Further, examining all four cases, we conclude that the multiphoton parameter $m$ plays a crucial role in surpassing the initial nonclassicalities of the photon number states. In particular, we find that the parameter $m$ should be at least greater than the mean photon number of one of the oscillators to achieve higher than the initial volume of the negative regions of the Wigner functions. 

\begin{figure*}[ht]
\centering
\includegraphics[width=0.95\textwidth,height=10.2cm]{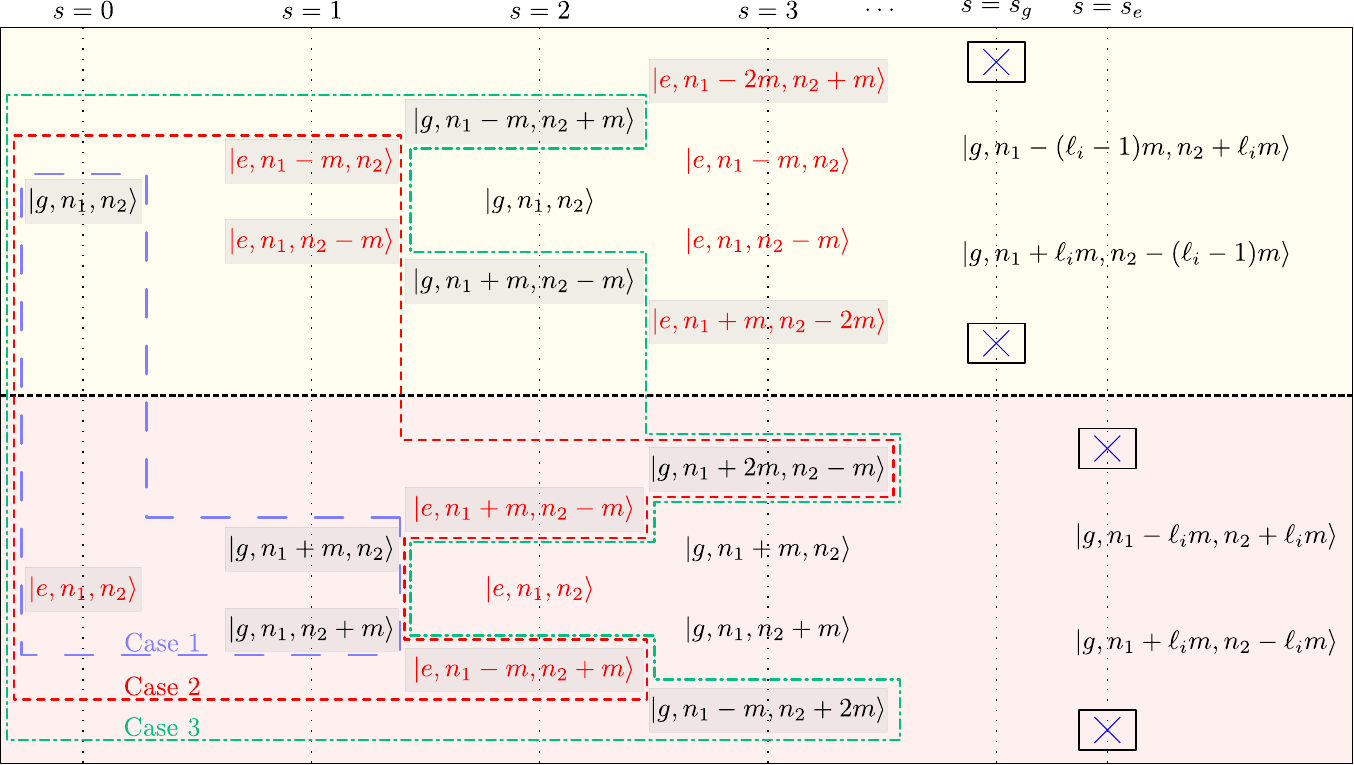}
\vspace{1ex}
\caption{For the tripartite $m$-photon Jaynes-Cummings model [$H$ in Eq.~\eqref{eqn_mpjcm}], featuring one qubit initially in a superposition state and two quantized harmonic oscillators occupying arbitrary Fock states $\ket{n_1}$ and $\ket{n_2}$, respectively, the linear increase in the number of basis states with increasing $n_1$ and $n_2$ is schematically illustrated. The basis states emerging from $\ket{g, n_1, n_2}$ and $\ket{e, n_1, n_2}$ constitute two complementary sets, as shown in the upper and lower panels. When $H$ acts on $\ket{g, n_1, n_2}$, it generates two new basis states $\ket{e, n_1-m, n_2}$ and $\ket{e, n_1, n_2-m}$, each contributing one additional basis state. This process results in two distinct arms, where at every second step (denoted by $s=1,\,3,\,5,\ldots$), the number of photons in one of the oscillators is reduced by $m$. The generation of one new basis state at each arm continues until one of the Fock states of that arm is completely annihilated (indicated by blue crosses corresponding to $s=s_g$ and $s_e$). For a fixed $n_i$, the total count of new basis states (shown within the rectangular boxes) is $2\ell_i-2$, where $\ell_i$ is the smallest positive integer for which $n_i-\ell_i m<0$ ($i=1,\,2$). Similarly, $\ket{e, n_1, n_2}$ gives rise to a total of $2\ell_i$ new basis states at each arm. Consequently, the cumulative number of basis states amounts to $4(\ell_1+\ell_2-1)$. The relevant basis states for the three special cases are depicted within the blue, red, and green dashed regions, respectively.}
\label{fig_schem_states}
\end{figure*}

The rest of the article is organized as follows: In the following section, we introduce the tripartite MPJC model Hamiltonian. In Sec.~\ref{sec:solution}, we analytically explore how a pure initial state of the full system temporally evolves under such a genuinely nonlinear system Hamiltonian. In Sec.~\ref{sec:state_transfer}, we obtain the reduced states of the oscillators and analyze the efficacy of excitation transfer between the two oscillators for various cases. In Sec.~\ref{sec:wigner}, we investigate the complex dynamics of the Wigner nonclassicalities of various initial oscillator states considering a wide range of system parameters. The crucial role of various environmentally induced effects is analyzed in Sec. ~\ref{subsec:wigner}. In Sec.~\ref{sec:conc} we summarize our article and indicate avenues for further research. Additionally, we include a set of Appendixes augmenting the results presented in this work.



\section{\label{sec:model}The tripartite MPJC  Hamiltonian}
As mentioned in the preceding section, we are going to analyze a tripartite quantum system comprising one qubit and two oscillators described by the usual free energy Hamiltonian $H_\text{free} =  \frac{\omega_0}{2}\sigma_z  + \omega_1 a_1^{\dagger}a_1  + \omega_2 a_2^{\dagger}a_2,$
where $\sigma_z$ is the standard Pauli spin operator and $\omega_0$ is the difference in energy between the two levels, $\ket{g}$ and $\ket{e}$ respectively, of the qubit. On the other hand, $a_1^\dagger$ and $a_1$ (similarly, $a_2^\dagger$ and $a_2$) are the creation and annihilation operators, respectively, of the first (second) oscillator with frequency $\omega_1$ ($\omega_2$). We use $\hbar=1$ throughout.

The qubit simultaneously interacts with both the oscillators through the $m$-photon JC interactions. The interaction Hamiltonian is given by $H_\text{int} = \sum_{i=1}^2 g_i\left(a_i^m\sigma_+ + a_i^{\dagger\, m} \sigma_-\right)$,
where $\sigma_+=\ket{e}\bra{g}$, $\sigma_-=\ket{g}\bra{e}$, and the parameter $g_1$ ($g_2$) determines the strength of the nonlinear MPJC interaction between the qubit and the first (second) oscillator. Further, the integer $m=1,\, 2,\, 3,\ldots$ tracks the multiphoton process. The total Hamiltonian for the tripartite system thus reads
\begin{align}
    H = H_\text{free} + H_\text{int}.
    \label{eqn_mpjcm}
\end{align}
Following standard convention, we define the two detuning parameters $\Delta_i = \omega_0 - m\omega_i$, where $i=1,\,2$. However, for simplicity, we will assume degenerate oscillators throughout this article, meaning they have equal frequencies, $\omega_1 = \omega_2 = \omega$. Therefore, $\Delta_1=\Delta_2=\Delta$. For this simple case, we can split $H = H_{I}+H_{II}$ such that $\left[H_I,\,H_{II}\right]=0$, where
\begin{align} 
\label{eqn_H_I}
  H_{I} &=  \frac{m\omega}{2}\sigma_z  + \omega( a_1^{\dagger}a_1  + a_2^{\dagger}a_2),\\
\label{eqn_H_II}
  H_{II} &= \frac{\Delta}{2} \sigma_z + \sum_{i=1}^2 g_i\left(a_i^m\sigma_+ + a_i^{\dagger\, m} \sigma_-\right).
\end{align}
Now, $H_I$ merely introduces a phase contribution in the temporal evolution of the full system. Conveniently, we work with $H_{II}$ and eliminate this trivial time dependence from the dynamics.
It is evident that when $m=1$, the system simplifies to the standard tripartite JC model. For a detailed practical implementation of the MPJC Hamiltonian in a specific quantum platform, we refer the reader to Ref.~\cite{laha_spinboson_2024}.

\section{\label{sec:solution}Solution: The state vector}
 Throughout this work, we assume that the qubit is prepared in a generic superposition state $\cos \phi \ket{g} + \sin \phi \ket{e}$ and the two oscillators are prepared in some arbitrary Fock states $\ket{n_1}$ and $\ket{n_2}$, respectively. In our notation, 
\begin{align}
  \ket{\psi(0)} = \cos \phi \ket{g, n_1, n_2} + \sin \phi \ket{e, n_1, n_2}. 
 \label{eqn_psi0}
\end{align}
We are interested in the state vector at a later time $\ket{\psi(t)}$ governed by the Hamiltonian $H_{II}$ in Eq.~\eqref{eqn_H_II}, starting with a completely pure initial state $\ket{\psi(0)}$.

In principle, we can obtain the state vector by solving the Schr\"odinger equation, i.e., $\ket{\psi(t)} = \exp\left(-iH_{II}t\right)\ket{\psi(0)}$.
A crucial aspect of this process is understanding how the two basis states $\ket{g, n_1, n_2}$ and $\ket{e, n_1, n_2}$ traverse the JC ladder. This entails determining the precise count of basis states contributing to $\ket{\psi(t)}$ across varying values of $n_1$, $n_2$, and~$m$. While the number of basis states is finite under generic circumstances, determining the exact quantity proves somewhat nontrivial due to the linear increase in required basis states with increasing $n_1$, $n_2$, and $m$ values, as illustrated schematically in Fig.~\ref{fig_schem_states}. It is important to note that such partitioning of Hilbert space is a standard technique in systems governed by JC interactions~\cite{Rodriguez_sci_rep_2018}. In our work, we extend this approach to include multiphoton interactions.

Let us begin by examining the initial basis state $\ket{g, n_1, n_2}$ in which the qubit is prepared in the ground state. As depicted in Fig.~\ref{fig_schem_states}, the application of the MPJC Hamiltonian to $\ket{g, n_1, n_2}$ yields two new states: $\ket{e, n_1-m, n_2}$ and $\ket{e, n_1, n_2-m}$. Each of these two new states would subsequently generate another new state. This recursive process of generating new states at each iteration continues until reaching a predetermined finite number of steps, at which point the corresponding Fock states will be annihilated and no further states are produced, thus concluding the JC ladder. By considering the specific values of $n_i$, we can identify the smallest positive integer $\ell_i$ such that $n_i-\ell_i m<0$, where $i=1,2$. It is straightforward to see that the state $\ket{n_i}$ is annihilated after $2\ell_i-1$ steps, giving rise to $2\ell_i-2$ new basis states. Consequently, the total count of states, including $\ket{g, n_1, n_2}$, is given by $2\ell_1-2+2\ell_2-2+1 = 2\ell_1+2\ell_2-3$.

Similarly, when considering the complementary initial state $\ket{e, n_1, n_2}$ in which the qubit is prepared in the excited state, the calculation proceeds analogously, albeit with a slight modification in the number of steps required for the annihilation of $\ket{n_i}$, which is now $2\ell_i$, generating $2\ell_i-1$ new basis states. Consequently, the total count of states, including $\ket{e, n_1, n_2}$, becomes $2\ell_1-1+2\ell_2-1+1 = 2\ell_1+2\ell_2-1$.

Now, in the most general scenario of initial superposition qubit state $\cos\phi\ket{g}+\sin\phi\ket{e}$, the aggregate number of basis states simply adds up to $2\ell_1+2\ell_2-3+ 2\ell_1+2\ell_2-1=4(\ell_1+\ell_2-1)$. This is because the basis states for the two marginal cases ($\phi=0$ and $\pi/2$) form two mutually independent sets. Therefore, the $4(\ell_1+\ell_2-1)$ coupled differential equations obtained from the Schr\"odinger equation can always be segregated into two sets: one comprising $2\ell_1+2\ell_2-3$ equations (originating from the initial ground qubit state), and the other containing $2\ell_1+2\ell_2-1$ equations (originating from the initial excited qubit state). Further, considering that each basis state can only be connected to a maximum of two new basis states (see Fig.~\ref{fig_schem_states}), the coupled differential equation governing each time-dependent coefficient will involve, at most, two additional coefficients aside from its own detuning term.

Now, depending upon the specific values of $n_1$, $n_2$, and $m$, we can identify four different cases:  (1) $n_1,\, n_2<m$, (2) $n_1<n_2=m$ and/or $n_2<n_1=m$, (3) $n_1=n_2=m$, and (4) $n_1,\,n_2>m$. For the first three cases, the corresponding $(\ell_1,\,\ell_2)$ values are (1,1), (1,2) and/or (2,1), and (2,2), respectively. In the following, we systematically investigate each of these cases individually. \\

\noindent{\bf Case 1:} In the constrained scenario where both $n_1$ and $n_2$ are smaller than $m$, a straightforward observation from Fig.~\ref{fig_schem_states} reveals that the state vector at any subsequent time $\ket{\psi(t)}_1$ will comprise only four basis states given by
\begin{align}
  \ket{\psi(t)}_1 &= x_1(t) \ket{g, n_1, n_2} + y_1(t) \ket{e, n_1, n_2}   \nonumber \\
                  &+ y_{2}(t) \ket{g, n_1+m, n_2}+ y_{3}(t)\ket{g, n_1, n_2+m}.
\label{eqn_psit_1}
\end{align}
As explained above, the four coupled differential equations from the Schr\"odinger equation can be partitioned into two distinct sets: one comprising $2\ell_1+2\ell_2-3=1$ state, and the other consisting of $2\ell_1+2\ell_2-1=3$ states (recall that $\ell_1=\ell_2=1$ in this case) with the initial condition $x_1(0)=\cos\phi$, $y_1(0)=\sin\phi$, and $y_2(0)=y_3(0)=0$. The solution to $x_1(t)$ is trivially obtained to be
\begin{align}
  x_{1}(t) &= \cos\phi\, e^{i \Delta' t}, 
  \label{eqn_x_coeff_case1}
\end{align} 
where $\Delta' = \Delta/2$.

On the other hand, the three coupled differential equations involving the $y$ coefficients can be efficiently expressed as
\begin{align}
 \frac{d}{dt} Y_1(t) = -i M_{1_y} Y_1(t),  
 \label{eqn_mat_case1}
\end{align}
where $Y_1(t) = \left(y_1(t),\, y_2(t),\, y_3(t)\right)^\top$, and the initial condition is $Y_1(0) = (\sin\phi,\, 0,\, 0)^\top$. It is straightforward to show that 
\begin{align}
    M_{1_y} =     
    \begin{pmatrix}
        \Delta' & g_{1_{n_1m}} & g_{2_{n_2m}} \\
        g_{1_{n_1m}} & - \Delta' & 0  \\
        g_{2_{n_2m}} & 0 & - \Delta' 
    \end{pmatrix},
  \label{my_mat_case1}
\end{align}
where $g_{1_{n_1m}}=\sqrt{(n_1+m)!/n_1!}\, g_1$ and $g_{2_{n_2m}}=\sqrt{(n_2+m)!/n_2!}\, g_2$.
Now, for the symmetric matrix $M_{1_y}$, we can always find an $S$ matrix that diagonalizes $M_{1_y}$ such that $M_{1_y}=S \,D\, S^{-1}$, with $D$ the diagonal matrix.  The standard solution of Eq.~\eqref{eqn_mat_case1} is given by $Y(t)= S\, e^{-iDt} S^{-1} Y(0)$. We used {\it Mathematica} software to obtain the solutions. The $y$ coefficients after simplifications can be expressed as
\begin{subequations}
 \begin{align}
  y_{1}(t) &= \sin\phi \left(\cos(\tilde{g}_1t) -i\tfrac{\Delta'}{\tilde{g}_1} \sin \left(\tilde{g}_1 t\right)\right),  \\
  y_{2}(t) &= -i\frac{\,g_{1_{n_1m}}}{\tilde{g}_1}\, \sin\phi\, \sin\left(\tilde{g}_1 t\right),\\
  y_{3}(t) &= -i\frac{g_{2_{n_2m}}}{\tilde{g}_1} \,  \sin\phi\,  \sin\left(\tilde{g}_1 t\right),
 \end{align} 
  \label{eqn_y_coeff_case1}
\end{subequations}
where $\tilde{g}_1=\sqrt{g_{1_{n_1m}}^2+g_{2_{n_2m}}^2 + \Delta'^2}$. \\

\noindent{\bf Case 2:} Next, we consider the case $n_1<n_2=m$. This corresponds to $\ell_1=1$ and $\ell_2=2$. Consequently, the state vector $\ket{\psi(t)}_2$ in this case will be a superposition of a total of eight basis states given by
\begin{align}
  \ket{\psi(t)}_2 &= x_1(t)\ket{g,n_1,m} + x_2(t)\ket{e,n_1,0} \nonumber \\
                  &+ x_3(t)\ket{g,n_1+m,0} + y_1(t)\ket{e,n_1,m} \nonumber \\
                  &+ y_2(t)\ket{g,n_1+m,m} + y_3(t)\ket{g,n_1,2m}, \nonumber \\
                  & + y_4(t)\ket{e,n_1+m,0} + y_5(t)\ket{g,n_1+2m,0}.
\label{eqn_psit_2}
\end{align}
Now, addressing the coupled differential equations involving the $x$ and $y$ coefficients independently, the two sets of equations can be expressed compactly (similar to Eq.~\eqref{eqn_mat_case1}) as
\begin{align}
  \frac{d}{dt} X_2(t) = -i M_{2_x} X_2(t),  \quad \frac{d}{dt} Y_2(t) = -i M_{2_y} Y_2(t), 
  \label{eqn_case2_mat}
\end{align}
where $ X_2(t) = \left(x_1(t),\, x_2(t),\, x_3(t)\right)^\top$, and $Y_2(t) = \left(y_1(t),\, y_2(t),\, y_3(t),\, y_4(t),\, y_5(t)\right)^\top$. The initial conditions are given by $X_2(0) = (\cos\phi,\, 0,\, 0)^\top$ and $Y_2(0) = (\sin\phi,\, 0,\, 0,\, 0,\, 0)^\top$.
The exact forms of the two symmetric matrices $M_{2_x}$ and $M_{2_y}$ can be found in Appendix~\ref{mat_elements}.  
As before, we have utilized {\it Mathematica} to obtain the solution. After simplification, the $x$ coefficients read 
\begin{subequations}
\begin{align}
  x_1(t) &= f_1 \left(\cos\left(\tilde{g}_2 t\right)+i \tfrac{\Delta'}{\tilde{g}_2} \sin\left(\tilde{g}_2 t\right) + f_0\, e^{i\Delta' t}\right),\\
  x_2(t) &= -if_2 \sin\left(\tilde{g}_2 t\right) , \\
  x_3(t) &= f_3 \left(\cos\left(\tilde{g}_2 t\right) + i \tfrac{\Delta'}{\tilde{g}_2} \sin\left(\tilde{g}_2 t\right) - e^{i\Delta' t} \right),
\end{align} 
  \label{xt_case2}
\end{subequations}
where  $\tilde{g}_2=\sqrt{g_{1_{n_1m}}^2+g_{2_m}^2+\Delta'^2}$\,, $g_{2_m} = \sqrt{m!}\,g_2$, $f_0=\tfrac{g_{1_{n_1m}}^2}{g_{2_m}^2}$, $f_1=\frac{g_{2_m}^2}{g_{1_{n_1m}}^2+g_{2_m}^2} \cos\phi$, $f_2=\frac{g_{2_m}}{\tilde{g}_2} \cos\phi$, and $f_3=\frac{g_{1_{n_1m}}g_{2_m}}{g_{1_{n_1m}}^2+g_{2_m}^2} \cos\phi$.

On the other hand, expressing the $y$ coefficients in simple algebraic forms appears challenging in the most general situation. However, ignoring the detuning, that is, setting $\Delta=0$ or considering only $H_{\text{int}}$, it is possible to express these coefficients succinctly. After simplifications, we obtain 
\begin{subequations}
\begin{align}
  y_1(t) &= \tfrac{\sin\phi}{s}\left(\tilde{\textsl{g}}_{s_+}'^2 \cos \tau_{+} - \tilde{\textsl{g}}_{s_-}'^2 \cos\tau_{-}\right),\\
  y_2(t) &= -i\tfrac{\sin\phi}{s}\left(f_{2_+} \sin\tau_{+} - f_{2_-} \sin\tau_{-}\right), \\
  y_3(t) &= i\tfrac{\sin\phi}{s}\left(f_{3_+}\sin\tau_{+} - f_{3_-} \sin\tau_{-}\right),\\
  y_4(t) &= \tfrac{\sin\phi}{s} \textsl{g}_1 \textsl{g}_3  \left(\cos \tau_{+} - \cos\tau_{-} \right),\\ 
  y_5(t) &= -i\tfrac{\sin\phi}{s} \textsl{g}_1 \textsl{g}_3 \textsl{g}_4 \left(\tfrac{\sin\tau_{+}}{\tilde{\textsl{g}}_{s_+}}-\tfrac{\sin \tau_{-}}{\tilde{\textsl{g}}_{s_-}}\right).
\end{align}    
  \label{yt_case2}
\end{subequations}
All unknown parameters corresponding to Eq.~\eqref{yt_case2} are neatly tabulated in Table~\ref{tab:table1}.

Before progressing to the subsequent scenario, we note that the case where $n_2 < n_1 = m$, corresponding to $\ell_1 = 2$ and $\ell_1 = 1$, is complementary to the situation where $n_1 < n_2 = m$.\\

\noindent{\bf Case 3:} Now, let us examine the scenario where $n_1=n_2=m$, corresponding to $\ell_1=\ell_2=2$. In this special case, the state vector comprises 12 basis states (partitioned into two sets consisting of five and seven states respectively) and is given by
\begin{align} 
\ket{\psi(t)}_3 &= x_1 \ket{g,m,m} + x_2 \ket{e,0,m} + x_3 \ket{e,m,0} \nonumber \\
&\,\,\,+ x_4 \ket{g,2m,0} + x_5\ket{g,0,2m} + y_1 \ket{e,m,m} \nonumber \\
&\,\,\,+ y_2 \ket{g,2m,m} + y_3 \ket{g,m,2m} + y_4 \ket{e,2m,0} \nonumber \\
&\,\,\,+ y_5 \ket{e,0,2m} + y_6 \ket{g,3m,0} + y_7 \ket{g,0,3m}.
\label{n1n2m_equal}
\end{align}
Note that for brevity, we ignore the explicit time dependencies of the coefficients $x$ and $y$ in Eq.~\eqref{n1n2m_equal}. Analogous to Eq.~\eqref{eqn_case2_mat}, we can express the coupled differential equations involving the coefficients $x$ and $y$ as
\begin{align}
 \frac{d}{dt} X_3(t) = -i M_{3_x} X_3(t),  \quad \frac{d}{dt} Y_3(t) = -i M_{3_y} Y_3(t).
 \label{eqn_case3_mat}
\end{align}
Here, $ X_3(t) = \left(x_1(t),\, x_2(t),\, x_3(t),\, x_4(t),\, x_5(t)\right)^\top$, and $Y_2(t) = \left(y_1(t),\, y_2(t),\, y_3(t),\, y_4(t),\, y_5(t),\, y_6(t),\, y_7(t)\right)^\top$. The initial conditions are given by $X_2(0) = (\cos\phi,\, 0,\, 0,\,0,\,0)^\top$ and $Y_2(0) = (\sin\phi,\, 0,\, 0,\, 0,\, 0,\,0,\,0)^\top$.
The elements of the symmetric matrices $M_{3_x}$ and $M_{3_y}$ are detailed in Appendix~\ref{mat_elements}. Analogously to the $y$ coefficients for case 2, we can express the solution for the $x$ coefficients in compact algebraic forms in the limit $\Delta=0$. Following a few simplification steps, we obtain
\begin{subequations}
\begin{align}
  x_1(t) &=  \tfrac{\cos\phi}{f_0} \left(f_{1_+} \cos\tau_{+} - f_{1_-} \cos\tau_{-} -  s \textsl{g}_3^2 \textsl{g}_4^2\right), \\
  x_2(t) &= -i \tfrac{\cos\phi}{f_0} \left(f_{2_+} \sin\tau_{+} - f_{2_-} \sin\tau_{-}\right),\\
  x_3(t) &= -i \tfrac{\cos\phi}{f_0} \left(f_{3_+} \sin\tau_{+} - f_{3_-} \sin\tau_{-}\right), \\
  x_4(t) &=  \tfrac{\cos\phi}{f_0} \left(f_{4_+} \cos\tau_{+} - f_{4_-} \cos\tau_{-}  -  s \textsl{g}_1 \textsl{g}_3^2 \textsl{g}_4 \right),  \\
  x_5(t) &=  \tfrac{\cos\phi}{f_0} \left(f_{5_+} \cos\tau_{+} - f_{5_-} \cos\tau_{-}  -  s \textsl{g}_2 \textsl{g}_3 \textsl{g}_4^2 \right).
\end{align}    
\label{xt_case3}
\end{subequations}
Similar to the previous case, all unknown parameters corresponding to Eq.~\eqref{xt_case3} can be found in Table~\ref{tab:table1}.

On the other hand, despite setting $\Delta=0$, it remains challenging to express the seven $y$ coefficients in simple algebraic terms, even with the assistance of {\it Mathematica}.\\


\begin{table*}[ht]
\caption{\label{tab:table1}%
Parameters corresponding to Eqs.~(\eqref{yt_case2}) and (\eqref{xt_case3}).}
\begin{ruledtabular}
\begin{tabular}{ccccc}
Parameter &  Eq.~\eqref{yt_case2} & Eq.~\eqref{xt_case3} \\
\colrule
\colrule
$s$  & $\sqrt{G^4 - 4\mathcal{G}^4}$ &  $\sqrt{G^4 - 4\mathcal{G}^4}$ \\
$\tilde{\textsl{g}}_{s_\pm}$  & $\sqrt{\tfrac{1}{2} \left(G^2\pm s\right)}$ &  $\sqrt{\tfrac{1}{2} \left(G^2\pm s\right)}$ \\
$\tilde{\textsl{g}}_{s_\pm}'^2$  & $\tfrac{1}{2} \left(G'^2\pm s\right)$ &  $\sqrt{\tfrac{1}{2} \left(G'^2\pm s\right)}$ \\
$G^2$  & $\textsl{g}_1^2+\textsl{g}_2^2+\textsl{g}_3^2+\textsl{g}_4^2$ &  $\textsl{g}_1^2 + \textsl{g}_2^2 + \textsl{g}_3^2 + \textsl{g}_4^2$ \\
$G'^2$  & $\textsl{g}_1^2+\textsl{g}_2^2-\textsl{g}_3^2-\textsl{g}_4^2$ &  $\textsl{g}_1^2-\textsl{g}_2^2-\textsl{g}_3^2+\textsl{g}_4^2$ \\
$\mathcal{G}^4$  & $\left(\textsl{g}_1^2 \textsl{g}_4^2+\textsl{g}_2^2 \textsl{g}_3^2+\textsl{g}_2^2 \textsl{g}_4^2\right)$ &  $\left(\textsl{g}_1^2 \textsl{g}_3^2+\textsl{g}_2^2 \textsl{g}_4^2+\textsl{g}_3^2 \textsl{g}_4^2\right)$ \\
$f_{0}$  &  &  $s\tilde{\textsl{g}}_{s_+}^2\tilde{\textsl{g}}_{s_-}^2$ \\
$f_{1\pm}$  &  &  $\textsl{g}_1^2\textsl{g}_3^4+\textsl{g}_2^2\textsl{g}_4^4- \left(\textsl{g}_1^2\textsl{g}_3^2+\textsl{g}_2^2\textsl{g}_4^2\right)\tilde{\textsl{g}}_{s_\pm}^2$ \\
$f_{2\pm}$  & $\textsl{g}_1\tilde{\textsl{g}}_{s_\pm} \left(\textsl{g}_2^2\textsl{g}_3^2+\textsl{g}_4^2 \tilde{\textsl{g}}_{s_\pm}'^2\right)/\mathcal{G}^4$ &  $\textsl{g}_1  \left(\textsl{g}_2^2\textsl{g}_4^2+\textsl{g}_3^2 \tilde{\textsl{g}}_{s_\pm}'^2 \right) \tilde{\textsl{g}}_{s_\pm}$ \\
$f_{3\pm}$  & $\textsl{g}_2 \tilde{\textsl{g}}_{s_-} \left(\textsl{g}_1^2\textsl{g}_4^2-\left(\textsl{g}_3^2+\textsl{g}_4^2 \right)\tilde{\textsl{g}}_{s_-}'^2\right)/\mathcal{G}^4 $ &  $\textsl{g}_2 \left(\textsl{g}_1^2 \textsl{g}_3^2-\textsl{g}_4^2\tilde{\textsl{g}}_{s_\mp}'^2\right)\tilde{\textsl{g}}_{s_\pm}$ \\
$f_{4\pm}$  &  &  $\textsl{g}_1 \textsl{g}_4 \left(\textsl{g}_2^2\textsl{g}_4^2+\textsl{g}_3^2 \tilde{\textsl{g}}_{s_\pm}'^2\right)$ \\
$f_{5\pm}$  &  &  $\textsl{g}_2 \textsl{g}_3\left(\textsl{g}_1^2\textsl{g}_3^2-\textsl{g}_4^2 \tilde{\textsl{g}}_{s_\mp}'^2\right)$ \\
$\tau_\pm$  & $\tilde{\textsl{g}}_{s_\pm} t$ &  $\tilde{\textsl{g}}_{s_\pm}t$\\
$\textsl{g}_1$  & $\sqrt{(n_1+m)!/n_1!}\,g_1$ &  $\sqrt{m!}\,g_1$ \\
$\textsl{g}_2$  & $\sqrt{(2m)!/m!}\,g_{2}$ &  $\sqrt{m!}\,g_2$ \\
$\textsl{g}_3$  & $\sqrt{(n_1+2m)!/(n_1+m)!}\,g_1$ &  $\sqrt{(2m)!/m!}\,g_1$ \\
$\textsl{g}_4$  & $\sqrt{m!}\,g_{2}$ &  $\sqrt{(2m)!/m!}\, g_2$ \\
\end{tabular}
\end{ruledtabular}
\end{table*}


\noindent{\bf Case 4:}
In the final case, where both $n_1$ and $n_2$ exceed $m$, the state vector at a later time, $\ket{\psi(t)}_4$, can be realized by following the JC ladder, as depicted in Fig.~\ref{fig_schem_states}. We employ a similar methodology for arbitrary values of $n_1$, $n_2$ and $m$, and construct two independent state vectors $\ket{\psi_x(t)}_4$ and $\ket{\psi_y(t)}_4$ (or equivalently $X_4(t)$ and $Y_4(t)$), such that $\ket{\psi(t)}_4=\ket{\psi_x(t)}_4+\ket{\psi_y(t)}_4$. The exact forms of $\ket{\psi_x(t)}_4$ and $\ket{\psi_y(t)}_4$ are provided in Appendix~\ref{mat_elements}.
Following the Schrödinger equation, we can derive the corresponding coupled differential equations of $X_4(t)$ and $Y_4(t)$, akin to Eqs.\eqref{eqn_case2_mat} and~\eqref{eqn_case3_mat}. This leads to the determination of two symmetric square matrices, denoted by $M_{4_x}$ and $M_{4_y}$, whose generic forms can be found in Appendix~\ref{mat_elements}. Both $M_x$ and $M_y$ exhibit a pentadiagonal structure and are, in principle, diagonalizable. Nevertheless, expressing the two generic state vectors $X_4(t)$ and $Y_4(t)$ in simple algebraic forms remains a formidable task.

\section{\label{sec:state_transfer}Time-evolved oscillator states}
In this section, we will extract the time-evolved states of each oscillator by tracing over the relevant subsystems from the corresponding state vectors for all cases. Additionally, we will assess the effectiveness of swapping Fock states or complete excitation transfer between the two oscillators for all scenarios. Recall that the two oscillators are initially in arbitrary Fock states $\ket{n_1}$ and $\ket{n_2}$ respectively, with the joint oscillator state expressed as $\ket{n_1,\, n_2}$.Perfect state swapping involves transforming $\ket{n_1, n_2}$ into $\ket{n_2, n_1}$, whereas complete excitation transfer entails converting $\ket{n_1, n_2}$ into $\ket{n_1+n_2, 0}$. Both processes are equivalent only when $n_1=0$ or $n_2=0$. For simplicity, we assume $\Delta=0$ for the analysis.\\

\noindent{\bf Case 1:}
Obtaining the reduced density matrices of the two oscillators $\rho_{1}^{(1)}(t)$ and $\rho_{2}^{(1)}(t)$ from $\ket{\psi(t)}_1$ in Eq.~\eqref{eqn_psit_1} is straightforward. Since only two basis states for each oscillator ($\ket{n_i}$ and $\ket{n_i+m}$ respectively, where $i=1,\, 2$) are involved in the dynamics, the reduced density matrices can be {\it effectively} represented as $2\times 2$ matrices and are given by
\begin{align}
\label{eqn_rho_os1_1}
    \rho_{1}^{(1)}(t) =     
    \begin{pmatrix}
        1-\vert y_2\vert^2 & x_1 y_2^* \\
        x_1^* y_2 & \vert y_2\vert^2 
    \end{pmatrix},
\end{align}
and
\begin{align}
\label{eqn_rho_os2_1}
    \rho_{2}^{(1)}(t) =    
    \begin{pmatrix}
        1-\vert y_3\vert^2 & x_1 y_3^* \\
        x_1^* y_3 & \vert y_3\vert^2 
    \end{pmatrix},
\end{align}
respectively. The $x$ and $y$ coefficients are given by Eqs.\eqref{eqn_x_coeff_case1} and~\eqref{eqn_y_coeff_case1}, respectively. Now, assuming $g_1=g_2=g$ (symmetrically coupled MPJC interactions), we have
$\tilde{g}_1=\sqrt{g_{1_{n_1m}}^2+g_{2_{n_2m}}^2}$, where $g_{n_im}= \sqrt{\frac{(n_i+m)!}{n_i!}} g$. Therefore, $\frac{g_{1_{n_1m}}}{\tilde{g}_1} = \left(1+\frac{n_1!}{n_2!}\right)^{-1/2}$, and $\frac{g_{2_{n_2m}}}{\tilde{g}_1} = \left(1+\frac{n_2!}{n_1!}\right)^{-1/2}$.
It is evident that if the qubit starts from the ground state ($\phi=0$), only the $x_1$ coefficient survives. Consequently, the state vector $\ket{\psi(t)}_1$ and thus both the oscillator Fock states merely acquire only an overall phase during the dynamics, implying $\ket{n_1,\, n_2}\longrightarrow e^{i\theta}\ket{n_1,\, n_2}$.

On the other hand, if the qubit is initially prepared in the excited state ($\phi=\pi/2$), then $x_1=0$. Therefore, the off-diagonal elements of both $\rho_{1}^{(1)}$ and $\rho_{1}^{(2)}$ become zero. Consequently, the time-evolved oscillator states transform into a trivial incoherent mixture of the two basis states $\ket{n_i}$ and $\ket{n_i+m}$, that is, $\rho_{1}^{(1)} = \text{diag}\left(1-|y_2|^2,\, |y_2|^2\right)$ and $\rho_{2}^{(1)} = \text{diag}\left(1-|y_3|^2,\, |y_3|^2\right)$.

Now, it is obvious from Eq.~\eqref{eqn_psit_1} that whenever $\vert y_2\vert^2=1$ or $\vert y_3\vert^2=1$ the joint oscillator state becomes $\ket{n_1+m,\, n_2}$ or $\ket{n_1,\, n_2+m}$. However, $|y_2|^2$ or $|y_3|^2$ oscillates between zero and $\left(1+\frac{n_1!}{n_2!}\right)^{-1}$ or zero and $\left(1+\frac{n_2!}{n_1!}\right)^{-1}$ in this case. Therefore, $|y_2|^2\to 1$ (or $|y_2|^2\to 1$) in the limit $n_2\to \infty$ (or $n_1\to \infty$) which is not feasible for this case, as both $n_1,\, n_2<m$. 
Finally, we note in passing that for other values of $\phi$, the two oscillator states remain a superposition state of the two effective basis states.\\

\noindent{\bf Case 2:}
Unlike the previous scenario, in this case, three basis states for each oscillator ($\ket{n_1}$, $\ket{n_1+m}$, and $\ket{n_1+2m}$ for the first oscillator and $\ket{0}$, $\ket{m}$, and $\ket{2m}$ for the second oscillator) are involved in the dynamics. Therefore, the reduced density matrices of both the oscillators  $\rho_{1}^{(2)}(t)$ and $\rho_{2}^{(2)}(t)$ can be {\em effectively} expressed as $3\times 3$ matrices. These are given by
\begin{align}
\label{eqn_rho_os1_2}
    \rho_{1}^{(2)}(t) &=     
    \begin{psmallmatrix}
        |x_1|^2+|x_2|^2 +|y_1|^2 +|y_3|^2 & x_1y_2^* + x_2 y_4^* & 0 \\
y_2x_1^* + y_4x_2^* & |x_3|^2+ |y_2|^2 +|y_4|^2 & x_3y_5^* \\
0 & y_5x_3^* & |y_5|^2
    \end{psmallmatrix},
\end{align}
and
\begin{align}    
\label{eqn_rho_os2_2}
    \rho_{2}^{(2)}(t) &=    
    \begin{psmallmatrix}
        |x_2|^2+|x_3|^2 + |y_4|^2 + |y_5|^2 & x_2 y_1^*+x_3y_2^* & 0 \\
y_1 x_2^* + y_2x_3^* & |x_1|^2+|y_1|^2+|y_2|^2 & x_1 y_3^* \\
0 & y_3x_1^* & |y_3|^2
    \end{psmallmatrix},
\end{align}
respectively.
Once again, for $\phi=0$ (the initial ground-state qubit), only the $x$ coefficients contribute. Consequently, the first oscillator becomes an incoherent mixture of only two basis states: $\ket{n_1}$ and $\ket{n_1+m}$, that is, $\rho_{1}^{(2)} = \text{diag}\left(1-|x_3|^2,\, |x_3|^2\right)$.
Similarly, the second oscillator state becomes an incoherent mixture of $\ket{0}$ and $\ket{m}$, leading to $\rho_{2}^{(2)} = \text{diag}\left(1-|x_1|^2,\, |x_1|^2\right)$. 
From Eq.~\eqref{xt_case2}, it is easy to see that 
\begin{align}
  \label{x3_case2}
  |x_3| &= A\, \sin^2\left(\tfrac{1}{2}\sqrt{g_{1_{n_1m}}^2+g_{2_m}^2} t\right),
\end{align}
where 
\begin{align}
  A &= \frac{2 \,g_{1_{n_1m}}g_{2_m}}{g_{1_{n_1m}}^2+g_{2_m}^2}.
  \label{eqn:A}
\end{align} 
Note that $g_{1_{n_1m}}=\sqrt{(n_1+m)!/n_1!}\, g_1$, and  $g_{2_m} = \sqrt{m!}\,g_2$.
Evidently, for the first oscillator state to be exactly $\ket{n_1+m}$ (or equivalently, the second oscillator state to be $\ket{0}$), the condition $|x_3|^2=1$ must be satisfied, which simply translates to $A=1$.

\begin{figure}[h]
    \centering
    \includegraphics[width=0.49\textwidth]{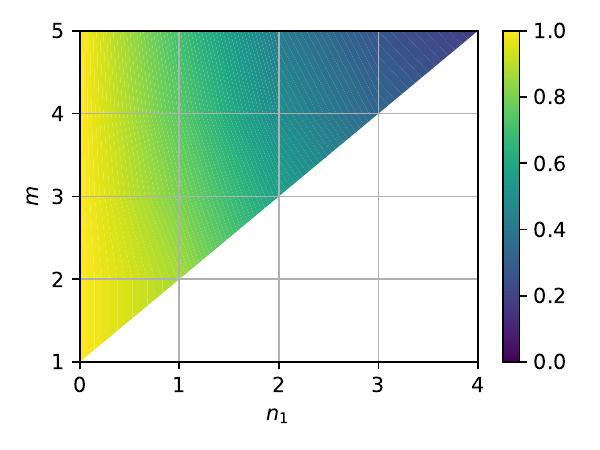}
    \vspace{-5ex} 
    \caption{(a) Parameter $A$ in Eq.~\eqref{eqn:a_g1_g2}, corresponding to $g_1=g_2$, as a function of $n_1$ and $m$ for case~2, that is, $n_1<n_2=m$, with $\Delta=0$.}
    \label{fig:contour_g1_g2_equal}
\end{figure}

Now, if we assume symmetric MPJC couplings ($g_1=g_2$), Eq.~\eqref{eqn:A} can be further simplified to 
\begin{align}
  A = \frac{2 \,\sqrt{(n_1+m)! n_1! m!}}{(n_1+m)! +n_1!m!}.
  \label{eqn:a_g1_g2}
\end{align} 
The parameter $A$ in Eq.~\eqref{eqn:a_g1_g2} is plotted as a function of $n_1$ and $m$ in Fig.~\ref{fig:contour_g1_g2_equal}. Unlike the previous scenario, both the analytical expression and Fig.~\ref{fig:contour_g1_g2_equal} clearly show that $A$ can reach its maximum value of unity, but only when $n_1 = 0$. This condition results in the perfect swapping of arbitrary Fock states between oscillators, specifically $\ket{0,\, m}\longrightarrow\ket{m,\, 0}$. It is noteworthy that with standard tripartite JC couplings, only the swapping of the first excited state with the ground state is achievable in this manner. However, we find that with MPJC interactions, swapping of arbitrary Fock states can be achieved in principle. We emphasize that a similar transfer can also be achieved using a beamsplitter or even in the standard tripartite JC model~\cite{Sun_PRA_2006}. However, in the latter case, this transfer was accomplished only in the large detuning limit (resulting in effective Hamiltonians) and by adiabatically fixing the qubit in one of its basis states.

On the other hand, for $n_1>0$, there will be a pronounced decay in the fidelity of the transition $\ket{n_1, m} \longrightarrow \ket{n_1+m,0}$ as the values of $n_1$ and $m$ increase. In other words, all excitations of the second oscillator cannot be completely transferred to the first oscillator if $n_1>0$ in this manner. This problem can be successfully tackled by relaxing the symmetric coupling assumptions in the MPJC couplings. The details of the analysis can be found in Appendix~\ref{asymm_coupoling}.

Now suppose the qubit is initially prepared in the excited state, i.e., $\phi=\pi/2$.  In this scenario, all $x$ coefficients vanish, resulting in $\rho_{1}^{(2)}$ and $\rho_{2}^{(2)}$ being strictly diagonal, comprising all three basis states in each case, that is, $\rho_{1}^{(2)} = \text{diag}\left(|y_1|^2 +|y_3|^2,\, |y_2|^2 +|y_4|^2,\, |y_5|^2\right)$, and $\rho_{2}^{(2)} = \text{diag}\left(|y_4|^2 +|y_5|^2,\, |y_1|^2 +|y_2|^2,\, |y_3|^2\right)$. Unlike all the previous subcases, here we find that the dynamics is controlled by two different frequencies $\tilde{\textsl{g}}_{s_\pm}$ [see Eq.~(\eqref{yt_case2})]. In the simplest case, when $n_1=0$ (along with $\Delta=0$ and $g_1=g_2=g$), we obtain 
\begin{subequations}
\begin{align}
    y_4 &= \frac{1}{\sqrt{2(2m)!}} \left(\cos(\tilde{\textsl{g}}_{s_+} t)-\cos(\tilde{\textsl{g}}_{s_-}t)\right), \\
    y_5 &= -i\tfrac{\sqrt{m!}}{2}\left(\frac{\sin(\tilde{\textsl{g}}_{s_+} t)}{\tilde{\textsl{g}}_{s_+}}-\frac{\sin(\tilde{\textsl{g}}_{s_-}t)}{\tilde{\textsl{g}}_{s_-}}\right),
\end{align}
\end{subequations}
where $\tilde{\textsl{g}}_{s_\pm}=\sqrt{m!+(2m)!/m!\pm\sqrt{(2m)!}}\,g$.
It can be shown that $|y_4|$ or $|y_5|$ can never reach the value of unity. Therefore, we can conclude that $\ket{0,\, m}\longrightarrow\ket{m,\, 0}$ or $\ket{0,\, m}\longrightarrow\ket{2m,\, 0}$ is also not feasible in this case.  For higher values of $n_1$, the analysis becomes even more intricate. However, we can see that the two oscillator states remain an incoherent mixture of the three corresponding basis states.\\

\noindent{\bf Case 3:} For this case, the reduced density matrices for the two oscillators can be effectively expressed in the $\ket{0}$, $\ket{m}$, $\ket{2m}$, and $\ket{3m}$ basis. The exact analytical expressions of $\rho_{1}^{(3)}(t)$ and $\rho_{2}^{(3)}(t)$ can be found in Appendix~\ref{mat_elements}. Specifically, for $\phi=0$ (for which we could obtain an exact analytical solution, assuming $\Delta=0$), the two oscillator states reduce to an incoherent mixture of the three basis states $\ket{0}$, $\ket{m}$, and $\ket{2m}$. These are given by $\rho_{1}^{(3)}=\text{diag}\left(|x_2|^2+|x_5|^2, |x_1|^2+|x_3|^2, |x_4|^2\right)$ and $\rho_{2}^{(3)}=\text{diag}\left(|x_3|^2+|x_4|^2, |x_1|^2+|x_2|^2, |x_5|^2\right)$, respectively. The feasibility of $\ket{m,\, m}\longrightarrow\ket{2m,\, 0}$ or $\ket{m,\, m}\longrightarrow\ket{0,\, 2m}$ can be obtained by calculating  $|x_4|$ and $|x_5|$ in Eq.~\eqref{xt_case3}. Now, even for the simplest case with $m=1$, it can be shown (assuming $g_1=g_2=g$) that $|x_1|^2$ oscillates between 0 and 1, $|x_2|^2$ and $|x_3|^2$ between 0 and $\frac{1}{4}$, and $|x_4|^2$ and $|x_5|^2$ between 0 and $\frac{1}{2}$. In particular, $|x_4|=|x_5| = \tfrac{1}{\sqrt{2}} \sin^2(\sqrt{2} g t)$, clearly indicating that even $\ket{1,\, 1}\longrightarrow\ket{2,\, 0}$ or $\ket{1,\, 1}\longrightarrow\ket{0,\, 2}$ is not feasible in this case. However, the fact that $|x_1|^2=1$ at periodic intervals, $\ket{1}$ returns to itself periodically. In fact, it can be easily shown that $|x_1|^2=1$ for all values of $m$, indicating the periodic return of $\ket{m}$ during the temporal evolution. However, the amplitude of the 
oscillations for other $x$ coefficients depends on the specific choice of the parameter $m$. For completeness, we note that for $\phi=\pi/2$, the two oscillator states are $\rho_{1}^{(3)}=\text{diag}\left(|y_5|^2+|y_7|^2, |y_1|^2+|y_3|^2, |y_2|^2+|y_4|^2,|y_6|^2\right)$ and $\rho_{2}^{(3)}=\text{diag}\left(|y_4|^2+|y_6|^2, |y_1|^2+|y_2|^2, |y_3|^2+|y_5|^2,|y_7|^2\right)$ in the $\ket{0}$, $\ket{m}$, $\ket{2m}$, and $\ket{3m}$ basis. \\


\noindent{\bf Case 4:}
Finally, for this case as well, we can derive general expressions for the reduced density matrices of the two oscillators $\rho_{1}^{(4)}$ and $\rho_{2}^{(4)}$ using a similar methodology (see Appendix~\ref{mat_elements} for details). Similar to previous cases, we find both matrices to be tridiagonal. Nevertheless, drawing general conclusions without knowledge of specific values for $n_1$, $n_2$, and $m$ remains challenging.

\section{\label{sec:wigner}Wigner nonclassicalities}
As noted in Sec.~\ref{intro}, Fock states exhibit pronounced nonclassical behavior. A commonly employed method to quantify the extent of nonclassicality in a given quantum state involves computing the volume of the negative region within its associated Wigner function, defined as~\cite{Anatole_joptb_2004,Arkhipov_sci_rep_2018,rosiek_arxiv_2023}
\begin{align}
  V_{W_{-}} = -\iint dx\, dp\, \min \left[W(x,p), 0\right],
    \label{eqn_wig_neg}
\end{align}
where the integration encompasses the entire phase space and $\iint dx\, dp\, W(x,p)=1$ is the normalization condition. As per definition, $V_{W_{-}}$ equals zero for all Gaussian states, including the vacuum state $\ket{0}$, coherent state $\ket{\alpha}$, or squeezed vacuum state $\ket{\xi}$. 

A natural question arises: How does this measure of nonclassicality evolve under such a nonlinear Hamiltonian? We address this question in the following, assuming the initial state of the tripartite system is $\ket{\psi(0)}$ in Eq.~\eqref{eqn_psi0}, that is ,the qubit is in a superposition state and the two oscillators are in arbitrary Fock states $\ket{n_1}$ and $\ket{n_2}$, respectively. For brevity, we discuss here only the evolution of $V_{W_{1-}}$ of the first oscillator Fock state $\ket{n_1}$, as similar results can also be obtained for the second oscillator.

\begin{figure}[ht]
\centering
\includegraphics{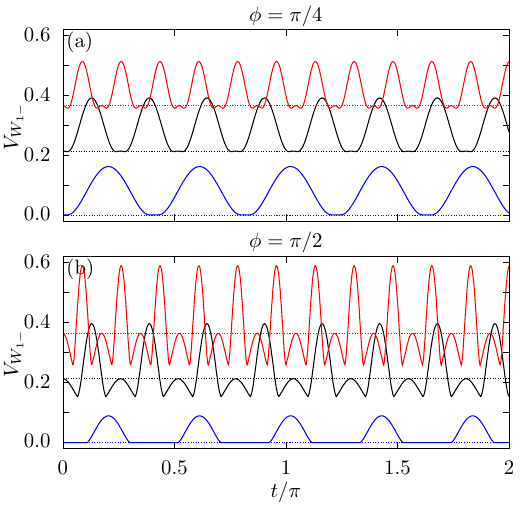}
\vspace{-3ex}
\caption{For the tripartite MPJC model, described by the Hamiltonian $H$ in Eq.~\eqref{eqn_mpjcm}, the temporal evolution of the Wigner nonclassicality [quantified by the volume of the negative region of the associated Wigner function $V_{W_{1-}}$ in Eq.~(\eqref{eqn_wig_neg})] of the initial Fock states $\ket{0}$, $\ket{1}$, and $\ket{2}$ (as represented by the blue, black, and red curves, respectively) of the first oscillator is illustrated for case 1, that is when $n_1,\, n_2<m$. Initially, the second oscillator is in the vacuum state $\ket{0}$ and the qubit is initially in a superposition state with (a) $\phi=\pi/4$ and (b) $\phi=\pi/2$. Further, $m=3$, and we set $g_1=g_2=1/\sqrt{2}$ and $\Delta=0$. In both panels, $V_{W_{1-}}$ surpasses its initial value (denoted by dashed horizontal lines) by significant amounts. More importantly, driven by the initial qubit energy alone, with both the oscillators initialized in the ground states, the MPJC interactions nontrivially produce negativity in the Wigner functions in both cases.}
\label{fig_wig_neg_1}
\end{figure}

To simplify the analysis, we set $g_1=g_2=1/\sqrt{2}$ and $\Delta=0$ (see, however, Appendix~\ref{sec:detuning} where the role of detuning is analyzed in detail), while restricting ourselves to $n_1,\, n_2,\, m\leqslant 3$. Additionally, we mainly consider three values for $\phi$, specifically, $0$, $\pi/4$, and $\pi/2$, corresponding to ground, maximally superposed, and excited initial qubit states, respectively. It is worth mentioning that the Wigner function of the photon number state $\ket{n}$ is given by $W(\alpha) = \frac{2}{\pi} (-1)^n e^{-2|\alpha|^2} L_n\left(4|\alpha|^2\right)$, where $\alpha=(x+ip)/\sqrt{2}$ and $L_{n}(\,\cdot\,)$ is the standard Laguerre polynomial (see Appendix~\ref{wig_appendix}).  \\


\noindent{\bf Case 1:} We begin by examining case 1 ($n_1,\, n_2 < m$). The time-evolved Wigner function of the first oscillator is found to be (see Appendix~\ref{wig_appendix} for details)
\begin{align}
  \tilde{W}_1(\alpha, t) &= (-1)^{n_1}   \Big[ \left(1-|y_2|^2\right) L_{n_1}(z)  + 2^{m} \sqrt{\tfrac{n_1!}{(n_1+m)!}}\nonumber \\
  &\hspace{0.75in} \times \big(x_1 y_2^* \alpha^{m} + y_2 x_1^* (\alpha^*)^{m}\big) L_{n_1}^{m}(z) \nonumber \\   
  &\hspace{0.85in} + (-1)^{m} |y_2|^2  L_{n_1+m}(z)\Big],
  \label{eqn_wig_case1_os1}
\end{align}
where $W_1(\alpha, t) = \frac{2}{\pi} e^{-2|\alpha|^2} \tilde{W}_1(\alpha, t)$, $z=4|\alpha|^2$, and $L_{k_1}^{k_2}(\,\cdot\,)$ is the associated Laguerre polynomial. For clarity, we consistently employ this notation in the equations for the Wigner functions throughout the remainder of this paper. The time-dependent coefficients $x$ and $y$ are given by Eq.~\eqref{eqn_y_coeff_case1}. As already discussed in the previous sections, the system undergoes trivial temporal evolution for $\phi=0$. Notably, the $y$ coefficients become zero, and $|x_1|=1$. It is straightforward to see that $W_{1}(\alpha, t) = \frac{2}{\pi} (-1)^{n_1} e^{-2|\alpha|^2}  L_{n_1} \left( 4|\alpha|^2 \right) = W_{1}(\alpha, 0)$ in this case. Thus, $V_{W_{1-}}$ remains constant over time. 

However, the evolution becomes nontrivial for other values of $\phi$, as the $y$ coefficients now contribute to the dynamics. Depending on the specific values of $n_1$, $n_2$, and $m$, significant changes manifest in the dynamics of $V_{W_{1-}}$, as illustrated in Fig.~\ref{fig_wig_neg_1}.
For example, consider the simplest case where the system is driven {\it solely} by the initial qubit energy, that is, both oscillators are initialized in the ground states characterized by Gaussian Wigner functions with no negative regions. Setting $n_1=n_2=0$ in Eq.~\eqref{eqn_wig_case1_os1}, the Wigner function of the first oscillator becomes
\begin{align} 
  \tilde{W}_1(\alpha, t) &=  1-|y_2|^2\left(1-(-1)^m L_m(z)\right)  \nonumber \\ 
  &\quad\quad+ \frac{2^m}{\sqrt{m!}} \left(x_1y_2^* \alpha^m + x_1^* y_2 (\alpha^*)^m\right) . 
  \label{eqn_wig_case1_os1_vac}
\end{align}

\begin{figure}[ht]
\centering
\includegraphics{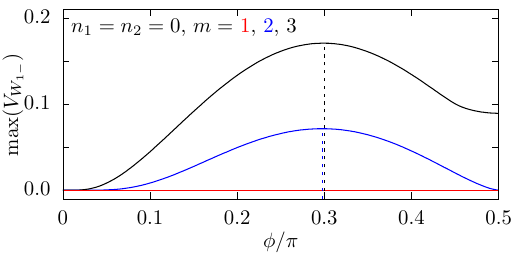}
\vspace{-3ex}
\caption{Maximum achievable $V_{W_{1-}}$ as a function of $\phi$ for the MPJC model with both oscillators initially in the ground states and where the qubit is initially in a superposition state $\cos\phi\ket{g}+\sin\phi\ket{e}$ with $m=$ 1 (red), 2 (blue), and 3 (black). Similar to Fig.~\ref{fig_wig_neg_1}, we set $g_1=g_2=1/\sqrt{2}$ and $\Delta=0$. For $m=1$ (tripartite JC interactions), the Wigner function of the oscillator remains positive at all times for all values of $\phi$. The dashed vertical lines indicate the values of $\phi$ (both very close to $\pi/6$) for which the corresponding curves attain the highest value.}
\label{fig_max_wig_neg_1_n1_0_n2_0}
\end{figure}

For the standard JC interactions ($m=1$), the above expression further simplifies to [using $L_1(z) = 1 - z$]
\begin{align} 
  \tilde{W}_1(\alpha, t) = 1 - 2  \left(1-2 |\alpha|^2\right) |y_2|^2 + 4 \text{Re}(x_1y_2^* \alpha). 
  \label{eqn_wig_case1_os1_vac_m_1}
\end{align}
Now, using $g_1=g_2=1/\sqrt{2}$ and $\Delta=0$, we get $x_1 = \cos\phi$, $y_2 = -\frac{i}{\sqrt{2}}\sin\phi \sin(t)$.  Substituting these in Eq.~\eqref{eqn_wig_case1_os1_vac_m_1}, we obtain
\begin{align} 
  \tilde{W}_1(\alpha, t) &=\sin^2 \phi \cos^2 t + \cos^2 \phi \cos^2\theta \nonumber \\
&+  \left(\sqrt{2}|\alpha| \sin \phi \sin t - \cos \phi \sin\theta\right)^2,
  \label{eqn_wig_case1_os1_vac_m_1_exact}
\end{align}
where we have used $\alpha = |\alpha| e^{i\theta}$. Evidently, $W_1(\alpha, t)$ is always positive resulting in $V_{W_{1-}}=0$ for all times and for all values of $\phi$, as confirmed by the numerical simulation (see the red line in Fig.~\ref{fig_max_wig_neg_1_n1_0_n2_0}).

On the other hand, substituting $m=2$ and $\phi=\pi/2$ into Eq.~\eqref{eqn_wig_case1_os1_vac} and using $L_2(z) = \frac{1}{2}(z^2 - 4z +2)$, we obtain, after a few steps of simplification,
\begin{align} 
  \tilde{W}_1(\alpha, t) =\cos^2 t +  \left(2|\alpha|^2-1 \right)^2 \sin^2 t. 
  \label{eqn_wig_case1_os1_vac_m_2_phi_pi2}
\end{align}
Again, the right-hand side is always positive.  Therefore, for $n_1=n_2=0$ and $m=2$ with $\phi=\pi/2$, we have $V_{W_{1-}}=0$. This is also borne out in the numerical simulation (see the endpoint of the blue line in Fig.~\ref{fig_max_wig_neg_1_n1_0_n2_0}).

For other values of $m$ and $\phi$, the nonlinear MPJC interactions yield substantial Wigner nonclassicalities in both oscillators periodically over time, as depicted by the blue curves in Fig.~\ref{fig_wig_neg_1} for the first oscillator with $m=3$. Additionally, comparing the blue curves in both Figs.~\ref{fig_wig_neg_1}(a) and \ref{fig_wig_neg_1}(b), we observe that $V_{W_{1-}}$ emerges after a relatively longer latent period for $\phi=\pi/2$. The impact of nonlinearity introduced by the multiphoton parameter $m$ significantly influences the attainment of higher nonclassicalities. In essence, the higher the value of $m$, the more pronounced the enhancements achieved in $V_{W_{1-}}$. These insights become more apparent in Fig.~\ref{fig_max_wig_neg_1_n1_0_n2_0}, where we depict the maximum achievable $V_{W_{1-}}$ as a function of $\phi$ for  $m=1$, 2, and 3, respectively. Interestingly, for both $m=2$ and 3, $V_{W_{1-}}$ does not reach its maximum value when $\phi=\pi/4$ (maximally superposed qubit), instead nearing $\phi=\pi/6$, as indicated by the vertical dashed lines in Fig.~\ref{fig_max_wig_neg_1_n1_0_n2_0}.

Furthermore, noteworthy enhancements in the initial Wigner negativity of Fock states $\ket{1}$ and $\ket{2}$ of the first oscillator are achieved during nonlinear evolution, as evidenced by the black and red curves in Fig.~\ref{fig_wig_neg_1} for both values of $\phi$. Interestingly, in contrast to $\phi=\pi/4$, the red and black curves for $\phi=\pi/2$ can dip below their initial value during the evolution. Comparing the red curves in both Fig.~\ref{fig_wig_neg_1}(a) and \ref{fig_wig_neg_1}(b) we observe that the extent of $V_{W_{1-}}$ is much higher for $\phi=\pi/2$ than for $\phi=\pi/4$. We have numerically verified that apart from these specific combinations, such enhancements in $V_{W_{1-}}$ are also present for other combinations of $n_1$, $n_2$, and $m$ that fall under case~1. Similar to $n_1=n_2=0$, we find that the nonlinearity introduced by $m$ significantly impacts whether the initial nonclassicality is surpassed.\\

\begin{figure}[ht]
\centering
\includegraphics{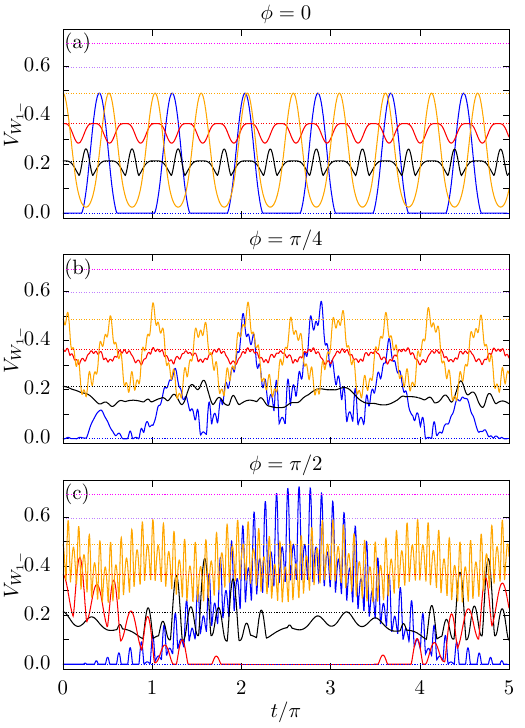}
\vspace{-3ex}
\caption{Nonlinear evolution of $V_{W_{1-}}$ of the first oscillator for case~2, that is, when $n_1<n_2=m$ or $n_2<n_1=m$, for (a) $\phi=0$, (b) $\phi=\pi/4$, and (c) $\phi=\pi/2$ (c), respectively. For all panels, the initial states of the first oscillator are $\ket{0}$ (blue), $\ket{1}$ (black), $\ket{2}$ (red), and $\ket{3}$ (yellow). Further, for the blue, black, red and yellow curves, the respective values of $(n_2,\, m)$ are (3,3), (3,3), (3,3), and (1,3); (b) (3,3), (2,2), (3,3), and (1,3); and (c) (3,3), (2,2), (0,2), and (1,3). Similar to Fig.~\ref{fig_wig_neg_1}, we set $g_1=g_2=1/\sqrt{2}$ and $\Delta=0$. The dashed horizontal lines correspond to $V_{W_{1-}}$ of the initial Fock states.}
\label{fig_wig_neg_2}
\end{figure}

\noindent{\bf Case 2:} Now, let us examine case 2, where $n_1<n_2=m$ or $n_2<n_1=m$. The time-evolved Wigner function of the first oscillator is found to be (see Appendix~\ref{wig_appendix} for details)
\begin{align}
  \tilde{W}_1(\alpha, t) &= (-1)^{n_1} \Big[\left(|x_1|^2+|x_2|^2 +|y_1|^2 +|y_3|^2\right) L_{n_1}(z) \nonumber \\
  &+ (-1)^{m} \left(|x_3|^2+ |y_2|^2 +|y_4|^2\right)  L_{n_1+m}(z)  \nonumber \\
  &+ 2^{m} \sqrt{\tfrac{n_1!}{(n_1+m)!}}  \big(\alpha^{m} \left(x_1 y_2^* + x_2 y_4^*\right) 
+\text{H.c.}\big)  L_{n_1}^{m}(z) \nonumber \\
&+ (-2)^{m} \sqrt{\tfrac{(n_1+m)!}{(n_1+2m)!}}  \left(x_3 y_5^* \alpha^{m}  +\text{H.c.}\right) L_{n_1+m}^{m}(z)\nonumber \\
&  + |y_5|^2 L_{n_1+2m}(z) \Big].   
  \label{wig_case2}
\end{align}    
Here, H.c.~stands for Hermitian conjugate.
Unlike the previous case, now the Wigner function does not remain constant for $\phi=0$, that is when all $y$ coefficients become zero, instead, it has the form
\begin{align}
  \tilde{W}_1(\alpha, t) &= (-1)^{n_1}\Big[\left(1-|x_3|^2 \right) L_{n_1}(z)   \nonumber \\ 
&\hspace{0.65in} + (-1)^{m} |x_3|^2 L_{n_1+m}(z)   \Big].   
  \label{wig_case2_phi_0}
\end{align}
This results in a nontrivial evolution of $V_{W_{1-}}$ in contrast to case~1, as depicted in Fig.~\ref{fig_wig_neg_2}(a). Now, if we assume that the first oscillator is prepared in the ground state $(n_1=0)$, we get
\begin{align}
 \tilde{W}_1(\alpha, t) = (-1)^{n_1} \left[1-|x_3|^2 + (-1)^{m} |x_3|^2 L_{m}(z)\right].   
  \label{wig_case2_phi_0_n1_0}
\end{align}
Therefore, whenever $|x_3|^2=1$, we obtain $W_1(\alpha, t) = \frac{2}{\pi}\,e^{-2|\alpha|^2} (-1)^{n_1} L_{m}\left(4|\alpha|^2\right)=W_2(\alpha, 0)$. 
As a result, $V_{W_{1-}}$ periodically attains the value of the initial $V_{W_{2-}}$ of the second oscillator (perfect Fock state swapping as discussed in the preceding section), as demonstrated by the blue curve in Fig.~\ref{fig_wig_neg_2}(a). Additionally, we observe that the enhancements in $V_{W_{1-}}$ diminish as the initial mean photon number of the first oscillator $n_1$ increases. Specifically, for $n_1=2$ and $n_2=m=3$ and also for $n_1=m=3$ and $n_2=2$ (corresponding to the situation $n_2<n_1=m$), no enhancements from their initial values are observed [as represented by the red and yellow curves respectively in Fig.~\ref{fig_wig_neg_2}(a)]. This occurs because, with a gradual increase in the value of $n_1$, the amplitude of $x_3$ decreases, suggesting a gradual reduction in the purity of the oscillator state, as explained in the preceding section.

On the other hand, when $\phi=\pi/2$, the $x$ coefficients vanish, significantly simplifying Eq. \eqref{wig_case2}, yielding
\begin{align}
  \tilde{W}_1(\alpha, t) &= (-1)^{n_1} \Big[\left(|y_1|^2+|y_3|^2 \right) L_{n_1}(z)   \nonumber \\ 
&\hspace{0.43in} + (-1)^{m} \left(|y_2|^2 +|y_4|^2\right) L_{n_1+m}(z) \nonumber\\
&\hspace{0.43in} + |y_5|^2 L_{n_1+2m}(z) \Big].   
  \label{wig_case2_phi_pi2}
\end{align}
Now, for the simple case when $n_1=0$, the above expression becomes
\begin{align}
  \tilde{W}_1(\alpha, t) &= |y_1|^2+|y_3|^2 + (-1)^{m} \left(|y_2|^2 +|y_4|^2\right) L_{m}(z)  \nonumber\\
&\hspace{0.3in} + |y_5|^2 L_{2m}(z) .   
  \label{wig_case2_phi_pi2_n1_0}
\end{align}
This clearly indicates that the initial ground state oscillator becomes nonclassical and the degree of the nonclassicality should increase with increasing $n_2=m$ value. We have numerically verified this, and the blue curve in Fig. \ref{fig_wig_neg_2}(c) illustrates the corresponding behavior when $n_2=m=3$. It is evident that the nonclassicality exceeds the corresponding value of $\ket{5}$ [dashed magenta line in Fig. \ref{fig_wig_neg_2}(c)]. For other combinations of $n_1$, $n_2$, and $m$, the dynamics appears to be similarly complex for this case, with comparatively less enhancement in their respective initial degree of nonclassicality.

Finally, for values of $\phi$ other than 0 and $\pi/2$, all the $x$ and $y$ coefficients contribute to the dynamics, leading to a notably intricate temporal evolution of $V_{W_{1-}}$ [see panel (b) of Fig.~\ref{fig_wig_neg_2}(b)]. We notice that the enhancements in $V_{W_{1-}}$ are smaller unless $n_1=0$. These inferences hold true for all combinations of $n_1$, $n_2$, and $m$ in this case, with the most significant ones depicted in Fig.~\ref{fig_wig_neg_2}.\\

\begin{figure}[ht]
\centering
\includegraphics{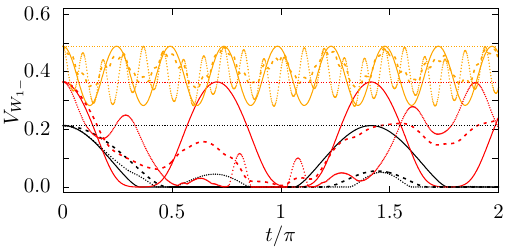}
\vspace{-3ex}
\caption{Nonlinear evolution of $V_{W_{1-}}$ of the first oscillator for case~3, that is, when $n_1=n_2=m=$ 1 (black), 2 (red), and 3 (yellow) with $\phi=0$ (solid), $\pi/4$ (dashed), and $\pi/2$ (dotted), respectively. Similar to Fig.~\ref{fig_wig_neg_1}, we set $g_1=g_2=1/\sqrt{2}$ and $\Delta=0$. Unlike the previous two cases, in this scenario, the nonlinear evolution fails to surpass the initial nonclassicality of the oscillator, at best $V_{W_{1-}}$ returns to its initial value at periodic intervals.
}
\label{fig_wig_neg_3}
\end{figure}

\noindent{\bf Cases 3 and 4:} We now move on to case 3, where $n_1=n_2=m$. The exact analytical expression for $W_1(\alpha, t)$ can be found in Appendix~\ref{wig_appendix}. The temporal evolution of $V_{W_{1-}}$ of the first oscillator is illustrated in Fig.~\ref{fig_wig_neg_3}. In complete contrast to the previous two cases, in this case, we do not observe any surpassing of the initial nonclassicality for any value of $\phi$. At best, $V_{W_{1-}}$ returns to its initial value at periodic intervals. This can be explained analytically for $\phi=0$. As mentioned in the preceding section, the oscillator state $\ket{m}$ (to be precise, $|x_1|^2$) returns to itself periodically for this scenario. The analytical expressions for the Wigner functions of the two oscillators for case 4 can also be found in Appendix~\ref{wig_appendix}. Similar to case 3, we have numerically verified that surpassing the initial nonclassicality for any value of $\phi$ is absent for this case as well.

In summary, after examining all four cases, we conclude that the multiphoton parameter $m$ plays a crucial role in surpassing the initial nonclassicalities of the photon-number states. In particular, we observe that $m$ should be at least greater than the mean photon number of one of the oscillators to achieve higher than the initial Wigner nonclassicality.

\begin{figure}[ht]
\centering
\includegraphics{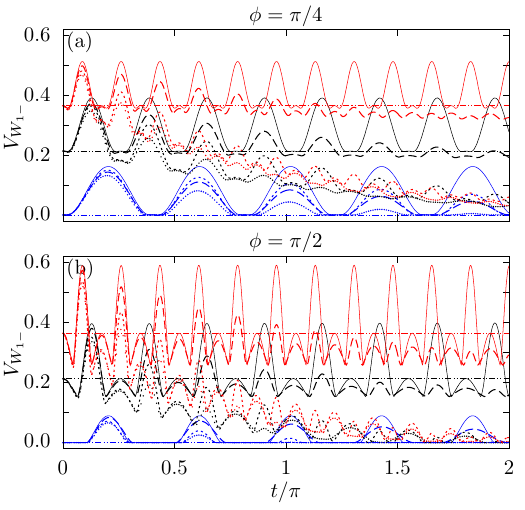}
\vspace{-3ex}
\caption{Role of environmentally induced decoherence and dephasing effects on $V_{W_{1-}}$ of the first oscillator for case~1 (as shown in Fig.~\ref{fig_wig_neg_1}) illustrated for $\bar{n}_{\text{th}}=0$ and (a) $\phi=\pi/4$ and $\phi=\pi/2$. In both panels, the long- (short-) dashed curves represent the temporal evolution in the presence of dissipation (dephasing) only, i.e., $\lambda_{d}=0$ ($\lambda_{r}=0$) with $\lambda_{r}=0.05$ ($\lambda_{d}=0.05$), while the dotted curves depict the cumulative effects of both dissipation and dephasing ($\lambda_{r}=\lambda_{d}=0.05$). The solid curves correspond to unitary dynamics (Fig.~\ref{fig_wig_neg_1}). The horizontal dash-double-dotted lines correspond to the initial nonclassicality of the Fock states.}
\label{fig_wig_neg_1_decoh_nth_0}
\end{figure}


\section{\label{subsec:wigner}Environmental effects}
Even with the tremendous experimental progress in harnessing and isolating quantum systems from environmentally induced effects, shielding the system entirely remains a challenge in any realistic quantum platform. So far in our analysis, we have ignored such contributions completely. In this section, we are going to numerically estimate the degree to which the nonclassicalities in Fock states are affected by considering realistic system-environment coupling parameters.
In the Lindblad formalism, the evolution of the tripartite system's density matrix $\rho_{S}(t)$ is described by the standard master equation
\begin{align}
 \displaystyle\frac{d\rho_{S}}{dt}  = -i \left[H, \rho_{S}\right] + \sum_{k} \lambda_{k}\left( A_{k} \rho_{S} A_{k}^{\dagger} -  \frac{1}{2}\left[\rho_{S}, A_{k}^{\dagger} A_{k}\right] \right).
 \label{rhot_decoh}
\end{align}
Here, the environment couples to the system via the operators $A_{k}$ with coupling rates $\lambda_{k}$. We assume a common thermal environment for the entire system with thermal energy $\bar{n}_{\text{th}}$ for simplicity. For dissipation, we consider Lindblad operators $a_i$ and $\sigma_-$ with dissipation rates $\sqrt{\lambda_{r}(1+\bar{n}_{\text{th}})}$, where $i=1,\,2$. We also set equal coupling coefficients to simplify the analysis. Similarly, for relaxation, the Lindblad operators are $a_i^\dagger$ and $\sigma_+$ with relaxation rates $\sqrt{\lambda_{r} \bar{n}_{\text{th}}}$.
Additionally, we include the effect of dephasing through environmental interactions. Here, the relevant Lindblad operators are $a_i^\dagger a_i$ and $\sigma_z$ with equal coupling rate $\sqrt{\lambda_{d}}$, which does not depend on $\bar{n}_{\text{th}}$. We keep in mind that, in the asymptotic limit, all states converge to a thermal state with no negativity in the Wigner function.


\begin{figure}[ht]
\centering
\includegraphics{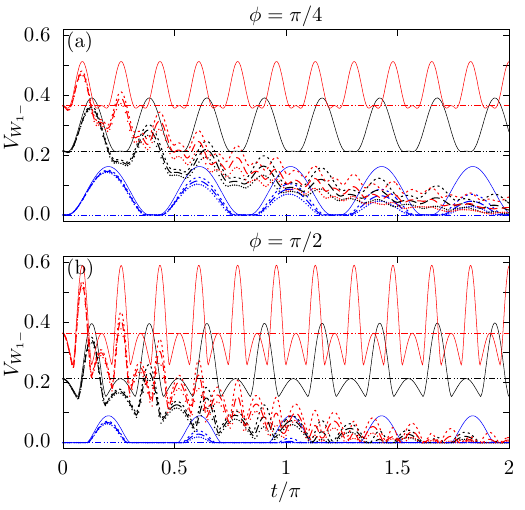}
\vspace{-3ex}
\caption{The role of thermal environment on $V_{W_{1-}}$ of the first oscillator for case~1 (as shown in Fig.~\ref{fig_wig_neg_1}) for (a) $\phi=\pi/4$  and $\phi=\pi/2$. In both panels, the short-dashed, the long-dashed, and dotted curves correspond to  $\bar{n}_{\text{th}}=0$, 0.1, and 0.2, respectively. Further, $\lambda_{r}=0.05$ and $\lambda_{d}=0$. The solid curves correspond to unitary dynamics (Fig.~\ref{fig_wig_neg_1}). 
}
\label{fig_wig_neg_1_decoh_nth_all}
\end{figure}

We begin by assuming $\bar{n}_{\text{th}}=0$, which indicates that the system interacts with a vacuum environment. The relevant Lindblad operators are $a_i$, $\sigma_-$, $a_i^\dagger a_i$ and $\sigma_z$. In Fig.~\ref{fig_wig_neg_1_decoh_nth_0}, we illustrate the detrimental effects of the environment on the temporal evolution of $V_{W_{1-}}$ corresponding to case~1. We consider three sources of losses: (i) pure dissipation ($\lambda_r=0.05$ and $\lambda_d=0$), (ii) pure dephasing ($\lambda_r=0$ and $\lambda_d=0.05$), and (iii) the combined effect of dissipation and dephasing ($\lambda_r=0.05$ and $\lambda_d=0.05$). These are represented by the long-dashed, short-dashed, and dotted curves, respectively, in Fig.~\ref{fig_wig_neg_1_decoh_nth_0}. It is evident that dissipation losses outweigh dephasing in most scenarios, except when $\phi=\pi/4$ and $n_1=0$ [blue dashed curves in Fig.~\ref{fig_wig_neg_1_decoh_nth_0}(a)]. Moreover, we find that the higher initial Fock states are more susceptible to noise, as expected.

The role of the temperature of the thermal bath $\bar{n}_{\text{th}}$ on the temporal evolution of $V_{W_{1-}}$ is investigated in Fig.~\ref{fig_wig_neg_1_decoh_nth_all}. Here all the Lindblad operators contribute to the dynamics. As expected, increasing the temperature of the thermal bath results in a faster reduction in nonclassicality. Similar to Fig.~\ref{fig_wig_neg_1_decoh_nth_0}, the nonclassicality of the higher Fock states degrades much faster.

\begin{figure}[ht]
\centering
\includegraphics{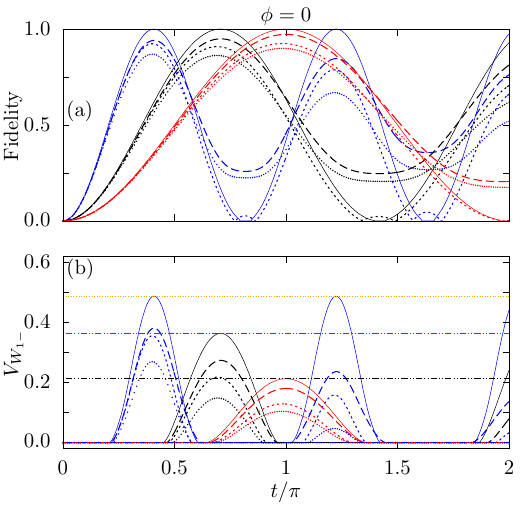}
\vspace{-3ex}
\caption{Robustness of the fidelity of the Fock state transfer and $V_{W_{1-}}$ of the first oscillator illustrated for case~2 against a vacuum environment, $\bar{n}_{\text{th}}=0$ for $\phi=0$. In both panels, the long- (short-) dashed curves correspond to $\lambda_{d}=0$ ($\lambda_{r}=0$) with $\lambda_{r}=0.05$ ($\lambda_{d}=0.05$), while the dotted curves correspond to $\lambda_{r}=\lambda_{d}=0.05$. The solid curves correspond to the unitary dynamics (Fig.~\ref{fig_wig_neg_1}).}
\label{fig_fid_wig_neg_1_decoh_nth_0_case2}
\end{figure}

Next, we focus on case~2 corresponding to $n_1<n_2=m$ and also test the robustness of the excitation transfer protocol as discussed earlier, in addition to the robustness of $V_{W_{1-}}$, against environmental interactions. The extent to which both these quantities get affected is depicted in  Fig.~\ref{fig_fid_wig_neg_1_decoh_nth_0_case2}, which corresponds to $\bar{n}_{\text{th}}=0$. It is evident that the qualitative degradation in both the fidelities and $V_{W_{1-}}$ is similar to the earlier case. We have verified that with the gradual increase in $\bar{n}_{\text{th}}$ these quantities deteriorate even further. Similar conclusions can also be drawn for cases~3 and~4.




\section{\label{sec:conc}Conclusions}
We have found and explored the analytical solution of the time-evolved state vector of the tripartite $m$-photon JC system considering a pure initial state in which the qubit is in a superposition state and the two quantized harmonic oscillators are in arbitrary Fock states $\ket{n_1}$ and $\ket{n_2}$, respectively. Depending on the specific values of $n_1$, $n_2$, and $m$, we identified four different cases and obtained exact analytical solutions for most of them. Furthermore, we analytically extracted the time-evolved oscillator states by tracing over the relevant subsystems and showed that perfect swapping of Fock states between oscillators $\ket{0,\, m}\longrightarrow\ket{m,\, 0}$ can be achieved under symmetrically coupled MPJC interactions in a specific case. This scheme is entirely different from the one described in~\cite{Sun_PRA_2006}, which involves the standard tripartite JC model where large detuning is required, followed by the adiabatic elimination of the qubit. Interestingly, such swapping of quantum states is also achievable through suitable beamsplitter interactions; here such single step swapping is the result of an exact analytical evolution of a fairly complicated, tripartite, nonlinear spin-boson Hamiltonian. Furthermore, by considering asymmetrically coupled MPJC interactions, we demonstrated that the $\ket{n_1, m} \longrightarrow \ket{n_1+m, 0}$ transition can be engineered within the tripartite MPJC model (see Appendix~\ref{asymm_coupoling}). Notably, such complete transfer of excitations is not achievable in a quantum beamsplitter (see Appendix~\ref{bs_output}).

In the latter half of the article, we carried out a detailed analysis of how the nonclassicalities of the initial oscillator Fock states evolve under such nonlinear Hamiltonian evolution considering diverse system parameters. Following previous work~\cite{Anatole_joptb_2004,Arkhipov_sci_rep_2018}, we quantified the degree of nonclassicality of a quantum state by the volume of the negative regions of its corresponding Wigner function. Besides producing substantial enhancements in the initial value for higher photon number states, our analysis revealed that the nonlinear MPJC interaction, driven solely by the initial qubit energy (with both oscillators initialized in the vacuum state), yields nontrivial Wigner negativities in the oscillators. Interestingly, it turns out that the additional nonlinearity of the multiphoton interactions $m$ dictates the eventual outcome of surpassing the initial nonclassicalities of the photon number states. 

For completeness, we also tested the robustness of such nonclassicalities under additional environmentally induced interactions and the role of imperfect matching of frequencies between the discrete and continuous variable quantum systems on the Wigner nonclassicalities. It would be interesting to conduct a similar analysis incorporating other initial qubit and oscillator states, including incoherent ones. The numerical results presented in this paper were obtained using the {\it QuTiP} library~\cite{qutip}.

It is important to note that the augmented nonlinearity due to the $m$-photon  generalization appears to be nontrivial, as evidenced by the fact that the additional constant of motion present in the standard tripartite JC model, ceases to hold for $m > 1$. The details of this observation are provided in Appendix~\ref{sec:const_of_motion}. As discussed in Sec.~\ref{intro}, several past studies exploited the additional symmetry in the standard tripartite JC model~\cite{Dutra_PRA_1993,Dutra_PRA_1994,Benivegna_JPhysA_1994,Abdalla_opt_commun_2002,Messina_j_mod_opt_2003,Wildfeuer_PRA_2003,larson_jmo_2006,Rodriguez_jphysa_2016,Rodriguez_sci_rep_2018,Alderete_2021}. This symmetry enables a canonical transformation of the degenerate Hamiltonian, simplifying it to a form involving only one JC interaction with the symmetric normal mode while the antisymmetric mode decouples. In Appendix~\ref{sec:canonical_H}, we demonstrate that such decoupling is not possible for $m > 1$. This finding could have significant implications for the results obtained for $m = 1$ and opens up an intriguing avenue for future research.

Finally, we also explored the squeezing properties of time-evolved oscillator states across all four distinct cases. Our findings reveal that none of the oscillators exhibit any conventional quadrature squeezing. However, the highly nonclassical nature of these states suggests the potential for squeezing extraction through various distillation techniques~\cite{laha_non-gaussian_2022}. Investigating the extent of distillable squeezing, if present, remains an interesting exercise. Additionally, from a theoretical standpoint, the next logical step would involve probing these phenomena within the multiphoton rendition of the standard double JC system, which includes two qubits and two oscillators~\cite{laha_josab_2023}.

Having obtained the reduced oscillator states by tracing over the relevant subsystems, which tends to reduce the purity of the quantum states, a natural next step would involve transitioning from this deterministic approach to a probabilistic one. This would entail obtaining the target oscillator state by measuring the qubit and the remaining oscillator.
Another facet of the study involves examining the dynamics of bosonic entanglement between the oscillators and qubit coherence, two of the most prominent resources in modern quantum technology. This aspect has already been extensively addressed in Ref.~\cite{laha_spinboson_2024}. 

Apart from advancing our theoretical understanding of the role of multiphoton interactions in various nonclassical phenomena, we believe that the results presented in this work will provide significant impetus to the rapidly evolving experimental exploration with multiphoton processes in various quantum platforms~\cite{Pan_RMP_2012,Zhang_adv_q_tech_2021,Yang_nature_photon_2022}, and potentially contribute to the ultimate applications in photonic quantum technology, involving universal and fault-tolerant processing, where most advanced, nonclassical, non-Gaussian optical quantum states are required.


\begin{acknowledgments}
We acknowledge funding from the BMBF in Germany (QR.X, PhotonQ, QuKuK, QuaPhySI), from the Deutsche Forschungsgemeinschaft (German Research Foundation) through Project No. 429529648, TRR 306 QuCoLiMa (Quantum Cooperativity of Light and Matter), and from the EU's Horizon Research and Innovation Actions (CLUSTEC).
\end{acknowledgments}

\appendix
\section{\label{mat_elements} The $M$ matrices }

\noindent{\bf Case 2:}
Given below are the expressions for the $M$ matrices for case 2:
\begin{align}
    M_{2_x} =     
    \begin{pmatrix}
        - \Delta' & g_{2_m} & 0 \\
        g_{2_m} &  \Delta' & g_{1_{n_1m}}  \\
        0 & g_{1_{n_1m}} & - \Delta' 
    \end{pmatrix},
  \label{mx_mat_case2}
\end{align}
where $\Delta'=\Delta/2$, $g_{1_{n_1m}}=\sqrt{(n_1+m)!/n_1!}\, g_1$, and $g_{2_m} = \sqrt{m!}\,g_2$, and
\begin{align}
    M_{2_y} =     
    \begin{pmatrix}
        \Delta' & g_{1_{n_1m}} & g'_{2_m} & 0 & 0 \\
        g_{1_{n_1m}} & -\Delta' & 0 & g_{2_m} & 0 \\
        g'_{2_m} & 0 & -\Delta' & 0 & 0 \\
        0 & g_{2_m} & 0 & \Delta' & g'_{1_{n_1m}} \\
        0 & 0 & 0 & g'_{1_{n_1m}} & -\Delta' 
    \end{pmatrix},
\label{my_mat_case2}
\end{align}
where $g'_{2_m}=\sqrt{(2m)!/m!}\, g_2$,  and $g'_{1_{n_1m}}=\sqrt{(n_1
+2m)!/(n_1+m)!}\, g_1$. \\

\noindent {\bf Case 3:}
Given below are the expressions for the $M$ matrices for case 3:
\begin{align}
    M_{3_x} =     
    \begin{pmatrix}
        -\Delta' & g_{1_m} & g_{2_m} & 0 & 0 \\
        g_{1_m} & \Delta' & 0 & g'_{2_m} & 0 \\
        g_{2_m} & 0 & \Delta' & 0 & g'_{1_m} \\
        0 & g'_{2_m} & 0 & -\Delta' & 0 \\
        0 & 0 & g'_{1_m} & 0 & -\Delta' 
    \end{pmatrix},
\label{mx_mat_case3}
\end{align}
where $g_{1_m}=\sqrt{m!} g_1$ and $g'_{1_m}=\sqrt{(2m)!/m!} g_1$, and
\begin{align}
  M_{3_y} =    
    \begin{pmatrix}
        \Delta' & g'_{1_m} & g'_{2_m} & 0 & 0 & 0 & 0 \\
        g'_{1_m} & -\Delta' & 0 & g_{2_m} & 0 & 0 & 0 \\
        g'_{2_m} & 0 & -\Delta' & 0 & g_{1_m} & 0 & 0 \\
        0 & g_{2_m} & 0 & \Delta' & 0 & g''_{1_m} & 0 \\
        0 & 0 & g_{1_m} & 0 & \Delta' & 0 & g''_{2_m} \\
        0 & 0 & 0 & g''_{1_m} & 0 & -\Delta' & 0 \\
        0 & 0 & 0 & 0 & g''_{2_m} & 0 & -\Delta' 
    \end{pmatrix},
  \label{my_mat_case3}
\end{align}
where $g''_{1_m}=\sqrt{(3m)!/(2m)!} g_1$ and $g''_{2_m}=\sqrt{(3m)!/(2m)!} g_2$.\\

The reduced density matrices for the two oscillators in the effective basis states $\ket{0}$, $\ket{m}$, $\ket{2m}$, and $\ket{3m}$ are given by
\begin{widetext}
    \begin{align}
    \label{eqn_rho1_3}
    \rho_{1}^{(3)} =     
    \begin{pmatrix}
        |x_2|^2+|x_5|^2+|y_5|^2+|y_7|^2 & x_2 y_1^* + x_5 y_3^* & 0 & 0 \\
        x_2^* y_1 + x_5^* y_3 & |x_1|^2+|x_3|^2+|y_1|^2+|y_3|^2 & x_1 y_2^* + x_3 y_4^* & 0 \\
        0 & x_1^* y_2 + x_3^* y_4 & |x_4|^2+|y_2|^2+|y_4|^2 & x_4 y_6^* \\
        0 & 0 & x_4^* y_6 & |y_6|^2
    \end{pmatrix},
\end{align}
and
\begin{align}
    \label{eqn_rho2_3}
    \rho_{2}^{(3)} =     
    \begin{pmatrix}
        |x_3|^2+|x_4|^2+|y_4|^2+|y_6|^2 & x_3 y_1^* + x_4 y_2^* & 0 & 0 \\
        x_3^* y_1 + x_4^* y_2 & |x_1|^2+|x_2|^2+|y_1|^2+|y_2|^2 & x_1 y_3^* + x_2 y_5^* & 0 \\
        0 & x_1^* y_3 + x_2^* y_5 & |x_5|^2+|y_3|^2+|y_5|^2 & x_5 y_7^* \\
        0 & 0 & x_5^* y_7 & |y_7|^2
    \end{pmatrix}.
\end{align}
\end{widetext}

\noindent {\bf Case 4:}
The exact forms of the state vectors $\ket{\psi_x(t)}_4$ and $\ket{\psi_y(t)}_4$ are given by
\begin{align}
  \ket{\psi_x(t)}_4 &= \sum_{k=0}^{\ell_2-1} \Big(x_{4k+1} \ket{g,\, n_1+km,\, n_2-km} \nonumber\\
  &+ x_{4k+3} \ket{e,\, n_1+k m,\, n_2-(k+1)m}\Big) \nonumber\\
  &+\sum_{k=0}^{\ell_1-2} \Big(x_{4k+2} \ket{e,\, n_1-(k+1)m,\, n_2+km} \nonumber\\
      &+ x_{4k+4} \ket{g,\, n_1-(k+1)m,\, n_2+(k+1)m} \Big),
\end{align}
and
\begin{align}
  \ket{\psi_y(t)}_4 &= \sum_{k=0}^{\ell_2-1}\Big(y_{4k+1} \ket{e,\, n_1-km,\, n_2+km} \nonumber\\
       &+ y_{4k+3} \ket{g,\, n_1-km,\, n_2+(k+1)m}\Big) \nonumber\\
       &+\sum_{k=0}^{\ell_1-1} \Big(y_{4k+2} \ket{g,\, n_1+(k+1)m,\, n_2-k m} \nonumber\\
       &+ y_{4k+4} \ket{e,\, n_1+(k+1)m,\, n_2-(k+1)m} \Big),
\end{align}
respectively.
Note that for this case $\ell_1,\,\ell_2>2$, as $\min(m)=1$. As mentioned earlier, $\ket{\psi_x(t)}_4$ and $\ket{\psi_y(t)}_4$ contain $2\ell_1+2\ell_2-3$  and $2\ell_1+2\ell_2-1$) basis states, respectively, with $\ell_i$ representing the smallest positive integer for which $n_i-\ell_i m<0$ ($i=1,\,2$).



The two $M$ matrices assume symmetric pentadiagonal structures. The nonzero elements of the upper half including the diagonal elements are 
\begin{align}
    \left(M_{4_x}\right)_{i,i} &= \left\{ 
    \begin{array}{rcl}
     -\Delta' & \mbox{for} & i=4k+1, 4k+4 \\ 
      \Delta' & \mbox{for} & i=4k+2, 4k+3, 
    \end{array}\right., \nonumber\\
    \left(M_{4_x}\right)_{0,1} &= \sqrt{\tfrac{n_1!}{(n_1-m)!}} g_1, \nonumber \\
    \left(M_{4_x}\right)_{i,i+2} &= \left\{ 
    \begin{array}{rcl}
     \sqrt{\frac{(n_2-j_1m)!}{(n_2-(j_1+1)m)!}} g_2 & \mbox{for}
       & i=4k+1, \\ 
     \sqrt{\frac{(n_2+j_2m)!}{(n_2+(j_2-1)m)!}} g_2 & \mbox{for}
       & i=4k+2, \\
      \sqrt{\frac{(n_1+j_2m)!}{(n_1+(j_2-1)m)!}} g_1 & \mbox{for}
       & i=4k+3, \\
      \sqrt{\frac{(n_1-j_1m)!}{(n_1-(j_1+1)m)!}} g_1 & \mbox{for}
       & i=4k+4, 
    \end{array}\right..
\end{align}
Here, $j_1=\lfloor \frac{i}{4} \rfloor$, $j_2=\lceil \frac{i}{4} \rceil$, and $k=0,\,1,\,2,\ldots$. Note that $\lfloor \cdot \rfloor$ and $\lceil \cdot \rceil$ denote floor and ceiling functions, respectively. Similarly, we can show that
\begin{align}
    \left(M_{4_y}\right)_{i,i} &= \left\{ 
    \begin{array}{rcl}
     \Delta'  & \mbox{for} & i=4k+1, 4k+4 \\ 
    -\Delta' & \mbox{for} & i=4k+2, 4k+3, 
    \end{array}\right., \nonumber\\
    \left(M_{4_y}\right)_{0,1} &= \sqrt{\tfrac{(n_1+m)!}{n_1!}} g_1, \nonumber \\
    \left(M_{4_y}\right)_{i,i+2} &= \left\{ 
    \begin{array}{rcl}
     \sqrt{\frac{(n_2+j_2m)!}{(n_2+(j_2-1)m)!}} g_2 & \mbox{for}
       & i=4k+1, \\ 
     \sqrt{\frac{(n_2-j_1m)!}{(n_2-(j_1+1)m)!}} g_2 & \mbox{for}
       & i=4k+2, \\
      \sqrt{\frac{(n_1-j_1m)!}{(n_1-(j_1+1)m)!}} g_1 & \mbox{for}
       & i=4k+3, \\
      \sqrt{\frac{(n_1+(j_1+1)m)!}{(n_1+j_1m)!}} g_1 & \mbox{for}
       & i=4k+4, 
    \end{array}\right..
\end{align}

The  time-evolved reduced state of the first oscillator for this case is given by
\begin{align}
  \label{c4r1}
  \rho_{1}^{(4)}(t) &= \sum_{k=0}^{k_\text{max}} \big[ \rho_{n_1+km,\,n_1+km} \ket{n_1+km} \bra{n_1+km} \nonumber\\
         &\,\,\,+ \rho_{n_1-km,\,n_1-km} \ket{n_1-km} \bra{n_1-km} \nonumber\\
         &\,\,\,+ \big(\rho_{n_1+km,\,n_1+(k+1)m} \ket{n_1+km} \bra{n_1+(k+1)m} \nonumber\\
         &\,\,\,+\rho_{n_1-(k+1)m,\,n_1-km} \ket{n_1-(k+1)m} \bra{n_1-km} \nonumber\\
         &\,\,\,+\text{H.c.}\big) \big],
\end{align}
where $\rho_{n_1+km,\,n_1+km}=|x_{4k+1}|^2 + |x_{4k+3}|^2 +|y_{4k-2}|^2 + |y_{4k}|^2$, 
$\rho_{n_1-km,\,n_1-km} = |x_{4k-2}|^2 + |x_{4k}|^2 + |y_{4k+1}|^2+|y_{4k+3}|^2$,
$\rho_{n_1+km,\,n_1+(k+1)m}= x_{4k+1} y^*_{4k+2}+x_{4k+3} y^*_{4k+4}$, and
$\rho_{n_1-(k+1)m,\,n_1-km}= x_{4k+4} y^*_{4k+3} + x_{4k+2} y^*_{4k+1}$. Further, $k_\text{max}=\max(\ell_1-1,\ell_2-1)$.

Similarly, for the second oscillator, we get
\begin{align}
  \label{c4r2}
  \rho_{2}^{(4)}(t) &= \sum_{k=0}^{k_\text{max}} \big[\rho_{n_2-km,\,n_2-km} \ket{n_2-km} \bra{n_2-km} \nonumber \\ 
&\,\,\,+ \rho_{n_2+km,\,n_2+km}  \ket{n_2+km} \bra{n_2+km} \nonumber \\ 
&\,\,\,+ \big(\rho_{n_2-(k+1)m,\,n_2-km} \ket{n_2-(k+1)m} \bra{n_2-k m} \nonumber \\ 
&\,\,\,+ \rho_{n_2+km,\,n_2+(k+1)m} \ket{n_2+km} \bra{n_2+(k+1)m}   \nonumber \\
&\,\,\,+ \text{H.c.}\big) \Big]+\big( \rho_{n_2,\,n_2+m} \ket{n_2} \bra{n_2+m} \nonumber\\
&\,\,\,+ \rho_{n_2,\,n_2-m} \ket{n_2} \bra{n_2-m} + \text{H.c.}\big),
\end{align}
where
$\rho_{n_2-km,\,n_2-km} = |x_{4k+1}|^2 + |y_{4k+2}|^2 + |x_{4k-1}|^2 + |y_{4k}|^2$,
$\rho_{n_2+km,\,n_2+km} = |x_{4k}|^2 + |y_{4k-1}|^2+|x_{4k+2}|^2 + |y_{4k+1}|^2$,
$\rho_{n_2-(k+1)m,\,n_2-km} = x_{4k+5} y_{4k+2}^* + x_{4k+3} y^*_{4k}$,
$\rho_{n_2+km,\,n_2+(k+1)m} =  x_{4k} y^*_{4k+3} + x_{4k+2} y^*_{4k+5}$, 
$\rho_{n_2,\,n_2+m}= x_1 y_3^* + x_4^* y_2$, and 
$\rho_{n_2,\,n_2-m}= x_2y_4^* + y_1x_3^*$. As before, $k_\text{max}=\max(\ell_1-1,\ell_2-1)$.

    

\section{\label{sec:detuning} The role of detuning $\Delta$}
Thus far in our analysis of the volume of the Wigner negativities $V_{W_{1-}}$, we have always assumed perfect matching of frequencies between the qubit and the oscillators, that is, $\Delta=0$. Here, we examine the changes that manifest in $V_{W_{1-}}$ when there is imperfect matching of frequencies. 

\begin{figure}
    \centering
    \includegraphics{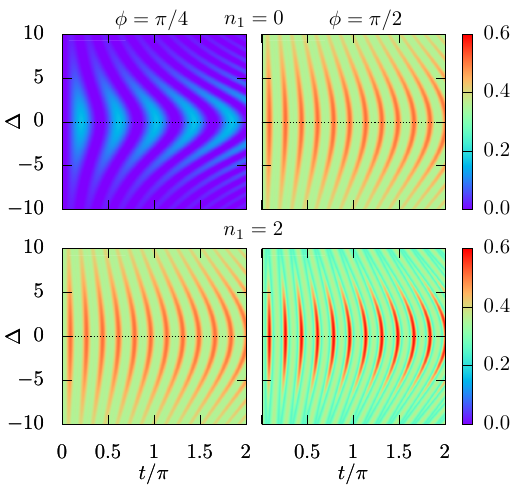}
    \caption{Wigner nonclassicality of the first oscillator $V_{W_{1-}}$ displayed as a function of time and the detuning $\Delta$ for case 1. In all panels, $m=3$ and $n_2=0$. The left and right columns represent $\phi=\pi/4$ and $\pi/2$, respectively, while the top and bottom rows correspond to $n_1=0$ and 2, respectively. As previously mentioned, $g_1=g_2=1/\sqrt{2}$.}
    \label{fig:detuning_case_1}
\end{figure}

For case 1, this is illustrated in Fig.~\ref{fig:detuning_case_1}. From both Eq.~\eqref{eqn_y_coeff_case1} and Fig.~\ref{fig:detuning_case_1}, it is evident that for fixed values of $n_1$, $n_2$ and $m$, the detuning $\Delta$  increases the periodicity and simultaneously decreases the amplitude of the oscillations. In particular, the farther we move away from perfect resonance ($\Delta=0$), the lower the degree of nonclassicality. Eventually, in the dispersive limit, i.e., when $\Delta\gg g_1, g_2$, the nonclassicality vanishes almost entirely. This is consistent with the expected behavior in the dispersive regime~\cite{gerry_knight_2004}.
Similar to case 1, we have numerically verified that the qualitative changes in the behavior of $V_{W_{1-}}$ due to nonzero detuning also remain similar for other cases.
\section{\label{wig_appendix} Derivations of the Wigner functions}
 In this appendix we give details on analytically deriving the Wigner functions of the time-evolved oscillator states. For any generic quantum state $\varrho$, characterized by its bosonic annihilation and creation operators $a$ and $a^\dagger$, the associated Wigner function $W(\alpha)$ is conventionally defined as~\cite{Cahill1_PhysRev_1969,Cahill2_PhysRev_1969}
\begin{align} 
  \label{wp1a}
  W(\alpha) &= \frac{1}{\pi} {\rm Tr}\,\left[\varrho\, T(\alpha)\right], 
\end{align}
where 
\begin{align}
  T(\alpha) = \int \frac{d^2\xi}{\pi}  \,\exp{(\alpha \xi^*-\alpha^* \xi)} D(\xi),
\end{align}
is a Hermitian operator and $D(\xi) = \exp (\xi a^\dagger-\xi^* a)$ is the standard displacement operator. The elements of the Hermitian matrix $T(\alpha)$ in the number basis are given by~\cite{Cahill1_PhysRev_1969}
\begin{align} 
  \braket{ k_1 | T(\alpha) | k_2 } = 2(-1)^{k_1}  e^{-2|\alpha|^2}& \sqrt{\tfrac{k_1!}{k_2!}}  \,(2\alpha^*)^{k_2-k_1} \nonumber\\
  &\times L_{k_1}^{k_2-k_1} \left(4|\alpha|^2\right),
  \label{wp2c}
\end{align}
for $k_2 \geq k_1$, and
\begin{align} 
  \braket{ k_1 | T(\alpha) | k_2 } = 2(-1)^{k_2} e^{-2|\alpha|^2} & \sqrt{\tfrac{k_2!}{k_1!}}\, (2\alpha)^{k_1-k_2} \nonumber\\
  &\times L_{k_2}^{k_1-k_2} \left(4|\alpha|^2\right),
  \label{wp2d}
\end{align}
for $k_2<k_1$. Here, $L_{k_1}^{k_2}(x)$ is the associated Laguerre polynomial. \\

\noindent{\bf Case 1:}
The reduced density matrices of the two oscillators in this case ($n_1,\,n_2 <m$) are given by Eqs.\eqref{eqn_rho_os1_1} and~\eqref{eqn_rho_os2_1}, respectively. Since only two basis states $\ket{n_i}$ and $\ket{n_i+m}$ are involved for both oscillators in this case ($i=1,\,2$), we can easily obtain the associated Wigner functions following the above procedure. For the second oscillator we have
\begin{align} 
\label{wp3a}
  W_2(\alpha,t) &= \frac{1}{\pi}\big[(|x_1|^2+|y_1|^2 +|y_2|^2) \langle n_2|T(\alpha)|n_2 \rangle \nonumber\\
  &+  \left(x_1 y_3^* \langle n_2+m|T(\alpha)|n_2 \rangle +\text{H.c.}\right)\nonumber\\
  &+|y_3|^2 \langle n_2+m|T(\alpha)|n_2+m \rangle\big].
\end{align}
Now, using Eqs. \eqref{wp2c} and \eqref{wp2d}, we finally obtain
\begin{align}
  \tilde{W}_2(\alpha, t) &= (-1)^{n_2} \left[ \left(|x_1|^2+|y_1|^2 +|y_2|^2\right)L_{n_2}(z) \right . \nonumber \\
    &+ 2^{m} \sqrt{\tfrac{n_2!}{(n_2+m)!}}   \left(x_1 y_3^* \alpha^{m} + y_3 x_1^* (\alpha^*)^{m}\right) L_{n_2}^{m}(z)   \nonumber \\
    &+ \left. (-1)^{m} |y_3|^2  L_{n_2+m}(z) \right],
\label{eqn_wig_case1_os2}
\end{align}
where $\tilde{W}_2(\alpha, t) = \frac{2}{\pi} e^{-2|\alpha|^2} W_2(\alpha, t)$, and $z=4|\alpha|^2$ as usual.
Following a similar procedure, we also obtain the Wigner function of the first oscillator and it is given by Eq.~\eqref{eqn_wig_case1_os1}.\\


\noindent{\bf Case 2:}
The reduced density matrices of the two oscillators in this case ($n_1<n_2=m$) are given by Eqs.~\eqref{eqn_rho_os1_2} and \eqref{eqn_rho_os2_2}, respectively. Here, only three basis states are involved for both oscillators. The associated Wigner function for the first oscillator is given by Eq.~\eqref{wig_case2}, while for the second oscillator, we have
\begin{align} \label{wp10}
\tilde{W}_2(\alpha, t) &= \left(|x_2|^2+|x_3|^2+ |y_4|^2 +|y_5|^2\right) L_{n_2}(z) \nonumber \\
  &+(-1)^{m}\left(|x_1|^2+|y_1|^2 +|y_2|^2\right)  L_{n_2+m}(z) \nonumber \\
  & +2^{m} \sqrt{\tfrac{n_2!}{n_2+m!}} \left\{\left(x_2 y_1^*+x_3y_2^*\right)  \alpha^{m} + \text{H.c.}\right\} L_{n_2}^{m}(z)\nonumber\\
  &+ (-2)^{m} \sqrt{\tfrac{n_2+m!}{(n_2+2m)!}} \left(x_1 y_3^* \alpha^{m} + \text{H.c.}\right) L_{n_2+m}^{m}(z)\nonumber\\
  &+ |y_3|^2 L_{n_2+2m}(z).
\end{align}

\noindent{\bf Case 3:} For this case, the reduced density matrices for the two oscillators are given by Eqs.\eqref{eqn_rho1_3} and \eqref{eqn_rho2_3}, respectively. The associated Wigner function for both oscillators can be expressed as
\begin{align} 
\tilde{W}(\alpha,t) &= \rho_{0,0} + \rho_{2m,2m} L_{2m}(z) + (-1)^m \big\{\rho_{m,m}  L_m(z) \nonumber\\
&\,\,\,+ \rho_{3m,3m} L_{3m}(z) \big\} + (2\alpha)^m\Big[ \tfrac{1}{\sqrt{m!}}\, \rho_{0,m} \nonumber \\
&\,\,\, + (-1)^m \sqrt{\tfrac{m!}{(2m)!}}\, \rho_{m,2m}  L_m^m(z)  \nonumber \\
&\,\,\,+ \sqrt{\tfrac{(2m)!}{(3m)!}}  \,\rho_{2m,3m} L_{2m}^m(z)   + {\rm H.c.} \Big],
\end{align}
where $\rho$ is identified with $\rho_{i}^{(3)}$ with $i=1,2$.\\

{\noindent \bf Case 4\,:} The reduced density matrix of the first oscillator $\rho_1^{(4)}$ for this case is given by in Eq.~\eqref{c4r1}. It can be easily shown that
\begin{align}
\tilde{W}_1(\alpha, t) &= \sum_{k=0}^{k_\text{max}} (-1)^{n_1}\big[ (-1)^{km}\rho_{n_1+km,\,n_1+km}  L_{n_1+km}(z) \nonumber\\
         &+ (-1)^{-km}\rho_{n_1-km,\,n_1-km} L_{n_1-km}(z) \nonumber\\
         &+ \big\{(2\alpha)^m\big((-1)^{km} \sqrt{\tfrac{(n_1+km)!}{(n_1+(k+1)m)!}} L_{n_1+km}^{m}(z)     \nonumber\\
         &\quad\times \rho_{n_1+km,\,n_1+(k+1)m}  \nonumber \\
         &+ (-1)^{-(k+1)m} \sqrt{\tfrac{(n_1-(k+1)m)!}{(n_1-km)!}} L_{n_1-(k+1)m}^{m}(z) \nonumber \\
         &\quad\times \rho_{n_1-(k+1)m,\,n_1-km}\big)+\text{H.c.} \big\}  \big].     
\end{align}
Similarly, for $\rho_2^{(4)}$ in Eq.~\eqref{c4r2} we obtain
\begin{align}
\tilde{W}_2(\alpha,t) &= \sum_{k=0}^{k_\text{max}}  (-1)^{n_2} \Big[ (-1)^{-km} \rho_{n_2-km,\,n_2-km}L_{n_2-km}(z) \nonumber \\ 
&+  (-1)^{km} \rho_{n_2+km,\,n_2+km} L_{n_2+km}(z) \nonumber \\ 
&+  \big\{(2\alpha)^m \big((-1)^{-(k+1)m} \sqrt{\tfrac{(n_2-(k+1)m)!}{(n_2-km)!}} \nonumber\\
&\quad\times\rho_{n_2-(k+1)m,\,n_2-km} L_{n_2-(k+1)m}^{m}(z) \nonumber \\ 
&+  (-1)^{km} \sqrt{\tfrac{(n_2+km)!}{(n_2+(k+1)m)!}} \rho_{n_2+km,\,n_2+(k+1)m}\nonumber \\
&\quad\times L_{n_2+km}^{m}(z)\big)  + \text{H.c.} \big\}\Big] \nonumber\\
&+ (-1)^{n_2}\big\{(2\alpha)^m \sqrt{\tfrac{(n_2)!}{(n_2+m)!}} \rho_{n_2,\,n_2+m} L_{n_1}^{m}(z) \nonumber \\ 
&+ \sqrt{\tfrac{(n_2-m)!}{(n_2)!}} (2\alpha^*)^m \rho_{n_2,\,n_2-m} L_{n_2-m}^m(z) + \text{H.c.} \big\}.   
\end{align}




\section{\label{asymm_coupoling} Complete excitation transfer in asymmetrically coupled MPJC models}
In this appendix we theoretically examine the feasibility of engineering arbitrary excitation transfers in case 2, i.e., when $n_1 < n_2 = m$, specifically focusing on the transition $\ket{n_1,\, m} \longrightarrow \ket{n_1+m,\, 0}$. We analyze this scenario for any finite values of $n_1$ and $m$ by manipulating the asymmetry in the coupling parameters of the MPJC interactions.

To demonstrate this, we consider the unitary dynamics of the system, set $\Delta = 0$, and initialize the qubit in the ground state ($\phi = 0$). Recall that complete excitation transfer is achievable whenever $A = \frac{2 \,g_{1_{n_1m}}g_{2_m}}{g_{1_{n_1m}}^2 + g_{2_m}^2} = 1$, where $g_{1_{n_1m}} = \sqrt{(n_1 + m)!/n_1!} \, g_1$, and $g_{2_m} = \sqrt{m!} \, g_2$.

Relaxing the symmetric coupling assumption while maintaining $g_1,\, g_2 \leq 1$ for practical purposes, we express $g_1 = \epsilon g_2$, where $0 < \epsilon \leq 1$. Noting that complete excitation transfer requires $g_{1_{n_1m}} = g_{2_m}$, we arrive at the condition 
\begin{align}
    \epsilon = \sqrt{\frac{n_1!\, m!}{(n_1+m)!}}.
    \label{eqn:epsilon_asym}
\end{align}
\begin{figure}[ht]
    \centering
    \includegraphics{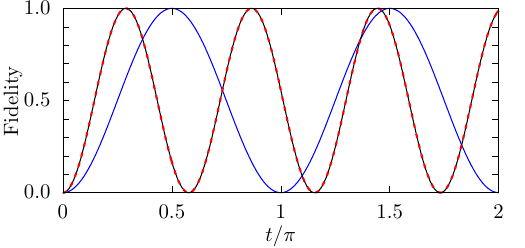}
    \caption{Asymmetrically coupled tripartite MPJC model: For case~2, the unitary evolution of the fidelity of the  $\ket{n_1,\, m} \longrightarrow \ket{n_1+m,\, 0}$ transfer is illustrated for nonzero $n_1$ values. We set $g_2=1$ and choose $g_1=\frac{1}{\sqrt{3}}$ (blue), $\frac{1}{2}$ (black), and $\frac{1}{\sqrt{10}}$ (red), respectively corresponding to $(n_1,\, m)=$ (1,2), (1,3), and (2,3). Further, $\phi=0$ and $\Delta=0$. }
    \label{fig:n1_nonzero_case2}
\end{figure}

\begin{figure}
    \centering
    \includegraphics[scale=0.9]{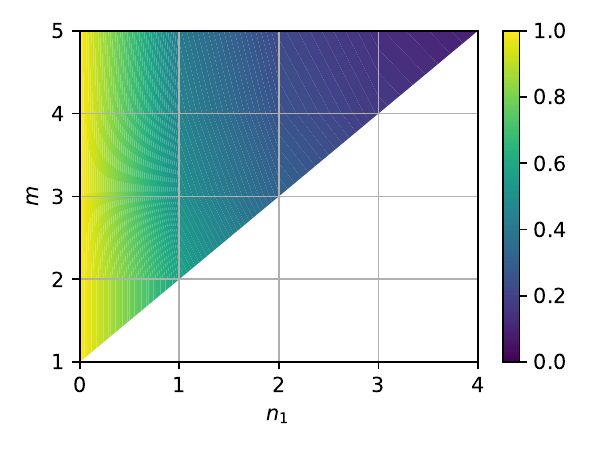}
    \vspace{-3ex}
    \caption{Asymmetrically coupled tripartite MPJC model. The skewness parameter $\epsilon$ in Eq.~\eqref{eqn:epsilon_asym} is plotted as a function of $n_1$ and $m$ for case 2 with $\Delta=0$.}
    \label{fig:contour}
\end{figure}

In Table~\ref{tab:table2}, we list the numerical values of $\epsilon$ for some smaller values of $n_1$ and $m$. 
\vskip2ex
\begin{table}[h!]
\centering
\begin{tabular}{c c c c}\hline\hline
        $n_1$ & $m$ & $\epsilon$ &  \\ 
        \hline
      1  & 2 & 0.58 & \\
      1  & 3 & 0.50 & $\frac{1}{\sqrt{m+1}}$ \\
      1  & 4 & 0.45 & \\
      \hline
      2  & 3 & 0.32 & $\frac{2}{\sqrt{(m+1)(m+2)}}$ \\
      2  & 4 & 0.26 & \\
       \hline\hline
    \end{tabular}
\caption{\label{tab:table2}%
Numerical values of the skewness parameter $\epsilon$ in Eq.~\eqref{eqn:epsilon_asym} for some specific values of $n_1$ and $m$.}
\end{table}

Numerical investigations support these analytical findings, as demonstrated in Fig.~\ref{fig:n1_nonzero_case2}, which shows the temporal evolution of fidelity for various values of $n_1$ and $m$. The oscillation frequency is governed by the factor $\sqrt{g_{1_{n_1m}}^2 + g_{2_m}^2}$. Notably, the black and red curves, corresponding to $(n_1,\, m) = (1, 3)$ and $(2, 3)$, exhibit identical frequencies.

Our analysis confirms the feasibility of arbitrarily transferring excitations from the second oscillator to the first, specifically $\ket{n_1,\, m} \longrightarrow \ket{n_1+m,\, 0}$ for any $n_1$ and $m$ for case 2 ($n_1 < n_2 = m$). However, as $n_1$ and $m$ increase, the required asymmetry between the coupling parameters $g_1$ and $g_2$ becomes more pronounced, as shown in Fig.~\ref{fig:contour}. Therefore, practical limitations may restrict the extent to which such excitation transfers can be engineered.


\section{\label{bs_output} Excitation transfer in a quantum beamsplitter}
In this appendix we analytically demonstrate that it is impossible to achieve the output state $\ket{n_1+n_2,\, 0}$ with unit probability in a quantum beamsplitter for arbitrary Fock input states $\ket{n_1}$ (entering through one port) and $\ket{n_2}$ (through the other), for any generic phase $\theta$, when $n_1,\, n_2 > 0$.

To establish this, we first express the output states of a beam splitter for arbitrary input Fock states $\ket{n_1,\, n_2}$. Let the input mode operators be denoted by $(a_1, a_1^\dagger)$ and $(a_2, a_2^\dagger)$, while $(b_1, b_1^\dagger)$ and $(b_2, b_2^\dagger)$ represent the mode operators of the two output ports. The unitary beamsplitter operator is given by $U_{\text{BS}} = e^{-\theta(a_1^\dagger a_2 - a_1 a_2^\dagger)}$, where $|R| = \sin\theta$ and $|T| = \cos\theta$ are the reflection and transmission coefficients, respectively~\cite{campos89}.
\begin{figure}[ht!]
    \centering
    \includegraphics[scale=0.35]{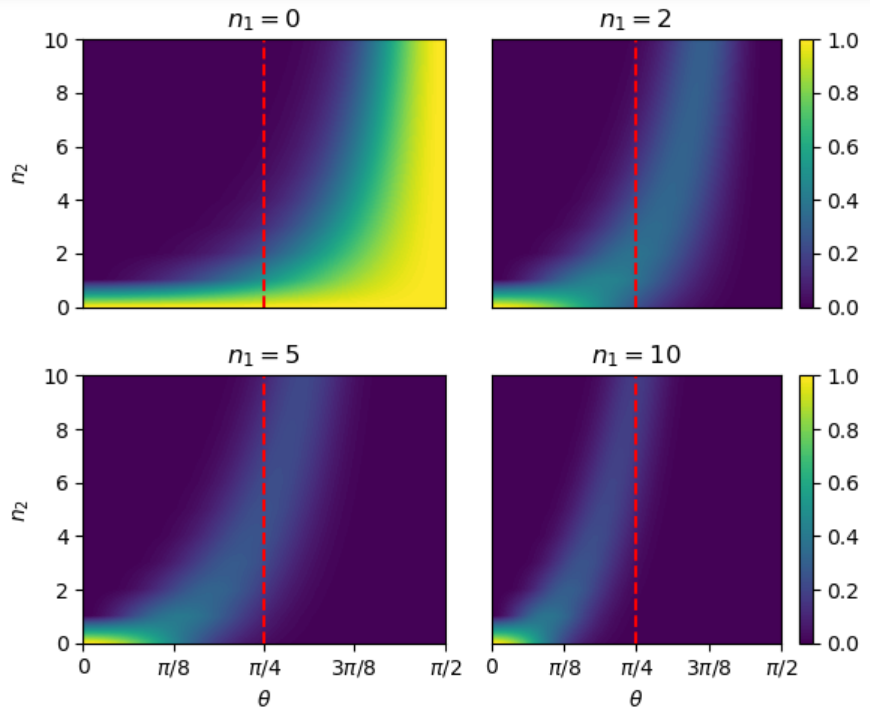}
    \vspace{-2ex}
    \caption{Probability $P$ in Eq.~\eqref{eqn_bs_prob} as a function of the beam-splitter phase $\theta$ and $n_2$, for $n_1 = 0$, 2, 5, and 10. The red dashed vertical lines in each panel correspond to $\theta=\pi/4$.}
    \label{fig:bs_prob}
\end{figure}
The output mode operators in terms of the input mode operators are
\begin{align}
   \begin{pmatrix}
        b_1^\dagger\\
        b_2^\dagger \\
    \end{pmatrix} = 
   \begin{pmatrix}
        U_{\text{BS}} a_1^\dagger\, U_{\text{BS}}^\dagger\\
        U_{\text{BS}} a_2^\dagger\, U_{\text{BS}}^\dagger \\
    \end{pmatrix} = 
    \begin{pmatrix}
        \cos\theta & \sin\theta\\
       -\sin\theta & \cos\theta \\
    \end{pmatrix} 
    \begin{pmatrix}
        a_1^\dagger\\
        a_2^\dagger \\
    \end{pmatrix}.
\label{eqn_b1_b2}
\end{align}
Using these relations, it is straightforward to write the output state of the beamsplitter, which is given by
\begin{align}
  \ket{\psi}_{\text{out}}=  \sum_{k=0}^{n_1} \sum_{l=0}^{n_2} C_{n_1,n_2,k,l}\, \ket{k+l,\, n_1+n_2-k-l}\,,
\label{eqn_bs_output}
\end{align}
where 
\begin{align}
  C_{n_1,n_2,k,l} = (-1)^l &\frac{\sqrt{(k+l)!(n_1+n_2-k-l)!\, n_1!\, n_2!}}{(n_1-k)!(n_2-l)!\, k!\, l!}\nonumber\\
  &\times\left(\sin \theta\right)^{n_1-k+l} \left(\cos\theta\right)^{n_2+k-l}.
\label{eqn_c_n1n2kl}
\end{align}
It is clear from Eq.~\eqref{eqn_bs_output} that the probability of obtaining the output state $\ket{n_1+n_2,\, 0}$ requires setting $n_1+n_2=k+l$. Since $k$ and $l$ are bounded above by $n_1$ and $n_2$, respectively, this condition simplifies to $k=n_1$ and $l=n_2$. Substituting these values into Eq.~\eqref{eqn_c_n1n2kl}, we obtain
\begin{align}
    C_{n_1,n_2,k,l} = (-1)^{n_2} \sqrt{\frac{(n_1+n_2)!}{n_1!\, n_2!}}\left(\sin \theta\right)^{n_2} \left(\cos\theta\right)^{n_1}.
\end{align}

Therefore, the probability of obtaining the output state $\ket{n_1+n_2,\, 0}$ is
\begin{align}
P = \left|C_{n_1,n_2,k,l}\right|^2 &= \frac{(n_1+n_2)!}{n_1!n_2!} (\sin \theta)^{2n_2} (\cos \theta)^{2n_1}.
\label{eqn_bs_prob}
\end{align}

It is evident that by setting $n_1=0$ and $\theta=\pi/2$ in Eq.~\eqref{eqn_bs_prob}, we obtain the output state $\ket{n_2,\, 0}$ with unit probability. This is further illustrated in Fig.~\ref{fig:bs_prob}. However, for any finite nonzero value of $n_1$, it is impossible to achieve an output state $\ket{n_1+n_2,\, 0}$ with unit probability.


\section{\label{sec:const_of_motion}Constant of motion}
The standard tripartite JC model, that is, $H$ in Eq.~\eqref{eqn_mpjcm} with $m=1$, is a special quantum system as it possesses an additional constant of motion given by
\begin{align}
    \mathcal{C} &= \tilde{g}_2^2 a_1^{\dagger}a_1  +  \tilde{g}_1^2 a_2^{\dagger}a_2 -  \tilde{g}_1 \tilde{g}_2 \left(a_1^{\dagger}a_2  +  a_1^{\dagger}a_2\right),
\end{align} 
where $\tilde{g}_j = g_j/\sqrt{g_1^2+g_2^2}$ with $j=1,\,2$.

In the following, we show that for $m>1$, $\mathcal{C}$ no longer remains a constant of motion. In particular, we find that
\begin{align}
\left[H, \,\mathcal{C}\right] &=\frac{m\, g_1 g_2}{g_1^2+g_2^2}\Bigg\{ g_2 \left( a_1^m \sigma_+ - a_1^{\dagger\,m} \sigma_- \right) \nonumber\\
&\quad\,\,\,\,- g_1  \left(a_1^{m-1} a_2\, \sigma_+ -  (a_1^\dagger)^{m-1} a_2^\dagger \,\sigma_-\right) \nonumber \\
&\quad\,\,\,\,+ g_1 \left( a_2^m \sigma_+ - a_2^{\dagger\,m} \sigma_- \right) \nonumber\\
&\quad\,\,\,\,- g_2 \left(  a_1 a_2^{m-1} \,\sigma_+ -  a_1^\dagger (a_2^\dagger)^{m-1}\,\sigma_- \right)\Bigg\}.
\end{align}
It is straightforward to see that $\left[H, \,\mathcal{C}\right]=0$ only when $m=1$.

\section{\label{sec:canonical_H}Canonical transformation of the MPJC Hamiltonian}
In this appendix we perform a canonical beamsplitter-like transformation on the MPJC Hamiltonian $H$ in Eq.~\eqref{eqn_mpjcm} so that $H\longrightarrow U_{\text{BS}}  H U_{\text{BS}}^\dagger$, where $U_{\text{BS}} = e^{-\theta(a_1^\dagger a_2 - a_1 a_2^\dagger)}$. As a result, the original mode operators transform to rotated or normal mode operators, as given by Eq.~\eqref{eqn_b1_b2}.

It can be easily shown that the free energy term $H_I$ in Eq.~\eqref{eqn_H_I} remains unchanged after the canonical transformation, as $a_1^{\dagger}a_1  + a_2^{\dagger}a_2=b_1^{\dagger}b_1  + b_2^{\dagger}b_2$.
On the other hand, the interaction Hamiltonian transforms nontrivially. It can be shown that 
\begin{align}
\label{eqn_H_II_can}
  H_{II} &=\sum_{k=0}^m \binom{m}{k} \bigg\{g_1\, (\cos\theta)^k (-\sin\theta)^{m-k} \nonumber\\
  &+ g_2 \, (\sin\theta)^k (\cos\theta)^{m-k}\bigg\} b_1^k\, b_2^{m-k}\sigma_+  + \text{H.c.}   
\end{align}

Now, we can split $H_{\text{int}}=H_{\text{bipartite}}+H_{\text{tripartite}}$.
The two bipartite interaction terms corresponding to $k=m$ and $k=0$ from Eq.~\eqref{eqn_H_II_can} are given by
\begin{align}
    H_{\text{bipartite}} = \tilde{g}\left(b_1^m \sigma_+ + b_1^{\dagger\,m}\sigma_-\right) + \tilde{\textsl{g}}\left(b_2^m\sigma_+ + b_2^{\dagger\,m}\sigma_-\right).
    \label{eqn_bipartite}
\end{align}
where $\tilde{g}=g_1 (\cos\theta)^m + g_2 (\sin\theta)^m$, and $\tilde{\textsl{g}}=g_1 (-\sin\theta)^m + g_2 (\cos\theta)^m$.
Now, the second MPJC interaction term in Eq.~\eqref{eqn_bipartite} disappears if we choose the angle of rotation such that $\tilde{\textsl{g}}=0$, which translates to $(-\tan\theta)^m=-g_2/g_1$, is satisfied. In that case, $H_{\text{bipartite}} = \tilde{g}\left(b_1^m \sigma_+ + b_1^{\dagger\,m}\sigma_-\right)$, where
\begin{align}
  \tilde{g} &=  g_1 (\sin\theta)^m + g_2 (\cos\theta)^m =  \frac{g_1^2 + (-1)^{m-1}g_2^{2}}{\left( g_1^{2/m}+g_2^{2/m} \right)^{m/2}} .
  \label{eqn_gtilde}
\end{align}

Apart from the two extreme cases [as shown in Eq.~\eqref{eqn_bipartite}], the aforementioned rotation introduces additional $m-1$ genuinely tripartite interaction terms given by
\begin{align}
 H_\text{tripartite} &= \sum_{k=1}^{m-1} \tilde{\texttt{g}}_{k} b_1^k\, b_2^{m-k}\sigma_+  + \text{H.c.}, 
 \label{eqn_tripartite}
\end{align}
where
\begin{align}
\tilde{\texttt{g}}_{k}  &=  \binom{m}{k} \left\{g_1\, (\cos \theta)^k (-\sin \theta)^{m-k}  + g_2 \, (\sin \theta)^k (\cos \theta)^{m-k}\right\}  \nonumber \\
 &= \tfrac{1}{\left(g_1^{2/m}+g_2^{2/m}\right)^{m/2}} \binom{m}{k} \left\{ g_1 g_1^{k/m} \bigg( (-1)^{m-1}g_2 \right)^{(m-k)/m} \nonumber\\
 &\quad\,\,\,+ (-1)^k g_2 \left( (-1)^{m-1}g_2 \right)^{k/m} g_1^{(m-k)/m} \bigg\}.   
\end{align}

It is straightforward to see that for $m=1$, there are no genuinely tripartite interaction terms in the transformed Hamiltonian. Further, if we set $\tan\theta=g_2/g_1$, we obtain $H_{II} = \tilde{g}\left(b_1 \sigma_+ + b_1^{\dagger}\sigma_-\right)$. 

In summary, the above analysis demonstrates that the transformation of the tripartite MPJC Hamiltonian via a beam-splitter unitary does not simplify the Hamiltonian beyond the case of $m=1$. Instead, it increases the complexity of the Hamiltonian, introducing $m-1$ additional genuinely tripartite interaction terms. Consequently, the multiphoton generalization of the standard tripartite JC Hamiltonian is indeed nontrivial, and the increased nonlinearity resulting from this generalization significantly complicates the overall analysis.

\bibliography{reference}

\begin{thebibliography}{82}%
\makeatletter
\providecommand \@ifxundefined [1]{%
 \@ifx{#1\undefined}
}%
\providecommand \@ifnum [1]{%
 \ifnum #1\expandafter \@firstoftwo
 \else \expandafter \@secondoftwo
 \fi
}%
\providecommand \@ifx [1]{%
 \ifx #1\expandafter \@firstoftwo
 \else \expandafter \@secondoftwo
 \fi
}%
\providecommand \natexlab [1]{#1}%
\providecommand \enquote  [1]{``#1''}%
\providecommand \bibnamefont  [1]{#1}%
\providecommand \bibfnamefont [1]{#1}%
\providecommand \citenamefont [1]{#1}%
\providecommand \href@noop [0]{\@secondoftwo}%
\providecommand \href [0]{\begingroup \@sanitize@url \@href}%
\providecommand \@href[1]{\@@startlink{#1}\@@href}%
\providecommand \@@href[1]{\endgroup#1\@@endlink}%
\providecommand \@sanitize@url [0]{\catcode `\\12\catcode `\$12\catcode
  `\&12\catcode `\#12\catcode `\^12\catcode `\_12\catcode `\%12\relax}%
\providecommand \@@startlink[1]{}%
\providecommand \@@endlink[0]{}%
\providecommand \url  [0]{\begingroup\@sanitize@url \@url }%
\providecommand \@url [1]{\endgroup\@href {#1}{\urlprefix }}%
\providecommand \urlprefix  [0]{URL }%
\providecommand \Eprint [0]{\href }%
\providecommand \doibase [0]{https://doi.org/}%
\providecommand \selectlanguage [0]{\@gobble}%
\providecommand \bibinfo  [0]{\@secondoftwo}%
\providecommand \bibfield  [0]{\@secondoftwo}%
\providecommand \translation [1]{[#1]}%
\providecommand \BibitemOpen [0]{}%
\providecommand \bibitemStop [0]{}%
\providecommand \bibitemNoStop [0]{.\EOS\space}%
\providecommand \EOS [0]{\spacefactor3000\relax}%
\providecommand \BibitemShut  [1]{\csname bibitem#1\endcsname}%
\let\auto@bib@innerbib\@empty
\bibitem [{\citenamefont {Jaynes}\ and\ \citenamefont
  {Cummings}(1963)}]{jaynes_cummings_1963}%
  \BibitemOpen
  \bibfield  {author} {\bibinfo {author} {\bibfnamefont {E.}~\bibnamefont
  {Jaynes}}\ and\ \bibinfo {author} {\bibfnamefont {F.}~\bibnamefont
  {Cummings}},\ }\bibfield  {title} {\bibinfo {title} {Comparison of quantum
  and semiclassical radiation theories with application to the beam maser},\
  }\href {https://doi.org/10.1109/PROC.1963.1664} {\bibfield  {journal}
  {\bibinfo  {journal} {Proceedings of the IEEE}\ }\textbf {\bibinfo {volume}
  {51}},\ \bibinfo {pages} {89} (\bibinfo {year} {1963})}\BibitemShut {NoStop}%
\bibitem [{\citenamefont {Larson}\ and\ \citenamefont
  {Mavrogordatos}(2021)}]{Larson_JCM_2021}%
  \BibitemOpen
  \bibfield  {author} {\bibinfo {author} {\bibfnamefont {J.}~\bibnamefont
  {Larson}}\ and\ \bibinfo {author} {\bibfnamefont {T.}~\bibnamefont
  {Mavrogordatos}},\ }\href {https://doi.org/10.1088/978-0-7503-3447-1} {\emph
  {\bibinfo {title} {The {J}aynes--{C}ummings Model and Its Descendants}}},\
  2053-2563\ (\bibinfo  {publisher} {IOP Publishing},\ \bibinfo {year}
  {2021})\BibitemShut {NoStop}%
\bibitem [{\citenamefont {Shore}\ and\ \citenamefont
  {Knight}(1993)}]{Shore_JCM_1993}%
  \BibitemOpen
  \bibfield  {author} {\bibinfo {author} {\bibfnamefont {B.~W.}\ \bibnamefont
  {Shore}}\ and\ \bibinfo {author} {\bibfnamefont {P.~L.}\ \bibnamefont
  {Knight}},\ }\bibfield  {title} {\bibinfo {title} {The {J}aynes-{C}ummings
  model},\ }\href {https://doi.org/10.1080/09500349314551321} {\bibfield
  {journal} {\bibinfo  {journal} {J. Mod. Opt.}\ }\textbf {\bibinfo {volume}
  {40}},\ \bibinfo {pages} {1195} (\bibinfo {year} {1993})}\BibitemShut
  {NoStop}%
\bibitem [{\citenamefont {Mischuck}\ and\ \citenamefont
  {M\o{}lmer}(2013)}]{mm2013}%
  \BibitemOpen
  \bibfield  {author} {\bibinfo {author} {\bibfnamefont {B.}~\bibnamefont
  {Mischuck}}\ and\ \bibinfo {author} {\bibfnamefont {K.}~\bibnamefont
  {M\o{}lmer}},\ }\bibfield  {title} {\bibinfo {title} {Qudit quantum
  computation in the {J}aynes-{C}ummings model},\ }\href
  {https://doi.org/10.1103/PhysRevA.87.022341} {\bibfield  {journal} {\bibinfo
  {journal} {Phys. Rev. A}\ }\textbf {\bibinfo {volume} {87}},\ \bibinfo
  {pages} {022341} (\bibinfo {year} {2013})}\BibitemShut {NoStop}%
\bibitem [{\citenamefont {Bina}(2012)}]{bina2012}%
  \BibitemOpen
  \bibfield  {author} {\bibinfo {author} {\bibfnamefont {M.}~\bibnamefont
  {Bina}},\ }\bibfield  {title} {\bibinfo {title} {The coherent interaction
  between matter and radiation: A tutorial on the {J}aynes-{C}ummings model},\
  }\href {https://doi.org/https://doi.org/10.1140/epjst/e2012-01541-3}
  {\bibfield  {journal} {\bibinfo  {journal} {Eur. Phys. J. Special Topics}\
  }\textbf {\bibinfo {volume} {203}},\ \bibinfo {pages} {163} (\bibinfo {year}
  {2012})}\BibitemShut {NoStop}%
\bibitem [{\citenamefont {Sasaki}\ \emph {et~al.}(1996)\citenamefont {Sasaki},
  \citenamefont {Usuda}, \citenamefont {Hirota},\ and\ \citenamefont
  {Holevo}}]{sasaki96}%
  \BibitemOpen
  \bibfield  {author} {\bibinfo {author} {\bibfnamefont {M.}~\bibnamefont
  {Sasaki}}, \bibinfo {author} {\bibfnamefont {T.~S.}\ \bibnamefont {Usuda}},
  \bibinfo {author} {\bibfnamefont {O.}~\bibnamefont {Hirota}},\ and\ \bibinfo
  {author} {\bibfnamefont {A.~S.}\ \bibnamefont {Holevo}},\ }\bibfield  {title}
  {\bibinfo {title} {Applications of the {J}aynes-{C}ummings model for the
  detection of nonorthogonal quantum states},\ }\href
  {https://doi.org/10.1103/PhysRevA.53.1273} {\bibfield  {journal} {\bibinfo
  {journal} {Phys. Rev. A}\ }\textbf {\bibinfo {volume} {53}},\ \bibinfo
  {pages} {1273} (\bibinfo {year} {1996})}\BibitemShut {NoStop}%
\bibitem [{\citenamefont {Georgescu}\ \emph {et~al.}(2014)\citenamefont
  {Georgescu}, \citenamefont {Ashhab},\ and\ \citenamefont {Nori}}]{GAF2014}%
  \BibitemOpen
  \bibfield  {author} {\bibinfo {author} {\bibfnamefont {I.~M.}\ \bibnamefont
  {Georgescu}}, \bibinfo {author} {\bibfnamefont {S.}~\bibnamefont {Ashhab}},\
  and\ \bibinfo {author} {\bibfnamefont {F.}~\bibnamefont {Nori}},\ }\bibfield
  {title} {\bibinfo {title} {Quantum simulation},\ }\href
  {https://doi.org/10.1103/RevModPhys.86.153} {\bibfield  {journal} {\bibinfo
  {journal} {Rev. Mod. Phys.}\ }\textbf {\bibinfo {volume} {86}},\ \bibinfo
  {pages} {153} (\bibinfo {year} {2014})}\BibitemShut {NoStop}%
\bibitem [{\citenamefont {Azuma}(2011)}]{azuma2011}%
  \BibitemOpen
  \bibfield  {author} {\bibinfo {author} {\bibfnamefont {H.}~\bibnamefont
  {Azuma}},\ }\bibfield  {title} {\bibinfo {title} {Quantum computation with
  the {J}aynes-{C}ummings model},\ }\href
  {https://doi.org/https://doi.org/10.1143/PTP.126.369} {\bibfield  {journal}
  {\bibinfo  {journal} {Prog. Theor. Phys.}\ }\textbf {\bibinfo {volume}
  {126}},\ \bibinfo {pages} {369} (\bibinfo {year} {2011})}\BibitemShut
  {NoStop}%
\bibitem [{\citenamefont {Li}\ \emph {et~al.}(2022)\citenamefont {Li},
  \citenamefont {Mei}, \citenamefont {Wu}, \citenamefont {Cai}, \citenamefont
  {Wang}, \citenamefont {Yao}, \citenamefont {Zhou},\ and\ \citenamefont
  {Duan}}]{Li2022}%
  \BibitemOpen
  \bibfield  {author} {\bibinfo {author} {\bibfnamefont {B.-W.}\ \bibnamefont
  {Li}}, \bibinfo {author} {\bibfnamefont {Q.-X.}\ \bibnamefont {Mei}},
  \bibinfo {author} {\bibfnamefont {Y.-K.}\ \bibnamefont {Wu}}, \bibinfo
  {author} {\bibfnamefont {M.-L.}\ \bibnamefont {Cai}}, \bibinfo {author}
  {\bibfnamefont {Y.}~\bibnamefont {Wang}}, \bibinfo {author} {\bibfnamefont
  {L.}~\bibnamefont {Yao}}, \bibinfo {author} {\bibfnamefont {Z.-C.}\
  \bibnamefont {Zhou}},\ and\ \bibinfo {author} {\bibfnamefont {L.-M.}\
  \bibnamefont {Duan}},\ }\bibfield  {title} {\bibinfo {title} {Observation of
  non-markovian spin dynamics in a {J}aynes-{C}ummings-{H}ubbard model using a
  trapped-ion quantum simulator},\ }\href
  {https://doi.org/10.1103/PhysRevLett.129.140501} {\bibfield  {journal}
  {\bibinfo  {journal} {Phys. Rev. Lett.}\ }\textbf {\bibinfo {volume} {129}},\
  \bibinfo {pages} {140501} (\bibinfo {year} {2022})}\BibitemShut {NoStop}%
\bibitem [{\citenamefont {Rempe}\ \emph {et~al.}(1987)\citenamefont {Rempe},
  \citenamefont {Walther},\ and\ \citenamefont {Klein}}]{Rempe_PRL_1987}%
  \BibitemOpen
  \bibfield  {author} {\bibinfo {author} {\bibfnamefont {G.}~\bibnamefont
  {Rempe}}, \bibinfo {author} {\bibfnamefont {H.}~\bibnamefont {Walther}},\
  and\ \bibinfo {author} {\bibfnamefont {N.}~\bibnamefont {Klein}},\ }\bibfield
   {title} {\bibinfo {title} {Observation of quantum collapse and revival in a
  one-atom maser},\ }\href {https://doi.org/10.1103/PhysRevLett.58.353}
  {\bibfield  {journal} {\bibinfo  {journal} {Phys. Rev. Lett.}\ }\textbf
  {\bibinfo {volume} {58}},\ \bibinfo {pages} {353} (\bibinfo {year}
  {1987})}\BibitemShut {NoStop}%
\bibitem [{\citenamefont {Boca}\ \emph {et~al.}(2004)\citenamefont {Boca},
  \citenamefont {Miller}, \citenamefont {Birnbaum}, \citenamefont {Boozer},
  \citenamefont {McKeever},\ and\ \citenamefont {Kimble}}]{Boca_PRL_2004}%
  \BibitemOpen
  \bibfield  {author} {\bibinfo {author} {\bibfnamefont {A.}~\bibnamefont
  {Boca}}, \bibinfo {author} {\bibfnamefont {R.}~\bibnamefont {Miller}},
  \bibinfo {author} {\bibfnamefont {K.~M.}\ \bibnamefont {Birnbaum}}, \bibinfo
  {author} {\bibfnamefont {A.~D.}\ \bibnamefont {Boozer}}, \bibinfo {author}
  {\bibfnamefont {J.}~\bibnamefont {McKeever}},\ and\ \bibinfo {author}
  {\bibfnamefont {H.~J.}\ \bibnamefont {Kimble}},\ }\bibfield  {title}
  {\bibinfo {title} {Observation of the vacuum {R}abi spectrum for one trapped
  atom},\ }\href {https://doi.org/10.1103/PhysRevLett.93.233603} {\bibfield
  {journal} {\bibinfo  {journal} {Phys. Rev. Lett.}\ }\textbf {\bibinfo
  {volume} {93}},\ \bibinfo {pages} {233603} (\bibinfo {year}
  {2004})}\BibitemShut {NoStop}%
\bibitem [{\citenamefont {Birnbaum}\ \emph {et~al.}(2005)\citenamefont
  {Birnbaum}, \citenamefont {Boca}, \citenamefont {Miller}, \citenamefont
  {Boozer}, \citenamefont {Northup},\ and\ \citenamefont
  {Kimble}}]{Birnbaum_nature_2005}%
  \BibitemOpen
  \bibfield  {author} {\bibinfo {author} {\bibfnamefont {K.~M.}\ \bibnamefont
  {Birnbaum}}, \bibinfo {author} {\bibfnamefont {A.}~\bibnamefont {Boca}},
  \bibinfo {author} {\bibfnamefont {R.}~\bibnamefont {Miller}}, \bibinfo
  {author} {\bibfnamefont {A.~D.}\ \bibnamefont {Boozer}}, \bibinfo {author}
  {\bibfnamefont {T.~E.}\ \bibnamefont {Northup}},\ and\ \bibinfo {author}
  {\bibfnamefont {H.~J.}\ \bibnamefont {Kimble}},\ }\bibfield  {title}
  {\bibinfo {title} {Photon blockade in an optical cavity with one trapped
  atom},\ }\href {https://doi.org/10.1038/nature03804} {\bibfield  {journal}
  {\bibinfo  {journal} {Nature}\ }\textbf {\bibinfo {volume} {436}},\ \bibinfo
  {pages} {87} (\bibinfo {year} {2005})}\BibitemShut {NoStop}%
\bibitem [{\citenamefont {Brune}\ \emph {et~al.}(1996)\citenamefont {Brune},
  \citenamefont {Schmidt-Kaler}, \citenamefont {Maali}, \citenamefont {Dreyer},
  \citenamefont {Hagley}, \citenamefont {Raimond},\ and\ \citenamefont
  {Haroche}}]{Brune_PRL_1996}%
  \BibitemOpen
  \bibfield  {author} {\bibinfo {author} {\bibfnamefont {M.}~\bibnamefont
  {Brune}}, \bibinfo {author} {\bibfnamefont {F.}~\bibnamefont
  {Schmidt-Kaler}}, \bibinfo {author} {\bibfnamefont {A.}~\bibnamefont
  {Maali}}, \bibinfo {author} {\bibfnamefont {J.}~\bibnamefont {Dreyer}},
  \bibinfo {author} {\bibfnamefont {E.}~\bibnamefont {Hagley}}, \bibinfo
  {author} {\bibfnamefont {J.~M.}\ \bibnamefont {Raimond}},\ and\ \bibinfo
  {author} {\bibfnamefont {S.}~\bibnamefont {Haroche}},\ }\bibfield  {title}
  {\bibinfo {title} {Quantum {R}abi oscillation: A direct test of field
  quantization in a cavity},\ }\href
  {https://doi.org/10.1103/PhysRevLett.76.1800} {\bibfield  {journal} {\bibinfo
   {journal} {Phys. Rev. Lett.}\ }\textbf {\bibinfo {volume} {76}},\ \bibinfo
  {pages} {1800} (\bibinfo {year} {1996})}\BibitemShut {NoStop}%
\bibitem [{\citenamefont {Walther}\ \emph {et~al.}(2006)\citenamefont
  {Walther}, \citenamefont {Varcoe}, \citenamefont {Englert},\ and\
  \citenamefont {Becker}}]{Walther_2006}%
  \BibitemOpen
  \bibfield  {author} {\bibinfo {author} {\bibfnamefont {H.}~\bibnamefont
  {Walther}}, \bibinfo {author} {\bibfnamefont {B.~T.~H.}\ \bibnamefont
  {Varcoe}}, \bibinfo {author} {\bibfnamefont {B.-G.}\ \bibnamefont
  {Englert}},\ and\ \bibinfo {author} {\bibfnamefont {T.}~\bibnamefont
  {Becker}},\ }\bibfield  {title} {\bibinfo {title} {Cavity quantum
  electrodynamics},\ }\href {https://doi.org/10.1088/0034-4885/69/5/R02}
  {\bibfield  {journal} {\bibinfo  {journal} {Rep. Prog. Phys.}\ }\textbf
  {\bibinfo {volume} {69}},\ \bibinfo {pages} {1325} (\bibinfo {year}
  {2006})}\BibitemShut {NoStop}%
\bibitem [{\citenamefont {Lee}\ \emph {et~al.}(2017)\citenamefont {Lee},
  \citenamefont {Martin}, \citenamefont {Jau}, \citenamefont {Keating},
  \citenamefont {Deutsch},\ and\ \citenamefont {Biedermann}}]{Lee_PRA_2017}%
  \BibitemOpen
  \bibfield  {author} {\bibinfo {author} {\bibfnamefont {J.}~\bibnamefont
  {Lee}}, \bibinfo {author} {\bibfnamefont {M.~J.}\ \bibnamefont {Martin}},
  \bibinfo {author} {\bibfnamefont {Y.-Y.}\ \bibnamefont {Jau}}, \bibinfo
  {author} {\bibfnamefont {T.}~\bibnamefont {Keating}}, \bibinfo {author}
  {\bibfnamefont {I.~H.}\ \bibnamefont {Deutsch}},\ and\ \bibinfo {author}
  {\bibfnamefont {G.~W.}\ \bibnamefont {Biedermann}},\ }\bibfield  {title}
  {\bibinfo {title} {Demonstration of the {J}aynes-{C}ummings ladder with
  {R}ydberg-dressed atoms},\ }\href
  {https://doi.org/10.1103/PhysRevA.95.041801} {\bibfield  {journal} {\bibinfo
  {journal} {Phys. Rev. A}\ }\textbf {\bibinfo {volume} {95}},\ \bibinfo
  {pages} {041801} (\bibinfo {year} {2017})}\BibitemShut {NoStop}%
\bibitem [{\citenamefont {Deppe}\ \emph {et~al.}(2008)\citenamefont {Deppe},
  \citenamefont {Mariantoni}, \citenamefont {Menzel}, \citenamefont {Marx},
  \citenamefont {Saito}, \citenamefont {Kakuyanagi}, \citenamefont {Tanaka},
  \citenamefont {Meno}, \citenamefont {Semba}, \citenamefont {Takayanagi},
  \citenamefont {Solano},\ and\ \citenamefont {Gross}}]{Deppe_nat_phys_2008}%
  \BibitemOpen
  \bibfield  {author} {\bibinfo {author} {\bibfnamefont {F.}~\bibnamefont
  {Deppe}}, \bibinfo {author} {\bibfnamefont {M.}~\bibnamefont {Mariantoni}},
  \bibinfo {author} {\bibfnamefont {E.~P.}\ \bibnamefont {Menzel}}, \bibinfo
  {author} {\bibfnamefont {A.}~\bibnamefont {Marx}}, \bibinfo {author}
  {\bibfnamefont {S.}~\bibnamefont {Saito}}, \bibinfo {author} {\bibfnamefont
  {K.}~\bibnamefont {Kakuyanagi}}, \bibinfo {author} {\bibfnamefont
  {H.}~\bibnamefont {Tanaka}}, \bibinfo {author} {\bibfnamefont
  {T.}~\bibnamefont {Meno}}, \bibinfo {author} {\bibfnamefont {K.}~\bibnamefont
  {Semba}}, \bibinfo {author} {\bibfnamefont {H.}~\bibnamefont {Takayanagi}},
  \bibinfo {author} {\bibfnamefont {E.}~\bibnamefont {Solano}},\ and\ \bibinfo
  {author} {\bibfnamefont {R.}~\bibnamefont {Gross}},\ }\bibfield  {title}
  {\bibinfo {title} {Two-photon probe of the {J}aynes--{C}ummings model and
  controlled symmetry breaking in circuit {QED}},\ }\href
  {https://doi.org/10.1038/nphys1016} {\bibfield  {journal} {\bibinfo
  {journal} {Nature Phys.}\ }\textbf {\bibinfo {volume} {4}},\ \bibinfo {pages}
  {686} (\bibinfo {year} {2008})}\BibitemShut {NoStop}%
\bibitem [{\citenamefont {Fink}\ \emph {et~al.}(2008)\citenamefont {Fink},
  \citenamefont {G{\"o}ppl}, \citenamefont {Baur}, \citenamefont {Bianchetti},
  \citenamefont {Leek}, \citenamefont {Blais},\ and\ \citenamefont
  {Wallraff}}]{Fink_nature_2008}%
  \BibitemOpen
  \bibfield  {author} {\bibinfo {author} {\bibfnamefont {J.~M.}\ \bibnamefont
  {Fink}}, \bibinfo {author} {\bibfnamefont {M.}~\bibnamefont {G{\"o}ppl}},
  \bibinfo {author} {\bibfnamefont {M.}~\bibnamefont {Baur}}, \bibinfo {author}
  {\bibfnamefont {R.}~\bibnamefont {Bianchetti}}, \bibinfo {author}
  {\bibfnamefont {P.~J.}\ \bibnamefont {Leek}}, \bibinfo {author}
  {\bibfnamefont {A.}~\bibnamefont {Blais}},\ and\ \bibinfo {author}
  {\bibfnamefont {A.}~\bibnamefont {Wallraff}},\ }\bibfield  {title} {\bibinfo
  {title} {Climbing the {J}aynes--{C}ummings ladder and observing its
  nonlinearity in a cavity {QED} system},\ }\href
  {https://doi.org/10.1038/nature07112} {\bibfield  {journal} {\bibinfo
  {journal} {Nature}\ }\textbf {\bibinfo {volume} {454}},\ \bibinfo {pages}
  {315} (\bibinfo {year} {2008})}\BibitemShut {NoStop}%
\bibitem [{\citenamefont {Hofheinz}\ \emph {et~al.}(2009)\citenamefont
  {Hofheinz}, \citenamefont {Wang}, \citenamefont {Ansmann}, \citenamefont
  {Bialczak}, \citenamefont {Lucero}, \citenamefont {Neeley}, \citenamefont
  {O'Connell}, \citenamefont {Sank}, \citenamefont {Wenner}, \citenamefont
  {Martinis},\ and\ \citenamefont {Cleland}}]{Hofheinz_nature_2009}%
  \BibitemOpen
  \bibfield  {author} {\bibinfo {author} {\bibfnamefont {M.}~\bibnamefont
  {Hofheinz}}, \bibinfo {author} {\bibfnamefont {H.}~\bibnamefont {Wang}},
  \bibinfo {author} {\bibfnamefont {M.}~\bibnamefont {Ansmann}}, \bibinfo
  {author} {\bibfnamefont {R.~C.}\ \bibnamefont {Bialczak}}, \bibinfo {author}
  {\bibfnamefont {E.}~\bibnamefont {Lucero}}, \bibinfo {author} {\bibfnamefont
  {M.}~\bibnamefont {Neeley}}, \bibinfo {author} {\bibfnamefont {A.~D.}\
  \bibnamefont {O'Connell}}, \bibinfo {author} {\bibfnamefont {D.}~\bibnamefont
  {Sank}}, \bibinfo {author} {\bibfnamefont {J.}~\bibnamefont {Wenner}},
  \bibinfo {author} {\bibfnamefont {J.~M.}\ \bibnamefont {Martinis}},\ and\
  \bibinfo {author} {\bibfnamefont {A.~N.}\ \bibnamefont {Cleland}},\
  }\bibfield  {title} {\bibinfo {title} {Synthesizing arbitrary quantum states
  in a superconducting resonator},\ }\href
  {https://doi.org/10.1038/nature08005} {\bibfield  {journal} {\bibinfo
  {journal} {Nature}\ }\textbf {\bibinfo {volume} {459}},\ \bibinfo {pages}
  {546} (\bibinfo {year} {2009})}\BibitemShut {NoStop}%
\bibitem [{\citenamefont {Gustafsson}\ \emph {et~al.}(2014)\citenamefont
  {Gustafsson}, \citenamefont {Aref}, \citenamefont {Kockum}, \citenamefont
  {Ekström}, \citenamefont {Johansson},\ and\ \citenamefont
  {Delsing}}]{Martin_science_2014}%
  \BibitemOpen
  \bibfield  {author} {\bibinfo {author} {\bibfnamefont {M.~V.}\ \bibnamefont
  {Gustafsson}}, \bibinfo {author} {\bibfnamefont {T.}~\bibnamefont {Aref}},
  \bibinfo {author} {\bibfnamefont {A.~F.}\ \bibnamefont {Kockum}}, \bibinfo
  {author} {\bibfnamefont {M.~K.}\ \bibnamefont {Ekström}}, \bibinfo {author}
  {\bibfnamefont {G.}~\bibnamefont {Johansson}},\ and\ \bibinfo {author}
  {\bibfnamefont {P.}~\bibnamefont {Delsing}},\ }\bibfield  {title} {\bibinfo
  {title} {Propagating phonons coupled to an artificial atom},\ }\href
  {https://doi.org/10.1126/science.1257219} {\bibfield  {journal} {\bibinfo
  {journal} {Science}\ }\textbf {\bibinfo {volume} {346}},\ \bibinfo {pages}
  {207} (\bibinfo {year} {2014})}\BibitemShut {NoStop}%
\bibitem [{\citenamefont {Manenti}\ \emph {et~al.}(2017)\citenamefont
  {Manenti}, \citenamefont {Kockum}, \citenamefont {Patterson}, \citenamefont
  {Behrle}, \citenamefont {Rahamim}, \citenamefont {Tancredi}, \citenamefont
  {Nori},\ and\ \citenamefont {Leek}}]{Manenti_nat_commun_2017}%
  \BibitemOpen
  \bibfield  {author} {\bibinfo {author} {\bibfnamefont {R.}~\bibnamefont
  {Manenti}}, \bibinfo {author} {\bibfnamefont {A.~F.}\ \bibnamefont {Kockum}},
  \bibinfo {author} {\bibfnamefont {A.}~\bibnamefont {Patterson}}, \bibinfo
  {author} {\bibfnamefont {T.}~\bibnamefont {Behrle}}, \bibinfo {author}
  {\bibfnamefont {J.}~\bibnamefont {Rahamim}}, \bibinfo {author} {\bibfnamefont
  {G.}~\bibnamefont {Tancredi}}, \bibinfo {author} {\bibfnamefont
  {F.}~\bibnamefont {Nori}},\ and\ \bibinfo {author} {\bibfnamefont {P.~J.}\
  \bibnamefont {Leek}},\ }\bibfield  {title} {\bibinfo {title} {Circuit quantum
  acoustodynamics with surface acoustic waves},\ }\href
  {https://doi.org/10.1038/s41467-017-01063-9} {\bibfield  {journal} {\bibinfo
  {journal} {Nat. Commun.}\ }\textbf {\bibinfo {volume} {8}},\ \bibinfo {pages}
  {975} (\bibinfo {year} {2017})}\BibitemShut {NoStop}%
\bibitem [{\citenamefont {Bienfait}\ \emph {et~al.}(2019)\citenamefont
  {Bienfait}, \citenamefont {Satzinger}, \citenamefont {Zhong}, \citenamefont
  {Chang}, \citenamefont {Chou}, \citenamefont {Conner}, \citenamefont {Dumur},
  \citenamefont {Grebel}, \citenamefont {Peairs}, \citenamefont {Povey},\ and\
  \citenamefont {Cleland}}]{Bienfait_science_2019}%
  \BibitemOpen
  \bibfield  {author} {\bibinfo {author} {\bibfnamefont {A.}~\bibnamefont
  {Bienfait}}, \bibinfo {author} {\bibfnamefont {K.~J.}\ \bibnamefont
  {Satzinger}}, \bibinfo {author} {\bibfnamefont {Y.~P.}\ \bibnamefont
  {Zhong}}, \bibinfo {author} {\bibfnamefont {H.~S.}\ \bibnamefont {Chang}},
  \bibinfo {author} {\bibfnamefont {M.~H.}\ \bibnamefont {Chou}}, \bibinfo
  {author} {\bibfnamefont {C.~R.}\ \bibnamefont {Conner}}, \bibinfo {author}
  {\bibfnamefont {{\'E}.}~\bibnamefont {Dumur}}, \bibinfo {author}
  {\bibfnamefont {J.}~\bibnamefont {Grebel}}, \bibinfo {author} {\bibfnamefont
  {G.~A.}\ \bibnamefont {Peairs}}, \bibinfo {author} {\bibfnamefont {R.~G.}\
  \bibnamefont {Povey}},\ and\ \bibinfo {author} {\bibfnamefont {A.~N.}\
  \bibnamefont {Cleland}},\ }\bibfield  {title} {\bibinfo {title}
  {Phonon-mediated quantum state transfer and remote qubit entanglement},\
  }\href {https://doi.org/10.1126/science.aaw8415} {\bibfield  {journal}
  {\bibinfo  {journal} {Science}\ }\textbf {\bibinfo {volume} {364}},\ \bibinfo
  {pages} {368} (\bibinfo {year} {2019})}\BibitemShut {NoStop}%
\bibitem [{\citenamefont {Meekhof}\ \emph {et~al.}(1996)\citenamefont
  {Meekhof}, \citenamefont {Monroe}, \citenamefont {King}, \citenamefont
  {Itano},\ and\ \citenamefont {Wineland}}]{Meekhof_PRL_1996}%
  \BibitemOpen
  \bibfield  {author} {\bibinfo {author} {\bibfnamefont {D.~M.}\ \bibnamefont
  {Meekhof}}, \bibinfo {author} {\bibfnamefont {C.}~\bibnamefont {Monroe}},
  \bibinfo {author} {\bibfnamefont {B.~E.}\ \bibnamefont {King}}, \bibinfo
  {author} {\bibfnamefont {W.~M.}\ \bibnamefont {Itano}},\ and\ \bibinfo
  {author} {\bibfnamefont {D.~J.}\ \bibnamefont {Wineland}},\ }\bibfield
  {title} {\bibinfo {title} {Generation of nonclassical motional states of a
  trapped atom},\ }\href {https://doi.org/10.1103/PhysRevLett.76.1796}
  {\bibfield  {journal} {\bibinfo  {journal} {Phys. Rev. Lett.}\ }\textbf
  {\bibinfo {volume} {76}},\ \bibinfo {pages} {1796} (\bibinfo {year}
  {1996})}\BibitemShut {NoStop}%
\bibitem [{\citenamefont {Leibfried}\ \emph {et~al.}(2003)\citenamefont
  {Leibfried}, \citenamefont {Blatt}, \citenamefont {Monroe},\ and\
  \citenamefont {Wineland}}]{Leibfried_RMP_2003}%
  \BibitemOpen
  \bibfield  {author} {\bibinfo {author} {\bibfnamefont {D.}~\bibnamefont
  {Leibfried}}, \bibinfo {author} {\bibfnamefont {R.}~\bibnamefont {Blatt}},
  \bibinfo {author} {\bibfnamefont {C.}~\bibnamefont {Monroe}},\ and\ \bibinfo
  {author} {\bibfnamefont {D.}~\bibnamefont {Wineland}},\ }\bibfield  {title}
  {\bibinfo {title} {Quantum dynamics of single trapped ions},\ }\href
  {https://doi.org/10.1103/RevModPhys.75.281} {\bibfield  {journal} {\bibinfo
  {journal} {Rev. Mod. Phys.}\ }\textbf {\bibinfo {volume} {75}},\ \bibinfo
  {pages} {281} (\bibinfo {year} {2003})}\BibitemShut {NoStop}%
\bibitem [{\citenamefont {Rodr\'{\i}guez-Lara}\ \emph
  {et~al.}(2005)\citenamefont {Rodr\'{\i}guez-Lara}, \citenamefont
  {Moya-Cessa},\ and\ \citenamefont {Klimov}}]{Lara_PRA_2005}%
  \BibitemOpen
  \bibfield  {author} {\bibinfo {author} {\bibfnamefont {B.~M.}\ \bibnamefont
  {Rodr\'{\i}guez-Lara}}, \bibinfo {author} {\bibfnamefont {H.}~\bibnamefont
  {Moya-Cessa}},\ and\ \bibinfo {author} {\bibfnamefont {A.~B.}\ \bibnamefont
  {Klimov}},\ }\bibfield  {title} {\bibinfo {title} {Combining
  {J}aynes-{C}ummings and anti-{J}aynes-{C}ummings dynamics in a trapped-ion
  system driven by a laser},\ }\href
  {https://doi.org/10.1103/PhysRevA.71.023811} {\bibfield  {journal} {\bibinfo
  {journal} {Phys. Rev. A}\ }\textbf {\bibinfo {volume} {71}},\ \bibinfo
  {pages} {023811} (\bibinfo {year} {2005})}\BibitemShut {NoStop}%
\bibitem [{\citenamefont {del Valle}\ \emph {et~al.}(2009)\citenamefont {del
  Valle}, \citenamefont {Laussy},\ and\ \citenamefont
  {Tejedor}}]{delValle_PRB_2009}%
  \BibitemOpen
  \bibfield  {author} {\bibinfo {author} {\bibfnamefont {E.}~\bibnamefont {del
  Valle}}, \bibinfo {author} {\bibfnamefont {F.~P.}\ \bibnamefont {Laussy}},\
  and\ \bibinfo {author} {\bibfnamefont {C.}~\bibnamefont {Tejedor}},\
  }\bibfield  {title} {\bibinfo {title} {Luminescence spectra of quantum dots
  in microcavities. ii. fermions},\ }\href
  {https://doi.org/10.1103/PhysRevB.79.235326} {\bibfield  {journal} {\bibinfo
  {journal} {Phys. Rev. B}\ }\textbf {\bibinfo {volume} {79}},\ \bibinfo
  {pages} {235326} (\bibinfo {year} {2009})}\BibitemShut {NoStop}%
\bibitem [{\citenamefont {Kasprzak}\ \emph {et~al.}(2010)\citenamefont
  {Kasprzak}, \citenamefont {Reitzenstein}, \citenamefont {Muljarov},
  \citenamefont {Kistner}, \citenamefont {Schneider}, \citenamefont {Strauss},
  \citenamefont {H{\"o}fling}, \citenamefont {Forchel},\ and\ \citenamefont
  {Langbein}}]{Kasprzak_nat_mat_2010}%
  \BibitemOpen
  \bibfield  {author} {\bibinfo {author} {\bibfnamefont {J.}~\bibnamefont
  {Kasprzak}}, \bibinfo {author} {\bibfnamefont {S.}~\bibnamefont
  {Reitzenstein}}, \bibinfo {author} {\bibfnamefont {E.~A.}\ \bibnamefont
  {Muljarov}}, \bibinfo {author} {\bibfnamefont {C.}~\bibnamefont {Kistner}},
  \bibinfo {author} {\bibfnamefont {C.}~\bibnamefont {Schneider}}, \bibinfo
  {author} {\bibfnamefont {M.}~\bibnamefont {Strauss}}, \bibinfo {author}
  {\bibfnamefont {S.}~\bibnamefont {H{\"o}fling}}, \bibinfo {author}
  {\bibfnamefont {A.}~\bibnamefont {Forchel}},\ and\ \bibinfo {author}
  {\bibfnamefont {W.}~\bibnamefont {Langbein}},\ }\bibfield  {title} {\bibinfo
  {title} {Up on the {J}aynes--{C}ummings ladder of a quantum-dot/microcavity
  system},\ }\href {https://doi.org/10.1038/nmat2717} {\bibfield  {journal}
  {\bibinfo  {journal} {Nature Mater.}\ }\textbf {\bibinfo {volume} {9}},\
  \bibinfo {pages} {304} (\bibinfo {year} {2010})}\BibitemShut {NoStop}%
\bibitem [{\citenamefont {Basset}\ \emph {et~al.}(2013)\citenamefont {Basset},
  \citenamefont {Jarausch}, \citenamefont {Stockklauser}, \citenamefont {Frey},
  \citenamefont {Reichl}, \citenamefont {Wegscheider}, \citenamefont {Ihn},
  \citenamefont {Ensslin},\ and\ \citenamefont {Wallraff}}]{Basset_PRB_2013}%
  \BibitemOpen
  \bibfield  {author} {\bibinfo {author} {\bibfnamefont {J.}~\bibnamefont
  {Basset}}, \bibinfo {author} {\bibfnamefont {D.-D.}\ \bibnamefont
  {Jarausch}}, \bibinfo {author} {\bibfnamefont {A.}~\bibnamefont
  {Stockklauser}}, \bibinfo {author} {\bibfnamefont {T.}~\bibnamefont {Frey}},
  \bibinfo {author} {\bibfnamefont {C.}~\bibnamefont {Reichl}}, \bibinfo
  {author} {\bibfnamefont {W.}~\bibnamefont {Wegscheider}}, \bibinfo {author}
  {\bibfnamefont {T.~M.}\ \bibnamefont {Ihn}}, \bibinfo {author} {\bibfnamefont
  {K.}~\bibnamefont {Ensslin}},\ and\ \bibinfo {author} {\bibfnamefont
  {A.}~\bibnamefont {Wallraff}},\ }\bibfield  {title} {\bibinfo {title}
  {Single-electron double quantum dot dipole-coupled to a single photonic
  mode},\ }\href {https://doi.org/10.1103/PhysRevB.88.125312} {\bibfield
  {journal} {\bibinfo  {journal} {Phys. Rev. B}\ }\textbf {\bibinfo {volume}
  {88}},\ \bibinfo {pages} {125312} (\bibinfo {year} {2013})}\BibitemShut
  {NoStop}%
\bibitem [{\citenamefont {D\'ora}\ \emph {et~al.}(2009)\citenamefont {D\'ora},
  \citenamefont {Ziegler}, \citenamefont {Thalmeier},\ and\ \citenamefont
  {Nakamura}}]{Dora_PRL_2009}%
  \BibitemOpen
  \bibfield  {author} {\bibinfo {author} {\bibfnamefont {B.}~\bibnamefont
  {D\'ora}}, \bibinfo {author} {\bibfnamefont {K.}~\bibnamefont {Ziegler}},
  \bibinfo {author} {\bibfnamefont {P.}~\bibnamefont {Thalmeier}},\ and\
  \bibinfo {author} {\bibfnamefont {M.}~\bibnamefont {Nakamura}},\ }\bibfield
  {title} {\bibinfo {title} {Rabi oscillations in {L}andau-quantized
  graphene},\ }\href {https://doi.org/10.1103/PhysRevLett.102.036803}
  {\bibfield  {journal} {\bibinfo  {journal} {Phys. Rev. Lett.}\ }\textbf
  {\bibinfo {volume} {102}},\ \bibinfo {pages} {036803} (\bibinfo {year}
  {2009})}\BibitemShut {NoStop}%
\bibitem [{\citenamefont {Sukumar}\ and\ \citenamefont
  {Buck}(1981)}]{SUKUMAR_pla_1981}%
  \BibitemOpen
  \bibfield  {author} {\bibinfo {author} {\bibfnamefont {C.}~\bibnamefont
  {Sukumar}}\ and\ \bibinfo {author} {\bibfnamefont {B.}~\bibnamefont {Buck}},\
  }\bibfield  {title} {\bibinfo {title} {Multi-phonon generalisation of the
  {J}aynes-{C}ummings model},\ }\href
  {https://doi.org/https://doi.org/10.1016/0375-9601(81)90825-2} {\bibfield
  {journal} {\bibinfo  {journal} {Phys. Lett. A}\ }\textbf {\bibinfo {volume}
  {83}},\ \bibinfo {pages} {211} (\bibinfo {year} {1981})}\BibitemShut
  {NoStop}%
\bibitem [{\citenamefont {Singh}(1982)}]{surendra_pra_1982}%
  \BibitemOpen
  \bibfield  {author} {\bibinfo {author} {\bibfnamefont {S.}~\bibnamefont
  {Singh}},\ }\bibfield  {title} {\bibinfo {title} {Field statistics in some
  generalized {J}aynes-{C}ummings models},\ }\href
  {https://doi.org/10.1103/PhysRevA.25.3206} {\bibfield  {journal} {\bibinfo
  {journal} {Phys. Rev. A}\ }\textbf {\bibinfo {volume} {25}},\ \bibinfo
  {pages} {3206} (\bibinfo {year} {1982})}\BibitemShut {NoStop}%
\bibitem [{\citenamefont {Shumovsky}\ \emph {et~al.}(1987)\citenamefont
  {Shumovsky}, \citenamefont {Kien},\ and\ \citenamefont
  {Aliskenderov}}]{SHUMOVSKY_pla_1987}%
  \BibitemOpen
  \bibfield  {author} {\bibinfo {author} {\bibfnamefont {A.}~\bibnamefont
  {Shumovsky}}, \bibinfo {author} {\bibfnamefont {F.~L.}\ \bibnamefont
  {Kien}},\ and\ \bibinfo {author} {\bibfnamefont {E.}~\bibnamefont
  {Aliskenderov}},\ }\bibfield  {title} {\bibinfo {title} {Squeezing in the
  multiphoton {J}aynes-{C}ummings model},\ }\href
  {https://doi.org/https://doi.org/10.1016/0375-9601(87)90025-9} {\bibfield
  {journal} {\bibinfo  {journal} {Phys. Lett. A}\ }\textbf {\bibinfo {volume}
  {124}},\ \bibinfo {pages} {351} (\bibinfo {year} {1987})}\BibitemShut
  {NoStop}%
\bibitem [{\citenamefont {Kien}\ \emph {et~al.}(1988)\citenamefont {Kien},
  \citenamefont {Kozierowski},\ and\ \citenamefont {Quang}}]{Kien_PRA_1988}%
  \BibitemOpen
  \bibfield  {author} {\bibinfo {author} {\bibfnamefont {F.~L.}\ \bibnamefont
  {Kien}}, \bibinfo {author} {\bibfnamefont {M.}~\bibnamefont {Kozierowski}},\
  and\ \bibinfo {author} {\bibfnamefont {T.}~\bibnamefont {Quang}},\ }\bibfield
   {title} {\bibinfo {title} {Fourth-order squeezing in the multiphoton
  {J}aynes-{C}ummings model},\ }\href {https://doi.org/10.1103/PhysRevA.38.263}
  {\bibfield  {journal} {\bibinfo  {journal} {Phys. Rev. A}\ }\textbf {\bibinfo
  {volume} {38}},\ \bibinfo {pages} {263} (\bibinfo {year} {1988})}\BibitemShut
  {NoStop}%
\bibitem [{\citenamefont {Huai-xin}\ and\ \citenamefont
  {Xiao-qin}(2000)}]{LuHuai_Chin_phys_2000}%
  \BibitemOpen
  \bibfield  {author} {\bibinfo {author} {\bibfnamefont {L.}~\bibnamefont
  {Huai-xin}}\ and\ \bibinfo {author} {\bibfnamefont {W.}~\bibnamefont
  {Xiao-qin}},\ }\bibfield  {title} {\bibinfo {title} {Multiphoton
  {J}aynes-{C}ummings model solved via supersymmetric unitary transformation},\
  }\href {https://doi.org/10.1088/1009-1963/9/8/003} {\bibfield  {journal}
  {\bibinfo  {journal} {Chin. Phys.}\ }\textbf {\bibinfo {volume} {9}},\
  \bibinfo {pages} {568} (\bibinfo {year} {2000})}\BibitemShut {NoStop}%
\bibitem [{\citenamefont {El-Orany}\ and\ \citenamefont
  {Obada}(2003)}]{El_Orany_job_2003}%
  \BibitemOpen
  \bibfield  {author} {\bibinfo {author} {\bibfnamefont {F.~A.~A.}\
  \bibnamefont {El-Orany}}\ and\ \bibinfo {author} {\bibfnamefont {A.-S.}\
  \bibnamefont {Obada}},\ }\bibfield  {title} {\bibinfo {title} {On the
  evolution of superposition of squeezed displaced number states with the
  multiphoton {J}aynes–{C}ummings model},\ }\href
  {https://doi.org/10.1088/1464-4266/5/1/309} {\bibfield  {journal} {\bibinfo
  {journal} {J. Opt. B Quantum Semiclass. Opt.}\ }\textbf {\bibinfo {volume}
  {5}},\ \bibinfo {pages} {60} (\bibinfo {year} {2003})}\BibitemShut {NoStop}%
\bibitem [{\citenamefont {El-Orany}(2004)}]{El_Orany_jpa_2004}%
  \BibitemOpen
  \bibfield  {author} {\bibinfo {author} {\bibfnamefont {F.~A.~A.}\
  \bibnamefont {El-Orany}},\ }\bibfield  {title} {\bibinfo {title} {The
  revival-collapse phenomenon in the fluctuations of quadrature field
  components of the multiphoton {J}aynes–{C}ummings model},\ }\href
  {https://doi.org/10.1088/0305-4470/37/38/007} {\bibfield  {journal} {\bibinfo
   {journal} {J. Phys. A Math. Theor.}\ }\textbf {\bibinfo {volume} {37}},\
  \bibinfo {pages} {9023} (\bibinfo {year} {2004})}\BibitemShut {NoStop}%
\bibitem [{\citenamefont {Villas-Boas}\ and\ \citenamefont
  {Rossatto}(2019)}]{Villas_PRL_2019}%
  \BibitemOpen
  \bibfield  {author} {\bibinfo {author} {\bibfnamefont {C.~J.}\ \bibnamefont
  {Villas-Boas}}\ and\ \bibinfo {author} {\bibfnamefont {D.~Z.}\ \bibnamefont
  {Rossatto}},\ }\bibfield  {title} {\bibinfo {title} {Multiphoton
  {J}aynes-{C}ummings model: Arbitrary rotations in fock space and quantum
  filters},\ }\href {https://doi.org/10.1103/PhysRevLett.122.123604} {\bibfield
   {journal} {\bibinfo  {journal} {Phys. Rev. Lett.}\ }\textbf {\bibinfo
  {volume} {122}},\ \bibinfo {pages} {123604} (\bibinfo {year}
  {2019})}\BibitemShut {NoStop}%
\bibitem [{\citenamefont {Maldonado-Villamizar}\ \emph
  {et~al.}(2021)\citenamefont {Maldonado-Villamizar}, \citenamefont
  {Gonz{\'a}lez-Guti{\'e}rrez}, \citenamefont {Villanueva-Vergara},\ and\
  \citenamefont {Rodr{\'\i}guez-Lara}}]{Rodriguez_sci_rep_2021}%
  \BibitemOpen
  \bibfield  {author} {\bibinfo {author} {\bibfnamefont {F.~H.}\ \bibnamefont
  {Maldonado-Villamizar}}, \bibinfo {author} {\bibfnamefont {C.~A.}\
  \bibnamefont {Gonz{\'a}lez-Guti{\'e}rrez}}, \bibinfo {author} {\bibfnamefont
  {L.}~\bibnamefont {Villanueva-Vergara}},\ and\ \bibinfo {author}
  {\bibfnamefont {B.~M.}\ \bibnamefont {Rodr{\'\i}guez-Lara}},\ }\bibfield
  {title} {\bibinfo {title} {Underlying {SUSY} in a generalized
  {J}aynes--{C}ummings model},\ }\href
  {https://doi.org/10.1038/s41598-021-95259-1} {\bibfield  {journal} {\bibinfo
  {journal} {Sci. Rep.}\ }\textbf {\bibinfo {volume} {11}},\ \bibinfo {pages}
  {16467} (\bibinfo {year} {2021})}\BibitemShut {NoStop}%
\bibitem [{\citenamefont {Dutra}\ \emph {et~al.}(1993)\citenamefont {Dutra},
  \citenamefont {Knight},\ and\ \citenamefont {Moya-Cessa}}]{Dutra_PRA_1993}%
  \BibitemOpen
  \bibfield  {author} {\bibinfo {author} {\bibfnamefont {S.~M.}\ \bibnamefont
  {Dutra}}, \bibinfo {author} {\bibfnamefont {P.~L.}\ \bibnamefont {Knight}},\
  and\ \bibinfo {author} {\bibfnamefont {H.}~\bibnamefont {Moya-Cessa}},\
  }\bibfield  {title} {\bibinfo {title} {Discriminating field mixtures from
  macroscopic superpositions},\ }\href
  {https://doi.org/10.1103/PhysRevA.48.3168} {\bibfield  {journal} {\bibinfo
  {journal} {Phys. Rev. A}\ }\textbf {\bibinfo {volume} {48}},\ \bibinfo
  {pages} {3168} (\bibinfo {year} {1993})}\BibitemShut {NoStop}%
\bibitem [{\citenamefont {Dutra}\ and\ \citenamefont
  {Knight}(1994)}]{Dutra_PRA_1994}%
  \BibitemOpen
  \bibfield  {author} {\bibinfo {author} {\bibfnamefont {S.~M.}\ \bibnamefont
  {Dutra}}\ and\ \bibinfo {author} {\bibfnamefont {P.~L.}\ \bibnamefont
  {Knight}},\ }\bibfield  {title} {\bibinfo {title} {Atomic probe for quantum
  states of the electromagnetic field},\ }\href
  {https://doi.org/10.1103/PhysRevA.49.1506} {\bibfield  {journal} {\bibinfo
  {journal} {Phys. Rev. A}\ }\textbf {\bibinfo {volume} {49}},\ \bibinfo
  {pages} {1506} (\bibinfo {year} {1994})}\BibitemShut {NoStop}%
\bibitem [{\citenamefont {Benivegna}\ and\ \citenamefont
  {Messina}(1994)}]{Benivegna_JPhysA_1994}%
  \BibitemOpen
  \bibfield  {author} {\bibinfo {author} {\bibfnamefont {G.}~\bibnamefont
  {Benivegna}}\ and\ \bibinfo {author} {\bibfnamefont {A.}~\bibnamefont
  {Messina}},\ }\bibfield  {title} {\bibinfo {title} {Canonical decoupling of
  different degrees of freedom for a two-level system coupled to phonons},\
  }\href {https://doi.org/10.1088/0305-4470/27/12/010} {\bibfield  {journal}
  {\bibinfo  {journal} {J. Phys. A: Math. Gen.}\ }\textbf {\bibinfo {volume}
  {27}},\ \bibinfo {pages} {L453} (\bibinfo {year} {1994})}\BibitemShut
  {NoStop}%
\bibitem [{\citenamefont {Abdalla}\ \emph {et~al.}(2002)\citenamefont
  {Abdalla}, \citenamefont {Abdel-Aty},\ and\ \citenamefont
  {Obada}}]{Abdalla_opt_commun_2002}%
  \BibitemOpen
  \bibfield  {author} {\bibinfo {author} {\bibfnamefont {M.}~\bibnamefont
  {Abdalla}}, \bibinfo {author} {\bibfnamefont {M.}~\bibnamefont {Abdel-Aty}},\
  and\ \bibinfo {author} {\bibfnamefont {A.-S.}\ \bibnamefont {Obada}},\
  }\bibfield  {title} {\bibinfo {title} {Quantum entropy of isotropic coupled
  oscillators interacting with a single atom},\ }\href
  {https://doi.org/https://doi.org/10.1016/S0030-4018(02)01854-0} {\bibfield
  {journal} {\bibinfo  {journal} {Opt. Commun.}\ }\textbf {\bibinfo {volume}
  {211}},\ \bibinfo {pages} {225} (\bibinfo {year} {2002})}\BibitemShut
  {NoStop}%
\bibitem [{\citenamefont {Messina}\ \emph {et~al.}(2003)\citenamefont
  {Messina}, \citenamefont {Maniscalco},\ and\ \citenamefont
  {Napoli}}]{Messina_j_mod_opt_2003}%
  \BibitemOpen
  \bibfield  {author} {\bibinfo {author} {\bibfnamefont {A.}~\bibnamefont
  {Messina}}, \bibinfo {author} {\bibfnamefont {S.}~\bibnamefont
  {Maniscalco}},\ and\ \bibinfo {author} {\bibfnamefont {A.}~\bibnamefont
  {Napoli}},\ }\bibfield  {title} {\bibinfo {title} {Interaction of bimodal
  fields with few-level atoms in cavities and traps},\ }\href
  {https://doi.org/10.1080/09500340308234530} {\bibfield  {journal} {\bibinfo
  {journal} {J. Mod. Opt.}\ }\textbf {\bibinfo {volume} {50}},\ \bibinfo
  {pages} {1} (\bibinfo {year} {2003})}\BibitemShut {NoStop}%
\bibitem [{\citenamefont {Wildfeuer}\ and\ \citenamefont
  {Schiller}(2003)}]{Wildfeuer_PRA_2003}%
  \BibitemOpen
  \bibfield  {author} {\bibinfo {author} {\bibfnamefont {C.}~\bibnamefont
  {Wildfeuer}}\ and\ \bibinfo {author} {\bibfnamefont {D.~H.}\ \bibnamefont
  {Schiller}},\ }\bibfield  {title} {\bibinfo {title} {Generation of entangled
  {$N$}-photon states in a two-mode {J}aynes-{C}ummings model},\ }\href
  {https://doi.org/10.1103/PhysRevA.67.053801} {\bibfield  {journal} {\bibinfo
  {journal} {Phys. Rev. A}\ }\textbf {\bibinfo {volume} {67}},\ \bibinfo
  {pages} {053801} (\bibinfo {year} {2003})}\BibitemShut {NoStop}%
\bibitem [{\citenamefont {Larson}(2006)}]{larson_jmo_2006}%
  \BibitemOpen
  \bibfield  {author} {\bibinfo {author} {\bibfnamefont {J.}~\bibnamefont
  {Larson}},\ }\bibfield  {title} {\bibinfo {title} {Scheme for generating
  entangled states of two field modes in a cavity},\ }\href
  {https://doi.org/10.1080/09500340600674291} {\bibfield  {journal} {\bibinfo
  {journal} {J. Mod. Opt.}\ }\textbf {\bibinfo {volume} {53}},\ \bibinfo
  {pages} {1867} (\bibinfo {year} {2006})}\BibitemShut {NoStop}%
\bibitem [{\citenamefont {Strauch}\ \emph {et~al.}(2010)\citenamefont
  {Strauch}, \citenamefont {Jacobs},\ and\ \citenamefont
  {Simmonds}}]{Strauch_PRL_2010}%
  \BibitemOpen
  \bibfield  {author} {\bibinfo {author} {\bibfnamefont {F.~W.}\ \bibnamefont
  {Strauch}}, \bibinfo {author} {\bibfnamefont {K.}~\bibnamefont {Jacobs}},\
  and\ \bibinfo {author} {\bibfnamefont {R.~W.}\ \bibnamefont {Simmonds}},\
  }\bibfield  {title} {\bibinfo {title} {Arbitrary control of entanglement
  between two superconducting resonators},\ }\href
  {https://doi.org/10.1103/PhysRevLett.105.050501} {\bibfield  {journal}
  {\bibinfo  {journal} {Phys. Rev. Lett.}\ }\textbf {\bibinfo {volume} {105}},\
  \bibinfo {pages} {050501} (\bibinfo {year} {2010})}\BibitemShut {NoStop}%
\bibitem [{\citenamefont {Li}\ \emph {et~al.}(2012)\citenamefont {Li},
  \citenamefont {Gao},\ and\ \citenamefont {Li}}]{Li_PRA_2012}%
  \BibitemOpen
  \bibfield  {author} {\bibinfo {author} {\bibfnamefont {P.-B.}\ \bibnamefont
  {Li}}, \bibinfo {author} {\bibfnamefont {S.-Y.}\ \bibnamefont {Gao}},\ and\
  \bibinfo {author} {\bibfnamefont {F.-L.}\ \bibnamefont {Li}},\ }\bibfield
  {title} {\bibinfo {title} {Engineering two-mode continuous-variable entangled
  states of distant atomic spin ensembles with superconducting quantum
  circuits},\ }\href {https://doi.org/10.1103/PhysRevA.85.014303} {\bibfield
  {journal} {\bibinfo  {journal} {Phys. Rev. A}\ }\textbf {\bibinfo {volume}
  {85}},\ \bibinfo {pages} {014303} (\bibinfo {year} {2012})}\BibitemShut
  {NoStop}%
\bibitem [{\citenamefont {Ma}\ \emph {et~al.}(2014)\citenamefont {Ma},
  \citenamefont {Li}, \citenamefont {Fang}, \citenamefont {Li}, \citenamefont
  {Gao},\ and\ \citenamefont {Li}}]{Ma_PRA_2014}%
  \BibitemOpen
  \bibfield  {author} {\bibinfo {author} {\bibfnamefont {S.-L.}\ \bibnamefont
  {Ma}}, \bibinfo {author} {\bibfnamefont {Z.}~\bibnamefont {Li}}, \bibinfo
  {author} {\bibfnamefont {A.-P.}\ \bibnamefont {Fang}}, \bibinfo {author}
  {\bibfnamefont {P.-B.}\ \bibnamefont {Li}}, \bibinfo {author} {\bibfnamefont
  {S.-Y.}\ \bibnamefont {Gao}},\ and\ \bibinfo {author} {\bibfnamefont {F.-L.}\
  \bibnamefont {Li}},\ }\bibfield  {title} {\bibinfo {title} {Controllable
  generation of two-mode-entangled states in two-resonator circuit {QED} with a
  single gap-tunable superconducting qubit},\ }\href
  {https://doi.org/10.1103/PhysRevA.90.062342} {\bibfield  {journal} {\bibinfo
  {journal} {Phys. Rev. A}\ }\textbf {\bibinfo {volume} {90}},\ \bibinfo
  {pages} {062342} (\bibinfo {year} {2014})}\BibitemShut {NoStop}%
\bibitem [{\citenamefont {Alderete}\ and\ \citenamefont
  {Rodríguez-Lara}(2016)}]{Rodriguez_jphysa_2016}%
  \BibitemOpen
  \bibfield  {author} {\bibinfo {author} {\bibfnamefont {C.~H.}\ \bibnamefont
  {Alderete}}\ and\ \bibinfo {author} {\bibfnamefont {B.~M.}\ \bibnamefont
  {Rodríguez-Lara}},\ }\bibfield  {title} {\bibinfo {title} {Cross-cavity
  quantum {R}abi model},\ }\href
  {https://doi.org/10.1088/1751-8113/49/41/414001} {\bibfield  {journal}
  {\bibinfo  {journal} {J. Phys. A: Math. Theor.}\ }\textbf {\bibinfo {volume}
  {49}},\ \bibinfo {pages} {414001} (\bibinfo {year} {2016})}\BibitemShut
  {NoStop}%
\bibitem [{\citenamefont {Huerta~Alderete}\ and\ \citenamefont
  {Rodr{\'\i}guez-Lara}(2018)}]{Rodriguez_sci_rep_2018}%
  \BibitemOpen
  \bibfield  {author} {\bibinfo {author} {\bibfnamefont {C.}~\bibnamefont
  {Huerta~Alderete}}\ and\ \bibinfo {author} {\bibfnamefont {B.~M.}\
  \bibnamefont {Rodr{\'\i}guez-Lara}},\ }\bibfield  {title} {\bibinfo {title}
  {Simulating para-{F}ermi oscillators},\ }\href
  {https://doi.org/10.1038/s41598-018-29771-2} {\bibfield  {journal} {\bibinfo
  {journal} {Sci. Rep.}\ }\textbf {\bibinfo {volume} {8}},\ \bibinfo {pages}
  {11572} (\bibinfo {year} {2018})}\BibitemShut {NoStop}%
\bibitem [{\citenamefont {Alderete}\ \emph {et~al.}(2021)\citenamefont
  {Alderete}, \citenamefont {Green}, \citenamefont {Nguyen}, \citenamefont
  {Zhu}, \citenamefont {Rodríguez-Lara},\ and\ \citenamefont
  {Linke}}]{Alderete_2021}%
  \BibitemOpen
  \bibfield  {author} {\bibinfo {author} {\bibfnamefont {C.~H.}\ \bibnamefont
  {Alderete}}, \bibinfo {author} {\bibfnamefont {A.~M.}\ \bibnamefont {Green}},
  \bibinfo {author} {\bibfnamefont {N.~H.}\ \bibnamefont {Nguyen}}, \bibinfo
  {author} {\bibfnamefont {Y.}~\bibnamefont {Zhu}}, \bibinfo {author}
  {\bibfnamefont {B.~M.}\ \bibnamefont {Rodríguez-Lara}},\ and\ \bibinfo
  {author} {\bibfnamefont {N.~M.}\ \bibnamefont {Linke}},\ }\href@noop {}
  {\bibinfo {title} {Experimental realization of para-particle oscillators}}
  (\bibinfo {year} {2021}),\ \Eprint {https://arxiv.org/abs/2108.05471}
  {arXiv:2108.05471 [quant-ph]} \BibitemShut {NoStop}%
\bibitem [{\citenamefont {Laha}\ \emph {et~al.}(2022)\citenamefont {Laha},
  \citenamefont {Slodi\v{c}ka}, \citenamefont {Moore},\ and\ \citenamefont
  {Filip}}]{laha_thermally_2022}%
  \BibitemOpen
  \bibfield  {author} {\bibinfo {author} {\bibfnamefont {P.}~\bibnamefont
  {Laha}}, \bibinfo {author} {\bibfnamefont {L.}~\bibnamefont {Slodi\v{c}ka}},
  \bibinfo {author} {\bibfnamefont {D.~W.}\ \bibnamefont {Moore}},\ and\
  \bibinfo {author} {\bibfnamefont {R.}~\bibnamefont {Filip}},\ }\bibfield
  {title} {\bibinfo {title} {Thermally induced entanglement of atomic
  oscillators},\ }\href {https://doi.org/10.1364/OE.449811} {\bibfield
  {journal} {\bibinfo  {journal} {Opt. Express}\ }\textbf {\bibinfo {volume}
  {30}},\ \bibinfo {pages} {8814} (\bibinfo {year} {2022})}\BibitemShut
  {NoStop}%
\bibitem [{\citenamefont {Kuzmich}\ and\ \citenamefont
  {Polzik}(2000)}]{Kuzmich_PRL_2000}%
  \BibitemOpen
  \bibfield  {author} {\bibinfo {author} {\bibfnamefont {A.}~\bibnamefont
  {Kuzmich}}\ and\ \bibinfo {author} {\bibfnamefont {E.~S.}\ \bibnamefont
  {Polzik}},\ }\bibfield  {title} {\bibinfo {title} {Atomic quantum state
  teleportation and swapping},\ }\href
  {https://doi.org/10.1103/PhysRevLett.85.5639} {\bibfield  {journal} {\bibinfo
   {journal} {Phys. Rev. Lett.}\ }\textbf {\bibinfo {volume} {85}},\ \bibinfo
  {pages} {5639} (\bibinfo {year} {2000})}\BibitemShut {NoStop}%
\bibitem [{\citenamefont {Wang}\ and\ \citenamefont
  {Clerk}(2012)}]{Wang_PRL_2012}%
  \BibitemOpen
  \bibfield  {author} {\bibinfo {author} {\bibfnamefont {Y.-D.}\ \bibnamefont
  {Wang}}\ and\ \bibinfo {author} {\bibfnamefont {A.~A.}\ \bibnamefont
  {Clerk}},\ }\bibfield  {title} {\bibinfo {title} {Using interference for high
  fidelity quantum state transfer in optomechanics},\ }\href
  {https://doi.org/10.1103/PhysRevLett.108.153603} {\bibfield  {journal}
  {\bibinfo  {journal} {Phys. Rev. Lett.}\ }\textbf {\bibinfo {volume} {108}},\
  \bibinfo {pages} {153603} (\bibinfo {year} {2012})}\BibitemShut {NoStop}%
\bibitem [{\citenamefont {Palomaki}\ \emph {et~al.}(2013)\citenamefont
  {Palomaki}, \citenamefont {Harlow}, \citenamefont {Teufel}, \citenamefont
  {Simmonds},\ and\ \citenamefont {Lehnert}}]{Palomaki_nature_2013}%
  \BibitemOpen
  \bibfield  {author} {\bibinfo {author} {\bibfnamefont {T.~A.}\ \bibnamefont
  {Palomaki}}, \bibinfo {author} {\bibfnamefont {J.~W.}\ \bibnamefont
  {Harlow}}, \bibinfo {author} {\bibfnamefont {J.~D.}\ \bibnamefont {Teufel}},
  \bibinfo {author} {\bibfnamefont {R.~W.}\ \bibnamefont {Simmonds}},\ and\
  \bibinfo {author} {\bibfnamefont {K.~W.}\ \bibnamefont {Lehnert}},\
  }\bibfield  {title} {\bibinfo {title} {Coherent state transfer between
  itinerant microwave fields and a mechanical oscillator},\ }\href
  {https://doi.org/10.1038/nature11915} {\bibfield  {journal} {\bibinfo
  {journal} {Nature}\ }\textbf {\bibinfo {volume} {495}},\ \bibinfo {pages}
  {210} (\bibinfo {year} {2013})}\BibitemShut {NoStop}%
\bibitem [{\citenamefont {Takeda}\ \emph {et~al.}(2015)\citenamefont {Takeda},
  \citenamefont {Fuwa}, \citenamefont {van Loock},\ and\ \citenamefont
  {Furusawa}}]{Takeda_PRL_2015}%
  \BibitemOpen
  \bibfield  {author} {\bibinfo {author} {\bibfnamefont {S.}~\bibnamefont
  {Takeda}}, \bibinfo {author} {\bibfnamefont {M.}~\bibnamefont {Fuwa}},
  \bibinfo {author} {\bibfnamefont {P.}~\bibnamefont {van Loock}},\ and\
  \bibinfo {author} {\bibfnamefont {A.}~\bibnamefont {Furusawa}},\ }\bibfield
  {title} {\bibinfo {title} {Entanglement swapping between discrete and
  continuous variables},\ }\href
  {https://doi.org/10.1103/PhysRevLett.114.100501} {\bibfield  {journal}
  {\bibinfo  {journal} {Phys. Rev. Lett.}\ }\textbf {\bibinfo {volume} {114}},\
  \bibinfo {pages} {100501} (\bibinfo {year} {2015})}\BibitemShut {NoStop}%
\bibitem [{\citenamefont {Maleki}\ and\ \citenamefont
  {Zheltikov}(2021)}]{Maleki_optics_commun_2021}%
  \BibitemOpen
  \bibfield  {author} {\bibinfo {author} {\bibfnamefont {Y.}~\bibnamefont
  {Maleki}}\ and\ \bibinfo {author} {\bibfnamefont {A.~M.}\ \bibnamefont
  {Zheltikov}},\ }\bibfield  {title} {\bibinfo {title} {Perfect swap and
  transfer of arbitrary quantum states},\ }\href
  {https://doi.org/https://doi.org/10.1016/j.optcom.2021.126870} {\bibfield
  {journal} {\bibinfo  {journal} {Opt. Commun.}\ }\textbf {\bibinfo {volume}
  {496}},\ \bibinfo {pages} {126870} (\bibinfo {year} {2021})}\BibitemShut
  {NoStop}%
\bibitem [{\citenamefont {Matsukevich}\ and\ \citenamefont
  {Kuzmich}(2004)}]{Matsukevich_science_2004}%
  \BibitemOpen
  \bibfield  {author} {\bibinfo {author} {\bibfnamefont {D.~N.}\ \bibnamefont
  {Matsukevich}}\ and\ \bibinfo {author} {\bibfnamefont {A.}~\bibnamefont
  {Kuzmich}},\ }\bibfield  {title} {\bibinfo {title} {Quantum state transfer
  between matter and light},\ }\href {https://doi.org/10.1126/science.1103346}
  {\bibfield  {journal} {\bibinfo  {journal} {Science}\ }\textbf {\bibinfo
  {volume} {306}},\ \bibinfo {pages} {663} (\bibinfo {year}
  {2004})}\BibitemShut {NoStop}%
\bibitem [{\citenamefont {Northup}\ and\ \citenamefont
  {Blatt}(2014)}]{Northup_nature_2014}%
  \BibitemOpen
  \bibfield  {author} {\bibinfo {author} {\bibfnamefont {T.~E.}\ \bibnamefont
  {Northup}}\ and\ \bibinfo {author} {\bibfnamefont {R.}~\bibnamefont
  {Blatt}},\ }\bibfield  {title} {\bibinfo {title} {Quantum information
  transfer using photons},\ }\href {https://doi.org/10.1038/nphoton.2014.53}
  {\bibfield  {journal} {\bibinfo  {journal} {Nat. Photonics}\ }\textbf
  {\bibinfo {volume} {8}},\ \bibinfo {pages} {356} (\bibinfo {year}
  {2014})}\BibitemShut {NoStop}%
\bibitem [{\citenamefont {Kurz}\ \emph {et~al.}(2014)\citenamefont {Kurz},
  \citenamefont {Schug}, \citenamefont {Eich}, \citenamefont {Huwer},
  \citenamefont {M{\"u}ller},\ and\ \citenamefont
  {Eschner}}]{Kurz_nat_commun_2014}%
  \BibitemOpen
  \bibfield  {author} {\bibinfo {author} {\bibfnamefont {C.}~\bibnamefont
  {Kurz}}, \bibinfo {author} {\bibfnamefont {M.}~\bibnamefont {Schug}},
  \bibinfo {author} {\bibfnamefont {P.}~\bibnamefont {Eich}}, \bibinfo {author}
  {\bibfnamefont {J.}~\bibnamefont {Huwer}}, \bibinfo {author} {\bibfnamefont
  {P.}~\bibnamefont {M{\"u}ller}},\ and\ \bibinfo {author} {\bibfnamefont
  {J.}~\bibnamefont {Eschner}},\ }\bibfield  {title} {\bibinfo {title}
  {Experimental protocol for high-fidelity heralded photon-to-atom quantum
  state transfer},\ }\href {https://doi.org/10.1038/ncomms6527} {\bibfield
  {journal} {\bibinfo  {journal} {Nat. Commun.}\ }\textbf {\bibinfo {volume}
  {5}},\ \bibinfo {pages} {5527} (\bibinfo {year} {2014})}\BibitemShut
  {NoStop}%
\bibitem [{\citenamefont {Kurpiers}\ \emph {et~al.}(2018)\citenamefont
  {Kurpiers}, \citenamefont {Magnard}, \citenamefont {Walter}, \citenamefont
  {Royer}, \citenamefont {Pechal}, \citenamefont {Heinsoo}, \citenamefont
  {Salath{\'e}}, \citenamefont {Akin}, \citenamefont {Storz}, \citenamefont
  {Besse}, \citenamefont {Gasparinetti}, \citenamefont {Blais},\ and\
  \citenamefont {Wallraff}}]{Kurpiers_nature_2018}%
  \BibitemOpen
  \bibfield  {author} {\bibinfo {author} {\bibfnamefont {P.}~\bibnamefont
  {Kurpiers}}, \bibinfo {author} {\bibfnamefont {P.}~\bibnamefont {Magnard}},
  \bibinfo {author} {\bibfnamefont {T.}~\bibnamefont {Walter}}, \bibinfo
  {author} {\bibfnamefont {B.}~\bibnamefont {Royer}}, \bibinfo {author}
  {\bibfnamefont {M.}~\bibnamefont {Pechal}}, \bibinfo {author} {\bibfnamefont
  {J.}~\bibnamefont {Heinsoo}}, \bibinfo {author} {\bibfnamefont
  {Y.}~\bibnamefont {Salath{\'e}}}, \bibinfo {author} {\bibfnamefont
  {A.}~\bibnamefont {Akin}}, \bibinfo {author} {\bibfnamefont {S.}~\bibnamefont
  {Storz}}, \bibinfo {author} {\bibfnamefont {J.~C.}\ \bibnamefont {Besse}},
  \bibinfo {author} {\bibfnamefont {S.}~\bibnamefont {Gasparinetti}}, \bibinfo
  {author} {\bibfnamefont {A.}~\bibnamefont {Blais}},\ and\ \bibinfo {author}
  {\bibfnamefont {A.}~\bibnamefont {Wallraff}},\ }\bibfield  {title} {\bibinfo
  {title} {Deterministic quantum state transfer and remote entanglement using
  microwave photons},\ }\href {https://doi.org/10.1038/s41586-018-0195-y}
  {\bibfield  {journal} {\bibinfo  {journal} {Nature}\ }\textbf {\bibinfo
  {volume} {558}},\ \bibinfo {pages} {264} (\bibinfo {year}
  {2018})}\BibitemShut {NoStop}%
\bibitem [{\citenamefont {Li}\ \emph {et~al.}(2018)\citenamefont {Li},
  \citenamefont {Ma}, \citenamefont {Han}, \citenamefont {Chen}, \citenamefont
  {Xu}, \citenamefont {Cai}, \citenamefont {Wang}, \citenamefont {Song},
  \citenamefont {Xue}, \citenamefont {Yin},\ and\ \citenamefont
  {Sun}}]{Li_PRApplied_2018}%
  \BibitemOpen
  \bibfield  {author} {\bibinfo {author} {\bibfnamefont {X.}~\bibnamefont
  {Li}}, \bibinfo {author} {\bibfnamefont {Y.}~\bibnamefont {Ma}}, \bibinfo
  {author} {\bibfnamefont {J.}~\bibnamefont {Han}}, \bibinfo {author}
  {\bibfnamefont {T.}~\bibnamefont {Chen}}, \bibinfo {author} {\bibfnamefont
  {Y.}~\bibnamefont {Xu}}, \bibinfo {author} {\bibfnamefont {W.}~\bibnamefont
  {Cai}}, \bibinfo {author} {\bibfnamefont {H.}~\bibnamefont {Wang}}, \bibinfo
  {author} {\bibfnamefont {Y.}~\bibnamefont {Song}}, \bibinfo {author}
  {\bibfnamefont {Z.-Y.}\ \bibnamefont {Xue}}, \bibinfo {author} {\bibfnamefont
  {Z.-q.}\ \bibnamefont {Yin}},\ and\ \bibinfo {author} {\bibfnamefont
  {L.}~\bibnamefont {Sun}},\ }\bibfield  {title} {\bibinfo {title} {Perfect
  quantum state transfer in a superconducting qubit chain with parametrically
  tunable couplings},\ }\href
  {https://doi.org/10.1103/PhysRevApplied.10.054009} {\bibfield  {journal}
  {\bibinfo  {journal} {Phys. Rev. Appl.}\ }\textbf {\bibinfo {volume} {10}},\
  \bibinfo {pages} {054009} (\bibinfo {year} {2018})}\BibitemShut {NoStop}%
\bibitem [{\citenamefont {Liu}\ \emph {et~al.}(2023)\citenamefont {Liu},
  \citenamefont {Liu},\ and\ \citenamefont {Xue}}]{Liu_jetp_lett_2023}%
  \BibitemOpen
  \bibfield  {author} {\bibinfo {author} {\bibfnamefont {X.~Q.}\ \bibnamefont
  {Liu}}, \bibinfo {author} {\bibfnamefont {J.}~\bibnamefont {Liu}},\ and\
  \bibinfo {author} {\bibfnamefont {Z.~Y.}\ \bibnamefont {Xue}},\ }\bibfield
  {title} {\bibinfo {title} {Robust and fast quantum state transfer on
  superconducting circuits},\ }\href
  {https://doi.org/10.1134/S0021364023601057} {\bibfield  {journal} {\bibinfo
  {journal} {JETP Lett.}\ }\textbf {\bibinfo {volume} {117}},\ \bibinfo {pages}
  {859} (\bibinfo {year} {2023})}\BibitemShut {NoStop}%
\bibitem [{\citenamefont {Weaver}\ \emph {et~al.}(2017)\citenamefont {Weaver},
  \citenamefont {Buters}, \citenamefont {Luna}, \citenamefont {Eerkens},
  \citenamefont {Heeck}, \citenamefont {de~Man},\ and\ \citenamefont
  {Bouwmeester}}]{Weaver_nature_commun_2017}%
  \BibitemOpen
  \bibfield  {author} {\bibinfo {author} {\bibfnamefont {M.~J.}\ \bibnamefont
  {Weaver}}, \bibinfo {author} {\bibfnamefont {F.}~\bibnamefont {Buters}},
  \bibinfo {author} {\bibfnamefont {F.}~\bibnamefont {Luna}}, \bibinfo {author}
  {\bibfnamefont {H.}~\bibnamefont {Eerkens}}, \bibinfo {author} {\bibfnamefont
  {K.}~\bibnamefont {Heeck}}, \bibinfo {author} {\bibfnamefont
  {S.}~\bibnamefont {de~Man}},\ and\ \bibinfo {author} {\bibfnamefont
  {D.}~\bibnamefont {Bouwmeester}},\ }\bibfield  {title} {\bibinfo {title}
  {Coherent optomechanical state transfer between disparate mechanical
  resonators},\ }\href {https://doi.org/10.1038/s41467-017-00968-9} {\bibfield
  {journal} {\bibinfo  {journal} {Nat. Commun.}\ }\textbf {\bibinfo {volume}
  {8}},\ \bibinfo {pages} {824} (\bibinfo {year} {2017})}\BibitemShut {NoStop}%
\bibitem [{\citenamefont {Ventura-Vel{\'a}zquez}\ \emph
  {et~al.}(2019)\citenamefont {Ventura-Vel{\'a}zquez}, \citenamefont
  {Jaramillo~{\'A}vila}, \citenamefont {Kyoseva},\ and\ \citenamefont
  {Rodr{\'\i}guez-Lara}}]{Ventura_sci_rep_2019}%
  \BibitemOpen
  \bibfield  {author} {\bibinfo {author} {\bibfnamefont {C.}~\bibnamefont
  {Ventura-Vel{\'a}zquez}}, \bibinfo {author} {\bibfnamefont {B.}~\bibnamefont
  {Jaramillo~{\'A}vila}}, \bibinfo {author} {\bibfnamefont {E.}~\bibnamefont
  {Kyoseva}},\ and\ \bibinfo {author} {\bibfnamefont {B.~M.}\ \bibnamefont
  {Rodr{\'\i}guez-Lara}},\ }\bibfield  {title} {\bibinfo {title} {Robust
  optomechanical state transfer under composite phase driving},\ }\href
  {https://doi.org/10.1038/s41598-019-40492-y} {\bibfield  {journal} {\bibinfo
  {journal} {Sci. Rep.}\ }\textbf {\bibinfo {volume} {9}},\ \bibinfo {pages}
  {4382} (\bibinfo {year} {2019})}\BibitemShut {NoStop}%
\bibitem [{\citenamefont {Qi}\ \emph {et~al.}(2020)\citenamefont {Qi},
  \citenamefont {Wang}, \citenamefont {Liu}, \citenamefont {Zhang},\ and\
  \citenamefont {Wang}}]{Qi_opt_lett_2020}%
  \BibitemOpen
  \bibfield  {author} {\bibinfo {author} {\bibfnamefont {L.}~\bibnamefont
  {Qi}}, \bibinfo {author} {\bibfnamefont {G.-L.}\ \bibnamefont {Wang}},
  \bibinfo {author} {\bibfnamefont {S.}~\bibnamefont {Liu}}, \bibinfo {author}
  {\bibfnamefont {S.}~\bibnamefont {Zhang}},\ and\ \bibinfo {author}
  {\bibfnamefont {H.-F.}\ \bibnamefont {Wang}},\ }\bibfield  {title} {\bibinfo
  {title} {Controllable photonic and phononic topological state transfers in a
  small optomechanical lattice},\ }\href {https://doi.org/10.1364/OL.388835}
  {\bibfield  {journal} {\bibinfo  {journal} {Opt. Lett.}\ }\textbf {\bibinfo
  {volume} {45}},\ \bibinfo {pages} {2018} (\bibinfo {year}
  {2020})}\BibitemShut {NoStop}%
\bibitem [{\citenamefont {Lei}\ \emph {et~al.}(2023)\citenamefont {Lei},
  \citenamefont {Wang}, \citenamefont {Li}, \citenamefont {Peng},\ and\
  \citenamefont {Xiong}}]{Lei_appl_phys_b_2023}%
  \BibitemOpen
  \bibfield  {author} {\bibinfo {author} {\bibfnamefont {S.}~\bibnamefont
  {Lei}}, \bibinfo {author} {\bibfnamefont {X.}~\bibnamefont {Wang}}, \bibinfo
  {author} {\bibfnamefont {H.}~\bibnamefont {Li}}, \bibinfo {author}
  {\bibfnamefont {R.}~\bibnamefont {Peng}},\ and\ \bibinfo {author}
  {\bibfnamefont {B.}~\bibnamefont {Xiong}},\ }\bibfield  {title} {\bibinfo
  {title} {High-fidelity and robust optomechanical state transfer based on
  pulse control},\ }\href {https://doi.org/10.1007/s00340-023-08135-3}
  {\bibfield  {journal} {\bibinfo  {journal} {Applied Physics B}\ }\textbf
  {\bibinfo {volume} {129}},\ \bibinfo {pages} {193} (\bibinfo {year}
  {2023})}\BibitemShut {NoStop}%
\bibitem [{\citenamefont {Sun}\ \emph {et~al.}(2006)\citenamefont {Sun},
  \citenamefont {Wei}, \citenamefont {Liu},\ and\ \citenamefont
  {Nori}}]{Sun_PRA_2006}%
  \BibitemOpen
  \bibfield  {author} {\bibinfo {author} {\bibfnamefont {C.~P.}\ \bibnamefont
  {Sun}}, \bibinfo {author} {\bibfnamefont {L.~F.}\ \bibnamefont {Wei}},
  \bibinfo {author} {\bibfnamefont {Y.-x.}\ \bibnamefont {Liu}},\ and\ \bibinfo
  {author} {\bibfnamefont {F.}~\bibnamefont {Nori}},\ }\bibfield  {title}
  {\bibinfo {title} {Quantum transducers: Integrating transmission lines and
  nanomechanical resonators via charge qubits},\ }\href
  {https://doi.org/10.1103/PhysRevA.73.022318} {\bibfield  {journal} {\bibinfo
  {journal} {Phys. Rev. A}\ }\textbf {\bibinfo {volume} {73}},\ \bibinfo
  {pages} {022318} (\bibinfo {year} {2006})}\BibitemShut {NoStop}%
\bibitem [{\citenamefont {Mei}\ \emph {et~al.}(2018)\citenamefont {Mei},
  \citenamefont {Chen}, \citenamefont {Tian}, \citenamefont {Zhu},\ and\
  \citenamefont {Jia}}]{Mei_PRA_2018}%
  \BibitemOpen
  \bibfield  {author} {\bibinfo {author} {\bibfnamefont {F.}~\bibnamefont
  {Mei}}, \bibinfo {author} {\bibfnamefont {G.}~\bibnamefont {Chen}}, \bibinfo
  {author} {\bibfnamefont {L.}~\bibnamefont {Tian}}, \bibinfo {author}
  {\bibfnamefont {S.-L.}\ \bibnamefont {Zhu}},\ and\ \bibinfo {author}
  {\bibfnamefont {S.}~\bibnamefont {Jia}},\ }\bibfield  {title} {\bibinfo
  {title} {Robust quantum state transfer via topological edge states in
  superconducting qubit chains},\ }\href
  {https://doi.org/10.1103/PhysRevA.98.012331} {\bibfield  {journal} {\bibinfo
  {journal} {Phys. Rev. A}\ }\textbf {\bibinfo {volume} {98}},\ \bibinfo
  {pages} {012331} (\bibinfo {year} {2018})}\BibitemShut {NoStop}%
\bibitem [{\citenamefont {Laha}\ \emph
  {et~al.}(2024{\natexlab{a}})\citenamefont {Laha}, \citenamefont {Yasir},\
  and\ \citenamefont {van Loock}}]{laha_spinboson_2024}%
  \BibitemOpen
  \bibfield  {author} {\bibinfo {author} {\bibfnamefont {P.}~\bibnamefont
  {Laha}}, \bibinfo {author} {\bibfnamefont {P.~A.~A.}\ \bibnamefont {Yasir}},\
  and\ \bibinfo {author} {\bibfnamefont {P.}~\bibnamefont {van Loock}},\
  }\href@noop {} {\bibinfo {title} {Genuine non-gaussian entanglement of light
  and quantum coherence for an atom from noisy multiphoton spin-boson
  interactions}} (\bibinfo {year} {2024}{\natexlab{a}}),\ \Eprint
  {https://arxiv.org/abs/2403.10207} {arXiv:2403.10207 [quant-ph]} \BibitemShut
  {NoStop}%
\bibitem [{\citenamefont {Kenfack}\ and\ \citenamefont
  {Życzkowski}(2004)}]{Anatole_joptb_2004}%
  \BibitemOpen
  \bibfield  {author} {\bibinfo {author} {\bibfnamefont {A.}~\bibnamefont
  {Kenfack}}\ and\ \bibinfo {author} {\bibfnamefont {K.}~\bibnamefont
  {Życzkowski}},\ }\bibfield  {title} {\bibinfo {title} {Negativity of the
  wigner function as an indicator of non-classicality},\ }\href
  {https://doi.org/10.1088/1464-4266/6/10/003} {\bibfield  {journal} {\bibinfo
  {journal} {J. Opt. B: Quantum Semiclass. Opt.}\ }\textbf {\bibinfo {volume}
  {6}},\ \bibinfo {pages} {396} (\bibinfo {year} {2004})}\BibitemShut {NoStop}%
\bibitem [{\citenamefont {Arkhipov}\ \emph {et~al.}(2018)\citenamefont
  {Arkhipov}, \citenamefont {Barasi{\'n}ski},\ and\ \citenamefont
  {Svozil{\'\i}k}}]{Arkhipov_sci_rep_2018}%
  \BibitemOpen
  \bibfield  {author} {\bibinfo {author} {\bibfnamefont {I.~I.}\ \bibnamefont
  {Arkhipov}}, \bibinfo {author} {\bibfnamefont {A.}~\bibnamefont
  {Barasi{\'n}ski}},\ and\ \bibinfo {author} {\bibfnamefont {J.}~\bibnamefont
  {Svozil{\'\i}k}},\ }\bibfield  {title} {\bibinfo {title} {Negativity volume
  of the generalized wigner function as an entanglement witness for hybrid
  bipartite states},\ }\href {https://doi.org/10.1038/s41598-018-35330-6}
  {\bibfield  {journal} {\bibinfo  {journal} {Sci. Rep.}\ }\textbf {\bibinfo
  {volume} {8}},\ \bibinfo {pages} {16955} (\bibinfo {year}
  {2018})}\BibitemShut {NoStop}%
\bibitem [{\citenamefont {Rosiek}\ \emph {et~al.}(2024)\citenamefont {Rosiek},
  \citenamefont {Rossi}, \citenamefont {Schliesser},\ and\ \citenamefont
  {S\o{}rensen}}]{rosiek_arxiv_2023}%
  \BibitemOpen
  \bibfield  {author} {\bibinfo {author} {\bibfnamefont {C.~A.}\ \bibnamefont
  {Rosiek}}, \bibinfo {author} {\bibfnamefont {M.}~\bibnamefont {Rossi}},
  \bibinfo {author} {\bibfnamefont {A.}~\bibnamefont {Schliesser}},\ and\
  \bibinfo {author} {\bibfnamefont {A.~S.}\ \bibnamefont {S\o{}rensen}},\
  }\bibfield  {title} {\bibinfo {title} {Quadrature squeezing enhances {W}igner
  negativity in a mechanical {D}uffing oscillator},\ }\href
  {https://doi.org/10.1103/PRXQuantum.5.030312} {\bibfield  {journal} {\bibinfo
   {journal} {PRX Quantum}\ }\textbf {\bibinfo {volume} {5}},\ \bibinfo {pages}
  {030312} (\bibinfo {year} {2024})}\BibitemShut {NoStop}%
\bibitem [{\citenamefont {Johansson}\ \emph {et~al.}(2013)\citenamefont
  {Johansson}, \citenamefont {Nation},\ and\ \citenamefont {Nori}}]{qutip}%
  \BibitemOpen
  \bibfield  {author} {\bibinfo {author} {\bibfnamefont {J.}~\bibnamefont
  {Johansson}}, \bibinfo {author} {\bibfnamefont {P.}~\bibnamefont {Nation}},\
  and\ \bibinfo {author} {\bibfnamefont {F.}~\bibnamefont {Nori}},\ }\bibfield
  {title} {\bibinfo {title} {Qutip 2: A python framework for the dynamics of
  open quantum systems},\ }\href
  {https://doi.org/https://doi.org/10.1016/j.cpc.2012.11.019} {\bibfield
  {journal} {\bibinfo  {journal} {Comp. Phys. Comm.}\ }\textbf {\bibinfo
  {volume} {184}},\ \bibinfo {pages} {1234} (\bibinfo {year}
  {2013})}\BibitemShut {NoStop}%
\bibitem [{\citenamefont {Laha}\ \emph
  {et~al.}(2024{\natexlab{b}})\citenamefont {Laha}, \citenamefont {Moore},\
  and\ \citenamefont {Filip}}]{laha_non-gaussian_2022}%
  \BibitemOpen
  \bibfield  {author} {\bibinfo {author} {\bibfnamefont {P.}~\bibnamefont
  {Laha}}, \bibinfo {author} {\bibfnamefont {D.~W.}\ \bibnamefont {Moore}},\
  and\ \bibinfo {author} {\bibfnamefont {R.}~\bibnamefont {Filip}},\ }\bibfield
   {title} {\bibinfo {title} {Entanglement growth via splitting of a few
  thermal quanta},\ }\href {https://doi.org/10.1103/PhysRevLett.132.210201}
  {\bibfield  {journal} {\bibinfo  {journal} {Phys. Rev. Lett.}\ }\textbf
  {\bibinfo {volume} {132}},\ \bibinfo {pages} {210201} (\bibinfo {year}
  {2024}{\natexlab{b}})}\BibitemShut {NoStop}%
\bibitem [{\citenamefont {Laha}(2023)}]{laha_josab_2023}%
  \BibitemOpen
  \bibfield  {author} {\bibinfo {author} {\bibfnamefont {P.}~\bibnamefont
  {Laha}},\ }\bibfield  {title} {\bibinfo {title} {Dynamics of a multipartite
  hybrid quantum system with beamsplitter, dipole-dipole, and {I}sing
  interactions},\ }\href {https://doi.org/10.1364/JOSAB.489223} {\bibfield
  {journal} {\bibinfo  {journal} {J. Opt. Soc. Am. B}\ }\textbf {\bibinfo
  {volume} {40}},\ \bibinfo {pages} {1911} (\bibinfo {year}
  {2023})}\BibitemShut {NoStop}%
\bibitem [{\citenamefont {Pan}\ \emph {et~al.}(2012)\citenamefont {Pan},
  \citenamefont {Chen}, \citenamefont {Lu}, \citenamefont {Weinfurter},
  \citenamefont {Zeilinger},\ and\ \citenamefont {\ifmmode~\dot{Z}\else
  \.{Z}\fi{}ukowski}}]{Pan_RMP_2012}%
  \BibitemOpen
  \bibfield  {author} {\bibinfo {author} {\bibfnamefont {J.-W.}\ \bibnamefont
  {Pan}}, \bibinfo {author} {\bibfnamefont {Z.-B.}\ \bibnamefont {Chen}},
  \bibinfo {author} {\bibfnamefont {C.-Y.}\ \bibnamefont {Lu}}, \bibinfo
  {author} {\bibfnamefont {H.}~\bibnamefont {Weinfurter}}, \bibinfo {author}
  {\bibfnamefont {A.}~\bibnamefont {Zeilinger}},\ and\ \bibinfo {author}
  {\bibfnamefont {M.}~\bibnamefont {\ifmmode~\dot{Z}\else \.{Z}\fi{}ukowski}},\
  }\bibfield  {title} {\bibinfo {title} {Multiphoton entanglement and
  interferometry},\ }\href {https://doi.org/10.1103/RevModPhys.84.777}
  {\bibfield  {journal} {\bibinfo  {journal} {Rev. Mod. Phys.}\ }\textbf
  {\bibinfo {volume} {84}},\ \bibinfo {pages} {777} (\bibinfo {year}
  {2012})}\BibitemShut {NoStop}%
\bibitem [{\citenamefont {Zhang}\ \emph {et~al.}(2021)\citenamefont {Zhang},
  \citenamefont {Huang}, \citenamefont {Liu}, \citenamefont {Li},\ and\
  \citenamefont {Guo}}]{Zhang_adv_q_tech_2021}%
  \BibitemOpen
  \bibfield  {author} {\bibinfo {author} {\bibfnamefont {C.}~\bibnamefont
  {Zhang}}, \bibinfo {author} {\bibfnamefont {Y.-F.}\ \bibnamefont {Huang}},
  \bibinfo {author} {\bibfnamefont {B.-H.}\ \bibnamefont {Liu}}, \bibinfo
  {author} {\bibfnamefont {C.-F.}\ \bibnamefont {Li}},\ and\ \bibinfo {author}
  {\bibfnamefont {G.-C.}\ \bibnamefont {Guo}},\ }\bibfield  {title} {\bibinfo
  {title} {Spontaneous parametric down-conversion sources for multiphoton
  experiments},\ }\href
  {https://doi.org/https://doi.org/10.1002/qute.202000132} {\bibfield
  {journal} {\bibinfo  {journal} {Adv. Quantum Technol.}\ }\textbf {\bibinfo
  {volume} {4}},\ \bibinfo {pages} {2000132} (\bibinfo {year}
  {2021})}\BibitemShut {NoStop}%
\bibitem [{\citenamefont {Yang}\ \emph {et~al.}(2022)\citenamefont {Yang},
  \citenamefont {Yu}, \citenamefont {Li}, \citenamefont {Jing}, \citenamefont
  {Bao},\ and\ \citenamefont {Pan}}]{Yang_nature_photon_2022}%
  \BibitemOpen
  \bibfield  {author} {\bibinfo {author} {\bibfnamefont {C.-W.}\ \bibnamefont
  {Yang}}, \bibinfo {author} {\bibfnamefont {Y.}~\bibnamefont {Yu}}, \bibinfo
  {author} {\bibfnamefont {J.}~\bibnamefont {Li}}, \bibinfo {author}
  {\bibfnamefont {B.}~\bibnamefont {Jing}}, \bibinfo {author} {\bibfnamefont
  {X.-H.}\ \bibnamefont {Bao}},\ and\ \bibinfo {author} {\bibfnamefont {J.-W.}\
  \bibnamefont {Pan}},\ }\bibfield  {title} {\bibinfo {title} {Sequential
  generation of multiphoton entanglement with a {R}ydberg superatom},\ }\href
  {https://doi.org/10.1038/s41566-022-01054-3} {\bibfield  {journal} {\bibinfo
  {journal} {Nat. Photonics}\ }\textbf {\bibinfo {volume} {16}},\ \bibinfo
  {pages} {658} (\bibinfo {year} {2022})}\BibitemShut {NoStop}%
\bibitem [{\citenamefont {Gerry}\ and\ \citenamefont
  {Knight}(2004)}]{gerry_knight_2004}%
  \BibitemOpen
  \bibfield  {author} {\bibinfo {author} {\bibfnamefont {C.}~\bibnamefont
  {Gerry}}\ and\ \bibinfo {author} {\bibfnamefont {P.}~\bibnamefont {Knight}},\
  }\href {https://doi.org/10.1017/CBO9780511791239} {\emph {\bibinfo {title}
  {Introductory Quantum Optics}}}\ (\bibinfo  {publisher} {Cambridge University
  Press},\ \bibinfo {year} {2004})\BibitemShut {NoStop}%
\bibitem [{\citenamefont {Cahill}\ and\ \citenamefont
  {Glauber}(1969{\natexlab{a}})}]{Cahill1_PhysRev_1969}%
  \BibitemOpen
  \bibfield  {author} {\bibinfo {author} {\bibfnamefont {K.~E.}\ \bibnamefont
  {Cahill}}\ and\ \bibinfo {author} {\bibfnamefont {R.~J.}\ \bibnamefont
  {Glauber}},\ }\bibfield  {title} {\bibinfo {title} {Ordered expansions in
  boson amplitude operators},\ }\href
  {https://doi.org/10.1103/PhysRev.177.1857} {\bibfield  {journal} {\bibinfo
  {journal} {Phys. Rev.}\ }\textbf {\bibinfo {volume} {177}},\ \bibinfo {pages}
  {1857} (\bibinfo {year} {1969}{\natexlab{a}})}\BibitemShut {NoStop}%
\bibitem [{\citenamefont {Cahill}\ and\ \citenamefont
  {Glauber}(1969{\natexlab{b}})}]{Cahill2_PhysRev_1969}%
  \BibitemOpen
  \bibfield  {author} {\bibinfo {author} {\bibfnamefont {K.~E.}\ \bibnamefont
  {Cahill}}\ and\ \bibinfo {author} {\bibfnamefont {R.~J.}\ \bibnamefont
  {Glauber}},\ }\bibfield  {title} {\bibinfo {title} {Density operators and
  quasiprobability distributions},\ }\href
  {https://doi.org/10.1103/PhysRev.177.1882} {\bibfield  {journal} {\bibinfo
  {journal} {Phys. Rev.}\ }\textbf {\bibinfo {volume} {177}},\ \bibinfo {pages}
  {1882} (\bibinfo {year} {1969}{\natexlab{b}})}\BibitemShut {NoStop}%
\bibitem [{\citenamefont {Campos}\ \emph {et~al.}(1989)\citenamefont {Campos},
  \citenamefont {Saleh},\ and\ \citenamefont {Teich}}]{campos89}%
  \BibitemOpen
  \bibfield  {author} {\bibinfo {author} {\bibfnamefont {R.~A.}\ \bibnamefont
  {Campos}}, \bibinfo {author} {\bibfnamefont {B.~E.}\ \bibnamefont {Saleh}},\
  and\ \bibinfo {author} {\bibfnamefont {M.~C.}\ \bibnamefont {Teich}},\
  }\bibfield  {title} {\bibinfo {title} {Quantum-mechanical lossless beam
  splitter: {SU} (2) symmetry and photon statistics},\ }\href
  {https://doi.org/https://doi.org/10.1103/PhysRevA.40.1371} {\bibfield
  {journal} {\bibinfo  {journal} {Phys. Rev. A}\ }\textbf {\bibinfo {volume}
  {40}},\ \bibinfo {pages} {1371} (\bibinfo {year} {1989})}\BibitemShut
  {NoStop}%
\end{thebibliography}%

\end{document}